\documentclass[aps,pra,twocolumn,showpacs,floatfix,longbibliography,superscriptaddress]{revtex4-2}
\usepackage{graphicx,amsfonts,amssymb,amsmath}
\usepackage{textcomp} 
\usepackage{esint}
\usepackage[english]{babel}
\usepackage{afterpage}
\usepackage{bbold}
\usepackage{soul}
\usepackage[colorlinks,linktocpage,citecolor=blue,linkcolor=red,urlcolor=blue]{hyperref}
\def\Zrpa{Z_{\rm RPA}}
\def\Gamrpa{\Gamma_{\rm RPA}}
\def\Gamloc{\Gamma_{\rm loc}}
\def\Gloc{G_{\rm loc}}
\def\Zqp{Z^\alpha_{\rm QP}}
\def\Zqpp{Z^+_{\rm QP}}
\def\Zqpm{Z^-_{\rm QP}}

\def\Pmi{P_{\rm MI}}

\def\Psing{P_{\rm sing}}
\def\nksing{n^{\rm sing}_{\bf k}}

\def\nkmi{n_\k^{\rm MI}}

\def\nmi{n_{\rm MI}}
\def\mlat{m_{\rm lat}} 
\def\alat{a_{\rm lat}} 
\def\cuni{C_{\rm univ}}
\def\csd{C_{\rm univ}^{\rm sd}}
\def\cfull{C}
\def\Vloc{V_{\rm loc}}

\begin{document}
    
\graphicspath{{figures_submit/}}
	
\allowdisplaybreaks
	

\newcommand{\oldnew}[2]{\marginpar{\scriptsize \textcolor{red}{correction}}{\textcolor{red}{#2}}}
\newcommand{\suppressed}[1]{\marginpar{\scriptsize \textcolor{red}{correction}}{\textcolor{red}{\st{#1}}}}
\newcommand{\correction}[1]{\marginpar{\textcolor{red}{\scriptsize #1}}}
\newcommand{\marge}[1]{\marginpar{\scriptsize #1}}
\newcommand{\remarque}[1]{\marginpar{\scriptsize Remarque}{\it [#1]}}

%

\def\rhoeq{\hat\rho_{\rm eq}}

\newcommand{\new}[1]{{\bf #1}}
\newlength{\textlarg}
\newcommand{\redbar}[1]{\textcolor{red}{\st{#1}}} 
\newcommand{\bluebar}[1]{\textcolor{blue}{\st{#1}}} 

\newcommand{\normord}[1]{:\mathrel{#1}:}

\newcommand{\beq}{\begin{equation}}
\newcommand{\eeq}{\end{equation}}
\newcommand{\bfig}{\begin{figure}}
\newcommand{\efig}{\end{figure}}
\newcommand{\bline}{\begin{multline}}
\newcommand{\eline}{\end{multline}}
\newcommand{\bremark}{\begin{quotation} \noindent \small }
\newcommand{\eremark}{\end{quotation}}
\newcommand{\llbrace}{\left\lbrace}  
\newcommand{\rrbrace}{\right\rbrace}
\newcommand{\lbraket}{\left[}
\newcommand{\rbraket}{\right]}
\newcommand{\llangle}{\left\langle}
\newcommand{\rrangle}{\right\rangle} 

\newcommand{\Tr}{{\rm Tr}} 
\newcommand{\tr}{{\rm tr}} 
\newcommand{\sgn}{\,{\rm sgn}} 
\newcommand{\mean}[1]{\langle #1 \rangle}
\newcommand{\commu}[2]{[#1,#2]} 
\newcommand{\bra}[1]{\langle#1|}
\newcommand{\ket}[1]{|#1\rangle}
\newcommand{\braket}[2]{\langle #1|#2\rangle}
\newcommand{\ketbra}[2]{|#1\rangle\langle#2|}
\newcommand{\dbraket}[3]{\langle #1|#2|#3\rangle}
\newcommand{\tens}[1]{\overleftrightarrow{#1}}  
\newcommand{\vac}{|{\rm vac}\rangle} 
\newcommand{\bravac}{\langle{\rm vac}|}
\newcommand{\const}{{\rm const}} 
\newcommand{\unif}{{\rm unif.}} 
\newcommand{\atanh}{\,{\rm atanh}}
\newcommand{\cotanh}{\,{\rm cotanh}}

\newcommand{\ie}{i.e.\xspace}
\newcommand{\iet}{i.e.}
\newcommand{\eg}{e.g.\xspace}
\newcommand{\cc}{{\rm c.c.}} 
\newcommand{\hc}{{\rm H.c.}} 
\newcommand{\etal}{{\it et al. }}
\newcommand\eme{$^{\mbox{\footnotesize ème}}$\xspace}

\newcommand{\jhatbf}{\hat {\textbf \jold}} 
\newcommand{\Jhatbf}{\hat {\textbf \J}} 
\newcommand{\jhat}{\hat {\jmath}} 
\newcommand{\Jhat}{\hat {J}} 
\newcommand{\jbf}{\textbf j}
\newcommand{\Jbf}{\textbf J}

\def\chibf{\boldsymbol{\chi}}
\def\down{\downarrow}
\def\eps{\epsilon}
\def\gam{\gamma} 
\def\alphabf{\boldsymbol{\alpha}}
\def\phibf{\boldsymbol{\phi}}
\def\varphibf{\boldsymbol{\varphi}}
\def\varphibfs{\boldsymbol{\varphi}_<}
\def\varphibfl{\boldsymbol{\varphi}_>}
\def\varphis{\varphi_{<}}
\def\varphil{\varphi_{>}}
\def\psibf{\boldsymbol{\psi}}
\def\thetabf{\boldsymbol{\theta}}
\def\Ome{\Omega}
\def\omeD{{\omega_D}} 
\def\bfOme{\boldsymbol{\Omega}} 
\def\Omebf{\boldsymbol{\Omega}} 
\def\lamb{\lambda}
\def\Lamb{\Lambda}
\def\sig{\sigma}
\def\Sig{\Sigma}
\def\sigp{{\sigma'}} 
\def\bfsig{\boldsymbol{\sigma}} 
\def\sigbf{\boldsymbol{\sigma}} 
\def\bfSig{\boldsymbol{\Sigma}} 
\def\The{\Theta} 
\def\up{\uparrow}

\def\epsk{\epsilon_{\bf k}} 
\def\epsp{\epsilon_{\bf p}} 
\def\xik{\xi_{\bf k}} 
\def\txik{\tilde\xi_{\bf k}} 
\def\xip{\xi_{\bf p}} 
\def\epsq{\epsilon_{\bf q}} 
\def\xiq{\xi_{\bf q}} 
\def\xikq{\xi_{{\bf k}+{\bf q}}} 
\def\Ek{E_{\bf k}} 
\def\Ep{E_{\bf p}}
\def\Eq{E_{\bf q}}
\def\Heff{\hat H_{\rm eff}}
\def\Hem{\hat H_{\rm em}}
\def\Hint{\hat H_{\rm int}}
\def\Hloc{\hat H_{\rm loc}}
\def\HMF{\hat H_{\rm MF}}
\def\HLL{\hat H_{\rm LL}}
\def\Hdis{\hat H_{\rm dis}}
\def\Sem{S_{\rm em}}
\def\SMF{S_{\rm MF}} 
\def\SHF{S_{\rm HF}} 
\def\SRPA{S_{\rm RPA}} 
\def\Sint{S_{\rm int}} 
\def\Sloc{S_{\rm loc}}
\def\TN{T_{\rm N}} 
\def\TNHF{T^{\rm HF}_{\rm N}} 
\def\Zloc{Z_{\rm loc}} 
\def\ZMF{Z_{\rm MF}} 
\def\ZHF{Z_{\rm HF}} 
\def\ZRPA{Z_{\rm RPA}} 
\def\RPA{{\rm RPA}}
\def\loc{{\rm loc}} 
\def\pp{{\rm pp}}
\def\ph{{\rm ph}} 
\def\ch{{\rm ch}}
\def\sp{{\rm sp}} 
\def\qtf{q_{\rm TF}}
\def\epstf{\eps^{}_{\rm TF}} 
\def\epsrpa{\eps^{}_{\rm RPA}} 
\def\chinnzpp{\chi_{nn}^{0}{}\!\!\!''}
\def\SigHF{\Sigma_{\rm HF}}
\def\psicl{\psi_{\rm cl}} 

\def\half{\frac{1}{2}}
\def\dhalf{\dfrac{1}{2}}
\def\third{\frac{1}{3}} 
\def\quarter{\frac{1}{4}}

\def\qr{{\bf q}\cdot{\bf r}}
\def\wt{\omega t} 

\def\a{{\bf a}}
\def\b{{\bf b}}
\newcommand{\cv}{{\bf c}} 
\def\e{{\bf e}}
\def\f{{\bf f}}
\def\g{{\bf g}}
\def\h{{\bf h}}
\def\jold{\char"11}
\def\j{{\bf j}}
\def\k{{\bf k}}
\def\l{{\bf l}}
\def\ellbf{\bm{\ell}} 
\def\m{{\bf m}}
\def\n{{\bf n}} 
\def\p{{\bf p}} 
\def\q{{\bf q}}
\def\r{{\bf r}}
\def\t{{\bf t}}
\def\u{{\bf u}}
\newcommand{\vv}{{\bf v}}
\def\x{{\bf x}}
\def\y{{\bf y}} 
\def\z{{\bf z}} 
\def\A{{\bf A}}
\def\B{{\bf B}}
\def\D{{\bf D}} 
\def\E{{\bf E}} 
\def\F{{\bf F}} 
\def\H{{\bf H}}  
\def\J{{\bf J}}
\def\K{{\bf K}} 

\def\G{{\bf G}}
\def\L{{\bf L}}
\def\M{{\bf M}}  
\def\O{{\bf O}} 
\def\P{{\bf P}} 
\def\Q{{\bf Q}} 
\def\R{{\bf R}}
\def\S{{\bf S}}
\def\U{{\bf U}} 
\def\V{{\bf V}} 
\def\X{{\bf X}} 
\def\Y{{\bf Y}} 
\def\epsbf{\boldsymbol{\epsilon}}
\def\betabf{\boldsymbol{\beta}}
\def\deltabf{\boldsymbol{\delta}}
\def\mubf{\boldsymbol{\mu}}
\def\nablabf{\boldsymbol{\nabla}}
\def\rhobf{\boldsymbol{\rho}}
\def\sigmabf{\boldsymbol{\sigma}} 
\def\Pibf{\boldsymbol{\Pi}}
\def\pibf{\boldsymbol{\pi}}

\def\para{\parallel}
\def\kpara{{k_\parallel}}
\def\kperp{{k_\perp}} 
\def\kperpp{{k_\perp'}} 
\def\qperp{{q_\perp}} 
\def\tperp{{t_\perp}} 

\def\w{\omega}
\def\wn{\omega_n}
\def\wm{\omega_m}
\def\wnu{\omega_\nu}
\def\wp{\omega_p} 
\def\dmu{{\partial_\mu}}
\def\dnu{{\partial_\nu}}
\def\dl{{\partial_l}}  
\def\dt{\partial_t} 
\def\tdt{\tilde\partial_t}
\def\dk{\partial_k}
\def\tdk{\tilde\partial_k}
\def\dx{\partial_x}
\def\dy{\partial_y} 
\def\dw{\partial_{\w}}
\def\dtau{{\partial_\tau}}  
\def\det{{\rm det}} 
\def\Pf{{\rm Pf}}
\def\diag{{\rm diag}}

\def\dsum{\displaystyle \sum}
\def\dint{\displaystyle \int} 
\def\intt{\int_{-\infty}^\infty dt} 
\def\inttp{\int_{-\infty}^\infty dt'} 
\def\intk{\int_{\bf k}} 
\def\intkd{\int \frac{d^dk}{(2\pi)^d}}
\def\intq{\int_{\bf q}} 
\def\intr{\int d^dr}  
\def\dintr{\displaystyle \int d^dr} 
\def\intrp{\int d^dr'}
\def\dinttau{\displaystyle \int_0^\beta d\tau}
\def\dinttaup{\displaystyle \int_0^\beta d\tau'}
\def\inttau{\int_0^\beta d\tau}
\def\inttaup{\int_0^\beta d\tau'}
\def\intx{\int d^{d+1}x} 
\def\inttaur{\int_0^\beta d\tau \int d^dr}
\def\intinf{\int_{-\infty}^\infty}
\def\dinttaur{\displaystyle \int_0^\beta d\tau \int d^dr}
\def\dintinf{\displaystyle \int_{-\infty}^\infty}
\def\intw{\int_{-\infty}^\infty \frac{d\w}{2\pi}}
\def\sumr{\sum_{\bf r}} 

\def\calA{{\cal A}}
\def\calAbf{\bm{{\cal A}}}
\def\calB{{\cal B}} 
\def\calC{{\cal C}} 
\def\dt{\partial_t}
\def\calD{{\cal D}}
\def\calE{{\cal E}}
\def\calF{{\cal F}} 
\def\calFbf{\bm{{\cal F}}}
\def\calG{{\cal G}}
\def\calH{{\cal H}}
\def\calI{{\cal I}}
\def\calJ{{\cal J}}
\def\calK{{\cal K}}
\def\calL{{\cal L}} 
\def\calM{{\cal M}} 
\def\calN{{\cal N}}
\def\calO{{\cal O}}
\def\calP{{\cal P}}  
\def\calR{{\cal R}} 
\def\calS{{\cal S}}
\def\calT{{\cal T}}
\def\calU{{\cal U}}
\def\calV{{\cal V}}
\def\calX{{\cal X}} 
\def\calY{{\cal Y}} 
\def\calW{{\cal W}} 
\def\calZ{{\cal Z}}

\def\tT{{\tilde T}}
\def\talpha{{\tilde\alpha}}
\def\tbeta{{\tilde\beta}}
\def\tchi{{\tilde\chi}}
\def\tdelta{{\tilde\delta}}
\def\tDelta{{\tilde\Delta}}
\def\teta{{\tilde\eta}} 
\def\tlamb{{\tilde\lambda}}
\def\tmu{{\tilde\mu}}
\def\tphibf{{\tilde\phibf}}
\def\trho{{\tilde\rho}}
\def\tvarphibf{{\tilde\varphibf}} 
\def\tq{\tilde q}
\def\tw{{\tilde\omega}}
\def\twn{{\tilde\omega_n}}
\def\twnu{{\tilde\omega_\nu}}

\def\asinh{{\rm asinh}} 
\def\Tbkt{T_{\rm BKT}}

\title{Strong-coupling RPA theory of a Bose gas near the superfluid--Mott-insulator transition: universal thermodynamics and two-body contact}
	
\author{Nicolas Dupuis}
\affiliation{Sorbonne Universit\'e, CNRS, Laboratoire de Physique Th\'eorique de la Mati\`ere Condens\'ee, LPTMC, F-75005 Paris, France}
	
\author{Moksh Bhateja}
\affiliation{Sorbonne Universit\'e, CNRS, Laboratoire de Physique Th\'eorique de la Mati\`ere Condens\'ee, LPTMC, F-75005 Paris, France}
\affiliation{Univ. Lille, CNRS, UMR 8523 – PhLAM – Laboratoire de Physique des Lasers Atomes et Molécules, F-59000 Lille, France}
	
\author{Adam Ran\c{c}on} 
\affiliation{Univ. Lille, CNRS, UMR 8523 – PhLAM – Laboratoire de Physique des Lasers Atomes et Molécules, F-59000 Lille, France}	
	
\date{September 29, 2025} 
	
\begin{abstract}
We present a strong-coupling expansion of the Bose-Hubbard model based on a mean-field treatment of the hopping term, while onsite fluctuations are taken into account exactly. This random phase approximation (RPA) describes the universal features of the generic Mott-insulator--superfluid transition (induced by a density change) and the superfluid state near the phase transition. The critical quasi-particles at the quantum critical point have a quadratic dispersion with an effective mass $m^*$ and their mutual interaction is described by an effective $s$-wave scattering length $a^*$. The singular part of the pressure takes the same form as in a dilute Bose gas, provided we replace the boson mass $m$ and the scattering length in vacuum $a$ by $m^*$ and $a^*$, and the density $n$ by the excess density $|n-n_{\rm MI}|$ of particles (or holes) with respect to the Mott insulator. We define a ``universal'' two-body contact $C_{\rm univ}$ that controls the high-momentum tail $\sim 1/|\k|^4$ of the singular part $n^{\rm sing}_{\bf k}$ of the momentum distribution. We also apply the strong-coupling RPA to a lattice model of hard-core bosons and find that the high-momentum distribution is controlled by a universal contact, in complete agreement with the Bose-Hubbard model. Finally, we discuss a continuum model of bosons in an optical lattice and define two additional two-body contacts: a short-distance ``universal'' contact $\csd$ which controls the high-momentum tail of $n^{\rm sing}_{\bf k}$ at scales larger than the inverse lattice spacing, and a ``full'' contact $C$ which controls the high-momentum tail of the full momentum distribution $n_{\bf k}$. 
\end{abstract}

\maketitle

\tableofcontents

\section{Introduction} 

The Mott transition is a paradigmatic example of a quantum phase transition induced by strong interactions between particles; it has become central in the field of quantum gases. Cold-atom experiments can implement both the fermionic (metal-insulator) Mott transition and its bosonic analog, by loading bosons into an optical lattice~\cite{Greiner02,Jimenez10,Becker10,Trotzky10,Mark11,Thomas17a,Herce21}.
By varying the strength of the optical lattice potential and/or the density, it is possible to induce a transition from a superfluid (SF) state to a Mott insulator (MI) where the mean number of bosons per site is integer. When the phase transition is induced by a density change, it belongs to the dilute-Bose-gas universality class, i.e., it is similar to the quantum phase transition between the vacuum state and the superfluid state obtained by varying the chemical potential from negative to positive values in a dilute Bose gas. 

The MI-SF transition is often studied in the framework of the Bose-Hubbard model, which describes bosons moving on a lattice with an onsite interaction~\cite{Fisher89,Jaksch98}. The main characteristics of the phase diagram are now well understood from various approaches:  strong-coupling expansion~\cite{Sheshadri93,Oosten01,Sengupta05,Ohashi06,Menotti08,Freericks09,Teichmann09a,Teichmann09b,Wang18,Kuebler19,SantAna19}, Green function method~\cite{Sajna15}, mapping on quantum rotor models~\cite{Polak07,Polak09,Zaleski11,Krzywicka22,Krzywicka24}, slave-boson technique~\cite{Dickerscheid03,Yu05,Huber07,Frerot16a}, time-dependent Gutzwiller approximation~\cite{Krutitsky11,DiLiberto18}, quantum Gutzwiller approach~\cite{Caleffi20}, dynamical mean-field theory~\cite{Byczuk08,Hu09,Anders10,Anders11,Panas15}, variational method~\cite{Capello07}, variational cluster approximation~\cite{Koller06,Knap10,Knap11,Arrigoni11}, nonperturbative functional renormalization group (FRG)~\cite{Rancon11a,Rancon11b,Rancon12a,Rancon12d,Rancon13b}, mapping on the quantum spherical model~\cite{Kopec24} and Monte Carlo simulations~\cite{Krauth91,Capogrosso07,Capogrosso08,Kato09,Pollet12}.

Of particular interest is the behavior of the superfluid phase near the SF-MI transition. The universality class of the phase transition implies that the additional particles (or holes) introduced in the Mott insulator behave as a dilute gas of quasi-particles with an effective mass $m^*$ and an effective $s$-wave scattering length $a^*$~\cite{Rancon12a,Rancon12d}. The singular part of the pressure (i.e. the part that is singular when crossing the transition by varying the chemical potential or the density) can be written in the same scaling form as in the dilute Bose gas ---as obtained, for example, from Bogoliubov's theory--- provided we replace the boson mass and the $s$-wave scattering length by $m^*$ and $a^*$, and the density $n$ by the excess density $|n-\nmi|$ of particles (or holes) with respect to the Mott insulator. The condensate and superfluid densities, expressed as a function of $m^*$, $a^*$ and $|n-\nmi|$, also take the same form as in a dilute Bose gas. 

In this paper, we describe a strong-coupling RPA theory of the Bose-Hubbard model~\cite{Sheshadri93,Oosten01,Sengupta05,Ohashi06,Menotti08,Freericks09,Teichmann09a,Teichmann09b,Wang18,Kuebler19,SantAna19} which captures the universal features of the MI-SF transition as well as the superfluid phase near the transition; it is essentially equivalent to the approach proposed in~\cite{Sengupta05} based on two successive Hubbard-Stratonovich transformations. In Refs.~\cite{Rancon11a,Rancon11b,Rancon12a,Rancon12b,Rancon12d,Rancon12d,Rancon13b}, the strong-coupling RPA theory was used as the initial condition of the flow in the nonperturbative FRG approach. In this paper, we show that many qualitative results can be obtained without integrating the nonperturbative flow equations. Moreover, the strong-coupling RPA allows one to explicitly compute a ``universal'' two-body contact $\cuni$ from the singular part of the pressure~\cite{not7}. We show that $\cuni$, which depends on $|n-\nmi|$ and the effective scattering length $a^*$, determines the high-momentum tail $\cuni/|\k|^4$ of the singular part of the momentum distribution $n_\k$ in the superfluid phase (as discussed in detail in the companion paper~\cite{Bhateja25}), a physical quantity which is not readily available~\cite{not4} from the nonperturbative FRG approach of Refs.~\cite{Rancon11a,Rancon11b,Rancon12a,Rancon12b,Rancon12d,Rancon12d}. We also apply the strong-coupling RPA to a lattice model of hard-core bosons and find that the high-momentum distribution is controlled by a universal contact, in complete agreement with the Bose-Hubbard model. In addition, we consider bosons in an optical lattice described by a continuum model. We argue that at length scales smaller than the optical lattice spacing, the boson system should be seen as a dilute Bose gas. This leads us to define two additional contacts, a short-distance ``universal''  contact $\csd$ and a ``full'' contact $\cfull$, which depend on the scattering length in vacuum $a$. 

The outline of the paper is as follows. The Bose-Hubbard model is discussed in Sec.~\ref{sec_bhm}. We first determine the Gibbs free energy (or effective action in the field-theory terminology) in the strong-coupling RPA, which is based on a mean-field treatment of the hopping term, while local (onsite) fluctuations are taken into account exactly (Sec.~\ref{sec_bhm:subsec_rpa}). We recover the phase diagram obtained in previous mean-field studies~\cite{Sheshadri93,Oosten01,Sengupta05}. We then determine the spectrum of the one-particle excitations in the Mott insulator. The critical excitations at the quantum critical point (QCP) that separates the Mott insulator from the superfluid state are quasi-particles with quadratic dispersion, effective mass $m^*$, and spectral weight $Z_{\rm QP}$. Their mutual interaction is characterized by an effective scattering length $a^*$ (Sec.~\ref{sec_bhm:subsec_mi}). The singular part $\Psing(n-\nmi,m^*,a^*)$ of the pressure in the superfluid state, as well as the condensate and superfluid densities, take the usual Bogoliubov expression. This leads us to define a universal contact $\cuni$ from the derivative of $\Psing$ with respect to $1/a^*$ (Sec.~\ref{sec_bhm:subsec_sf}). In Sec.~\ref{sec_bhm:subsec_spectrum}, we determine the spectrum in the superfluid phase and show that the singular part $\nksing=n_\k-\nkmi$ of the momentum distribution exhibits a high-momentum tail $\cuni/|\k|^4$ over a wide range of momenta in the Brillouin zone provided that the system is near the SF-MI transition ($\nkmi$ denotes the momentum distribution in the Mott insulator). In Sec.~\ref{sec_bhm:subsec_universal}, we discuss the universal properties of the superfluid phase from a broader perspective based on the universality class of the MI-SF transition. This allows us to express the contact in terms of the universal scaling function that determines the pressure and obtain the Lee-Huang-Yang correction which is not included in the strong-coupling RPA. In Sec.~\ref{sec_hard_core} we discuss the strong-coupling RPA in a lattice model of hard-core bosons and recover the momentum distribution obtained from a spin-wave analysis of the equivalent XY model~\cite{Coletta12}. We find that the high-momentum limit exhibits a $1/|\k|^4$ tail whose strength is determined by a universal contact $\cuni$ which can be defined from the pressure. This confirms the result obtained in the Bose-Hubbard model, but without the limitations of the strong-coupling RPA in the latter, due to a slight violation of the sum rule relating the momentum distribution to the density. 
In Sec.~\ref{sec_optical_lattice}, we consider bosons in an optical lattice described by a continuum model and briefly discuss the link to the Bose-Hubbard model. We define a short-distance universal contact $\csd$, which extends the definition of the previously defined contact $\cuni$ to length scales smaller than the lattice spacing, and a full contact $\cfull$ from the full pressure (including both singular and regular parts). Contrary to the contact $\cuni$ considered in the framework of the Bose-Hubbard model, $\csd$ and $\cfull$ are defined from a derivative of the pressure with respect to the inverse of the scattering length $a$ in vacuum. Using the effective description provided by the Bose-Hubbard model to compute the pressure of the boson system in the optical lattice, we obtain the expression of the contacts $\csd$ and $\cfull$ in various cases (low density limit, near the Mott insulator $\nmi=1$, etc.).

\section{Bose-Hubbard model} 
\label{sec_bhm}

The Bose-Hubbard model is defined by the (grand canonical) Hamiltonian 
\beq
\hat H = \sum_{\r,\r'} t_{\r,\r'} \hat\psi^\dagger_\r \hat\psi_{\r'} 
+  \sum_\r \Bigl( - \mu \hat\psi^\dagger_\r \hat\psi_\r + \frac{U}{2} \hat\psi^\dagger_\r \hat\psi^\dagger_\r \hat\psi_\r  \hat\psi_\r \Bigr) ,
\label{hamBHM} 
\eeq
where $\{\r\}$ denotes the $N$ sites of a cubic lattice. We set the lattice spacing $\ell$ to unity (so that we do not distinguish between the total number of sites $N$ and the volume $\calV=N\ell^3$). The hopping matrix is defined by $t_{\r,\r'}=-t$ if $\r$ and $\r'$ are nearest neighbors and $t_{\r,\r'}=0$ otherwise. $U$ is the on-site repulsion between bosons and $\mu$ is the chemical potential. The partition function can be written as a functional integral over a complex field $\psi_\r(\tau)$ with the action
\beq 
S[\psi^*,\psi] =\inttau \sum_{\r,\r'} t_{\r,\r'} \psi^*_\r \psi_{\r'} +  \Sloc[\psi^*,\psi] ,
\label{Sbhm}  
\eeq 
where 
\beq 
\Sloc[\psi^*,\psi] = \inttau \sum_\r \biggl( - \mu \psi^*_\r \psi_\r + \frac{U}{2} |\psi_\r|^4 \biggr) 
\eeq 
is the local part of the action. $\tau$ is an imaginary time and $\beta=1/T\to\infty$ the inverse temperature. We set $\hbar=k_B=1$ throughout.

\subsection{RPA effective action} 
\label{sec_bhm:subsec_rpa} 

In the presence of an external (complex) source $J_\r$, the partition function is given by 
\beq 
Z[J^*,J] = \int \calD[\psi^*,\psi]\, e^{-S[\psi^*,\psi] + \inttau \sum_\r (J^*_\r \psi_\r + \cc) } .
\eeq 
We consider the inter-site hopping term at the mean-field level, i.e. we replace $t_{\r,\r'}\psi^*_\r \psi_{\r'}$ by $t_{\r,\r'}(\psi^*_\r \phi_{\r'}+ \phi^*_\r \psi_{\r'} - \phi^*_\r \phi_{\r'})$ where 
\beq 
\begin{split}
\phi_\r(\tau) &= \mean{\psi_\r(\tau)} = \frac{\delta\ln Z[J^*,J]}{\delta J^* _\r(\tau)} , \\
\phi^*_\r(\tau) &= \mean{\psi^*_\r(\tau)} = \frac{\delta\ln Z[J^*,J]}{\delta J _\r(\tau)}
\end{split} 
\label{phidef} 
\eeq
are the expectation values of the boson field computed in the presence of the external source. This mean-field decoupling of the intersite hopping term is familiar in the context of mean-field studies of the Ising model. More formally, we can decouple the hopping term by means of a Hubbard-Stratonovich transformation and perform a saddle-point approximation on the auxiliary field, but this leads to the same partition function (see Appendix~\ref{app_HST}). We thus obtain the action
\begin{align}
\SRPA[\psi^*,\psi] ={}& \inttau \sum_{\r,\r'} t_{\r,\r'}(\psi^*_\r \phi_{\r'}+ \phi^*_\r \psi_{\r'} - \phi^*_\r \phi_{\r'}) \nonumber \\ & + \Sloc[\psi^*,\psi] 
\label{Srpa} 
\end{align}
and the corresponding partition function 
\begin{align}
Z_{\rm RPA}[J^*,J] &= \int \calD[\psi^*,\psi] \, e^{-\SRPA[\psi^*,\psi] + \inttau \sum_\r (J^*_\r \psi_\r + \cc)  } \nonumber \\ 
&= \Zloc[\tilde J^*,\tilde J] \, e^{\inttau \sum_{\r,\r'} t_{\r,\r'} \phi^*_\r \phi_{\r'} } 
\label{Zrpa} 
\end{align}
where 
\beq 
\tilde J^*_\r = J^*_\r - \sum_{\r'} t_{\r,\r'} \phi^*_{\r'} , \quad 
\tilde J_\r = J_\r - \sum_{\r'} t_{\r,\r'} \phi_{\r'} ,
\eeq 
and $\Zloc[\tilde J^*,\tilde J]$ is the partition function in the local ($t=0$) limit in the presence of the external source $\tilde J_\r$. In the following, we will refer to $\SRPA$ and $\ZRPA$ as the RPA action and RPA partition function, respectively.

The RPA effective action, or Gibbs free energy, is defined as the Legendre transform of $\ln Z_{\rm RPA}[J^*,J]$, 
\begin{align}
\Gamrpa[\phi^*,\phi] ={}& - \ln Z_{\rm RPA}[J^*,J] + \inttau \sum_\r (J_\r^*\phi_\r + \cc) \nonumber\\ 
={}& - \ln \Zloc[\tilde J^*,\tilde J]  
+ \inttau \sum_{\r,\r'} t_{\r,\r'} \phi^*_\r \phi_{\r'} \nonumber\\ 
& + \inttau \sum_\r (\tilde J_\r^*\phi_\r + \cc),
\label{gam1} 
\end{align} 
where the order parameter $\phi_\r^{(*)}$ is defined by~(\ref{phidef}) with the partition function $Z[J^*,J]$ approximated by $\Zrpa[J^*,J]$. Using~(\ref{Zrpa}), we can relate the order parameter to the local partition function~\cite{not3},  
\beq
\begin{split} 
	\phi_\r(\tau) &= \frac{\delta\ln \Zloc[\tilde J^*, \tilde J]}{\delta \tilde J^*_\r(\tau)} , \\ 
	\phi^*_\r(\tau) &= \frac{\delta\ln \Zloc[\tilde J^*, \tilde J]}{\delta \tilde J_\r(\tau)} . 
\end{split}
\label{op} 
\eeq 
From Eqs.~(\ref{gam1}-\ref{op}), we deduce that 
\beq 
\Gamrpa[\phi^*,\phi] = \Gamloc[\phi^*,\phi] + \inttau \sum_{\r,\r'} t_{\r,\r'} \phi^*_\r \phi_{\r'} , 
\label{gam1a} 
\eeq 
where 
\beq 
\Gamloc[\phi^*,\phi] = - \ln \Zloc[\tilde J^*,\tilde J] + \inttau \sum_\r (\tilde J_\r^*\phi_\r + \cc)
\eeq  
is the effective action in the local limit, defined as the Legendre transform of $\ln \Zloc[\tilde J^*,\tilde J]$. 

The state of the system in the absence of an external source is obtained from the equation of state
\beq
\frac{\delta\Gamrpa[\phi^*,\phi]}{\delta\phi_\r(\tau)} =
\frac{\delta\Gamrpa[\phi^*,\phi]}{\delta\phi^*_\r(\tau)} = 0 .
\label{eos} 
\eeq 
The order parameter takes a nonzero value $\phi_\r(\tau)=\phi_0$ in the superfluid state and vanishes in the Mott insulator. The grand potential is given by $\Omega=\Gamrpa[\phi^*_0,\phi_0]/\beta$. In the vicinity of the SF-MI transition, the value of the order parameter is small  and we can expand $\Gamloc[\phi^*,\phi]$ to quartic order, 
\begin{multline}
\Gamloc[\phi^*,\phi] = \Gamloc[0,0] 
+  \sum_\r \biggl\{ \int_{\tau,\tau'}\phi^*_\r(\tau) \Gamma^{(2)}_{\rm loc}(\tau-\tau')   \phi_\r(\tau') \\ + \quarter \int_{\{\tau_i\}} \Gamma_{\rm loc}^{(4)}(\tau_1,\tau_2,\tau_3,\tau_4) 
\phi^*_{\r}(\tau_1) \phi^*_{\r}(\tau_2) \phi_\r(\tau_3) \phi_\r(\tau_4) \biggr\} ,
\label{gamrpa1} 
\end{multline}
where we use the notation $\int_\tau\equiv \inttau$, and 
\beq
\frac{1}{\beta N}\Gamloc[0,0]=-\mu \nmi + \frac{U}{2} \nmi(\nmi-1)
\eeq 
is the grand potential (per site) of the Mott insulator with $\nmi\equiv \nmi(\mu)$ the mean density (i.e. the mean number of bosons per site): $\nmi=0$ if $\mu<0$ and $\nmi-1<\mu/U<\nmi$ if $\mu>0$. The two-point vertex $\Gamma_{\rm loc}^{(2)}=-G_{\rm loc}^{-1}$ is the inverse of the local Green function (in the absence of source),
\begin{align}
G_{\rm loc}(i\wn) &= - \mean{\psi_\r(i\wn)\psi^*_\r(i\wn)}_{\rm loc} \nonumber\\ 
&= \frac{\nmi+1}{i\wn+\mu-U\nmi} - \frac{\nmi}{i\wn+\mu-U(\nmi-1)} ,
\label{Gloc}
\end{align}
expressed here in Fourier space with $\wn=2n\pi T$ ($n\in\mathbb{Z}$) a Matsubara frequency. The four-point vertex 
\beq 
\Gamma^{(4)}_{\rm loc}(i\w_{n_1},i\w_{n_2},i\w_{n_3},i\w_{n_4}) = -\frac{G_{\rm loc}^{(4)}(i\w_{n_1},i\w_{n_2},i\w_{n_3},i\w_{n_4})}{\prod_{j=1}^4 G_{\rm loc}(i\w_{n_j}) }
\eeq 
is related to the two-particle local Green function $G_{\rm loc}^{(4)}$, its expression is given in Appendix~\ref{sec_local}. We approximate the four-point vertex by its static limit $\Gamma_{\rm loc}^{(4)}(\{i\w_{n_j}=0\})=2g$, which leads to 
\begin{multline} 
\Gamrpa[\phi^*,\phi] = \Gamloc[0,0] + \int_{\tau,\tau'} \sum_{\r,\r'} \phi^*_\r(\tau) \bigl[ t_{\r,\r'} \delta(\tau-\tau') \\ - \delta_{\r,\r'} G^{-1}_{\rm loc}(\tau-\tau') \bigr]  \phi_{\r'}(\tau') + \frac{g}{2} \sum_\r \int_{\tau} |\phi_\r|^4 .   
\label{gamrpa2} 
\end{multline}
The RPA effective action~(\ref{gamrpa1},\ref{gamrpa2}) coincides with the effective Wilsonian action obtained in Ref.~\cite{Sengupta05} from two successive Hubbard-Stratonovich transformations~\cite{not1}. When the latter is treated at the mean-field level, or by including Gaussian fluctuations about the saddle-point approximation, it leads to the same results as the RPA effective action.

\subsection{Mott insulator and MI-SF transition}
\label{sec_bhm:subsec_mi} 

Since the order parameter  $\phi_\r$ vanishes in the Mott insulator, the pressure is given by 
\begin{align}
\Pmi &= - \frac{1}{\beta N}\Gamloc[0,0] \nonumber\\ 
&= \mu\nmi- \frac{U}{2} \nmi(\nmi-1) .
\end{align} 
The mean density $n=\dmu\Pmi=\nmi$ does not depend on the chemical potential and the compressibility $\kappa=\dmu n=\partial^2_\mu\Pmi$ vanishes: The system is incompressible. 

\begin{figure}
	\centerline{\includegraphics[width=7cm]{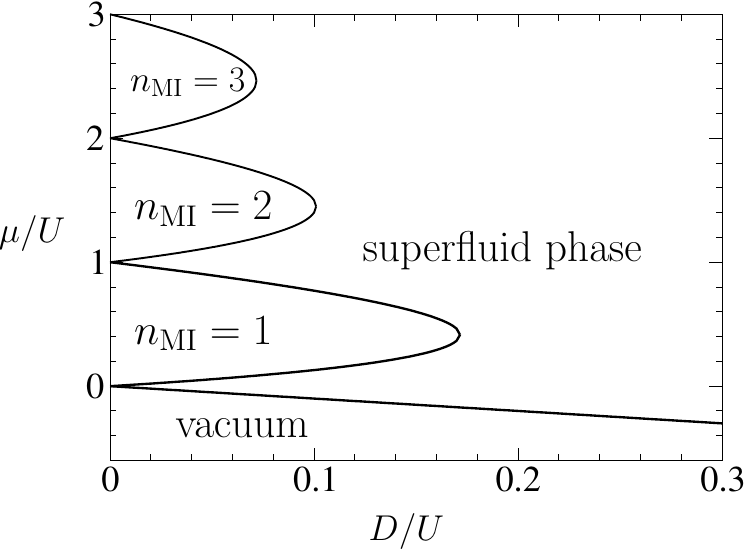}}
	\caption{Phase diagram of the three-dimensional Bose-Hubbard model obtained from the criterion $\Gloc^{-1}(i\wn=0)+D=0$ ($D=-t_{\k=0}=6t$). Each Mott lobe is labeled by the integer $\nmi$ giving the mean number of bosons per site. The trivial Mott insulator $\nmi=0$ corresponds to the vacuum.} 
	\label{fig_phase_dia} 
\end{figure}

The boson propagator is determined by the inverse of the two-point vertex, i.e. 
\begin{align}
G(\k,i\wn) &= - \biggl( \frac{\delta^2 \Gamma[\phi^*,\phi]}{\delta\phi^*_\k(i\wn) \delta\phi_\k(i\wn)} \biggl|_{\phi=\phi^*=0} \biggr)^{-1} \nonumber\\ 
&= \frac{\Gloc(i\wn)}{1-t_\k \Gloc(i\wn)} , 
\label{Gkw} 
\end{align} 
where \beq 
t_\k=-2t(\cos k_x+\cos k_y+\cos k_z)
\eeq 
is the Fourier transform of the hopping matrix $t_{\r,\r'}$ and $\k$ belongs to the Brillouin zone $[-\pi,\pi]^3$. The stability of the Mott insulator requires $-G(0,0)>0$, so the MI-SF transition is obtained from the criterion $\Gloc^{-1}(0)+D=0$ where $D=-t_{\k=0}=6t$. It is convenient to use the notation $\delta\mu=\mu-U(x-1)$ and $x=\nmi+1/2$. When $\nmi\neq 0$, the transition occurs when $\delta\mu=\delta\mu_\pm$ with 
\beq
\delta\mu_\pm = - \frac{D}{2} \pm \half \bigl( D^2 - 4DUx + U^2 \bigr)^{1/2} . 
\label{deltamupm} 
\eeq 
The Mott insulator is stable for $\delta\mu_-\leq\delta\mu\leq\delta\mu_+$ or, equivalently, $\mu_-\leq\mu\leq\mu_+$. For $D=0$, $\mu_+=U\nmi$ and $\mu_-=U(\nmi-1)$. The two solutions $\delta\mu_\pm$ merge when $D=D_c=U[2\nmi+1-2(\nmi^2+\nmi)^{1/2}]$ and are then equal to $\delta\mu_c=-D_c/2$. For $D>D_c$, there is no region of stability for the Mott insulator. Thus, we obtain a series of Mott lobes, labeled by the integer $\nmi$, as shown in Fig.~\ref{fig_phase_dia} and in agreement with previous mean-field studies~\cite{Sheshadri93,Oosten01,Sengupta05}. For $\nmi=0$, the equation $\Gloc^{-1}(0)+D=0$ has a single solution $\mu_+=-D$ that corresponds to the transition between vacuum and superfluid. 

\begin{figure}
	\centerline{\includegraphics[width=4cm]{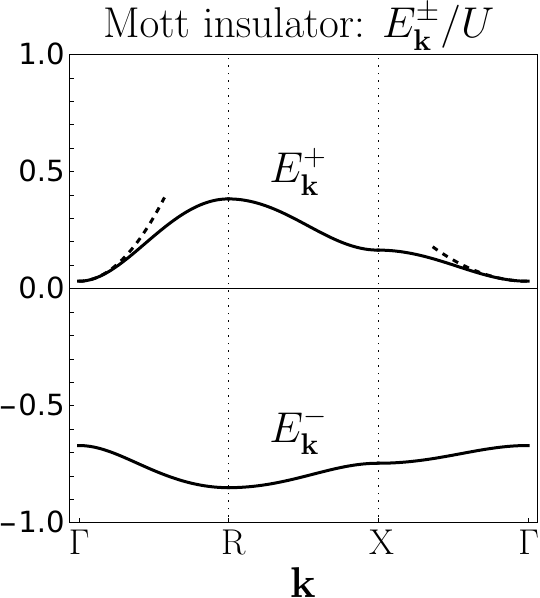}
	\hspace{0.3cm}
	\includegraphics[width=4cm]{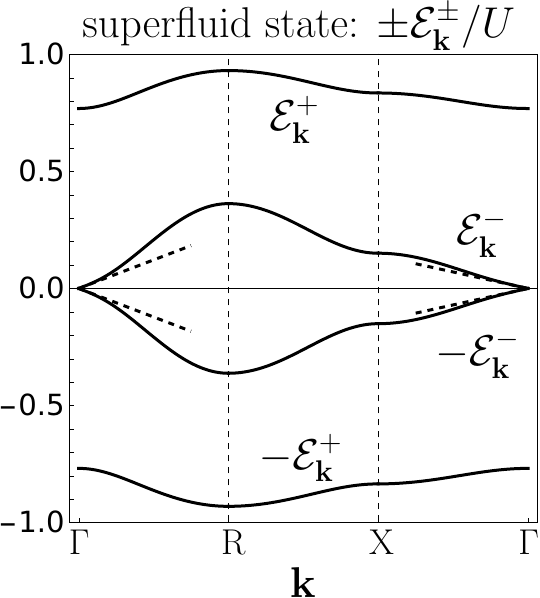}}
	\caption{Excitation energies in the Mott insulator ($\mu=0.9\,\mu_+$, left) and the superfluid state ($\mu=1.1\,\mu_+$, right) for $\nmi=1$. The dashed lines show the approximate low-energy forms, valid near $\k=0$, $E_\k^+= \k^2/2m^*_+ + \mu_+ - \mu$ and $\calE_\k^-=c|\k|$ (with $c$ the sound velocity~(\ref{cdef}) in the superfluid state). $\Gamma=(0,0,0)$, R$=(\pi,\pi,\pi)$ and X$=(\pi,0,0)$.}
	\label{fig_spectrum} 
\end{figure}

The spectrum in the Mott insulator is obtained from the poles of $G(\k,i\wn)$ after analytic continuation $i\wn\to\w+i0^+$ to real frequency. For $\nmi\neq 0$, this gives two bands, 
\beq 
E^\pm_\k = - \delta\mu + \frac{t_\k}{2} \pm \half \bigl(t_\k^2+4Uxt_\k+U^2 \bigr)^{1/2} , 
\eeq 
one with positive energy ($E^+_\k$) and the other with negative energy ($E^-_\k$); see Fig.~\ref{fig_spectrum}. For $t_\k=0$ (i.e. $t=0$), $E^+_\k=-\mu+U\nmi$ and $E^-_\k=-\mu+U(\nmi-1)$; one recovers the poles of the local propagator $\Gloc(\w+i0^+)$ corresponding to particle and hole excitation on an isolated site. The energy $E_\k^+$ is minimum and $E_\k^-$ maximum for $\k=0$. Since $E^\pm_{\k=0}=-\delta\mu+\delta\mu_\pm=-\mu+\mu_\pm$, the transition to the superfluid state occurs when one of the two excitation bands becomes gapless. At the tip of the Mott lobe, the two bands become gapless simultaneously. In the following, we focus on the generic transition where $\mu_+-\mu$ or $\mu-\mu_-$ vanishes (but not the two of them). This implies that $D^2-4DUx+U^2>0$ (see Eq.~(\ref{deltamupm})) and the dispersion is quadratic in the small-$\k$ limit,
\beq 
E^\alpha_\k = \alpha \left( \Delta_\alpha + \frac{\k^2}{2m^*_\alpha} \right) \qquad (\k\to 0), 
\eeq 
where 
\begin{align}
\Delta_\alpha &= - \alpha (\mu-\mu_\alpha) , \label{Delta1} \\ 
\frac{\mlat}{m^*_\alpha} &= \half \left[ \alpha + \frac{2Ux-D}{(D^2-4DUx+U^2)^{1/2}} \right] .
\label{m1} 
\end{align}
Each band $\alpha=\pm$ is characterized by an excitation gap $\Delta_\alpha$ and an effective mass $m^*_\alpha$. We denote the effective mass of the free bosons moving on the cubic lattice by $\mlat=1/2t$ (the free dispersion $t_\k=\k^2/2\mlat-D$ is quadratic for $\k\to 0$). It is easy to verify that $\Delta_\alpha$ and $m^*_\alpha$ are positive. At the QCP $\mu=\mu_\alpha$, $\Delta_\alpha$ vanishes and the critical mode has a quadratic dispersion law, $E_\k^\alpha=\alpha\k^2/2m^*_\alpha$, while the other band remains gapped ($\Delta_{-\alpha}>0$). We conclude that the dynamical critical exponent takes the value $z=2$. Furthermore, since the gap vanishes linearly with $\mu-\mu_\alpha$, the correlation-length exponent $\nu$ satisfies $z\nu=1$, which implies $\nu=1/2$. These exponents are characteristic of the dilute-Bose-gas universality class~\cite{Sachdev_book,NDbook1}. For $\nmi=0$, there is a single excitation branch $E_\k^+=t_\k-\mu$, $\Delta_+=-\mu+\mu_+=-\mu-D$ and $m^*_+=\mlat$. 

To obtain the quasi-particle weight $\Zqp$ associated with the critical quasi-particle excitations, we expand the propagator~(\ref{Gkw}) for small $\k$ and $\wn$,  
\beq  
G(\k,i\wn) \simeq \alpha \frac{\Zqp}{i\wn - \alpha \left( \frac{\k^2}{2m^*_\alpha} + \Delta_\alpha \right) } , 
\label{Gqp} 
\eeq 
where 
\begin{align}
\Delta_\alpha &= - \alpha \frac{1+D\Gloc(0)}{D\Gloc'(0)} , \label{Delta2}  \\
\Zqp &= \frac{\mlat}{m^*_\alpha} = \alpha  \frac{\Gloc(0)}{D\Gloc'(0)} ,
\label{m2} 
\end{align}
with $\Gloc'(0)=\partial_{i\w}\Gloc(i\w)|_{\w=0}$ (since $T=1/\beta\to 0$, the Matsubara frequency $\wn\equiv\w$ is a continuous variable). In Appendix~\ref{app_Delta_mstar}, we show that the expression of $\Delta_\alpha$ and $m^*_\alpha$ in~(\ref{Delta2}) and (\ref{m2}) agree with the expressions~(\ref{Delta1}) and (\ref{m1}) obtained from the energy $E^\alpha_\k$.

\begin{figure}
	\centerline{\includegraphics[width=4cm]{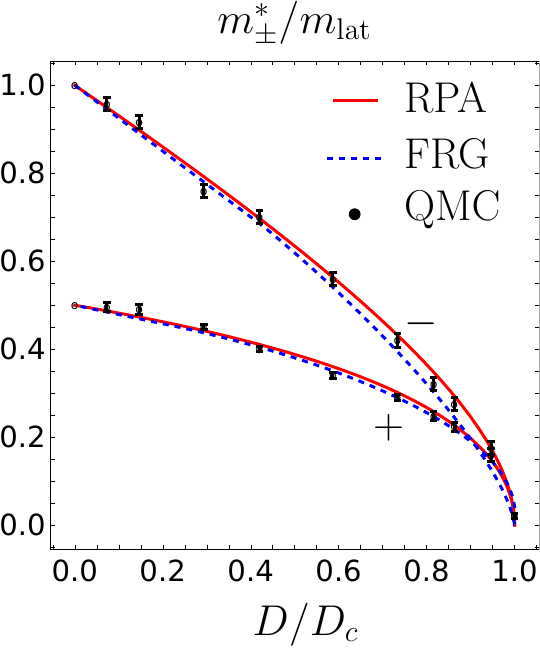}
		\hspace{0.3cm}
		\includegraphics[width=4cm]{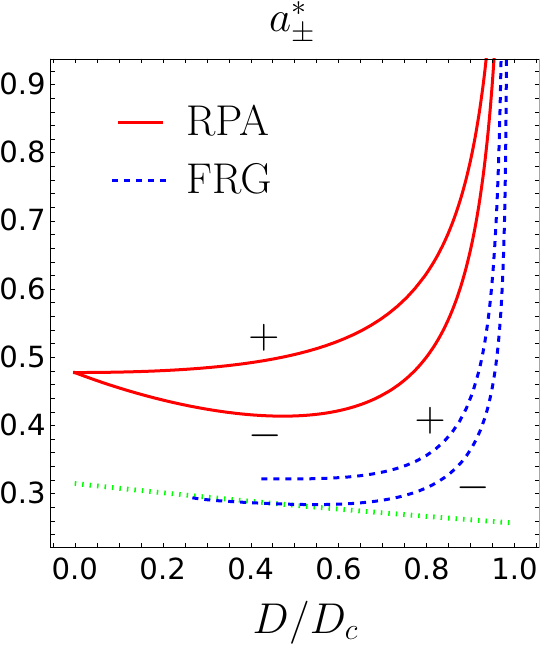}}
	\caption{Effective mass $m^*_\alpha/\mlat=1/\Zqp$ (left) and effective scattering length $a^*_\alpha$ (right) vs $D/D_c$ at the quantum critical point between the Mott insulator $\nmi=1$ and the superfluid state, obtained from strong-coupling RPA, nonperturbative functional renormalization group (FRG)~\cite{Rancon12d} and quantum Monte Carlo simulations (QMC)~\cite{Capogrosso07}. The green dotted line in the right panel shows the (vacuum) scattering length $\alat$ of the bosons moving on the lattice [Eq.~(\ref{alat})].}
	\label{fig_mastar}.  
\end{figure}

Having identified the critical quasi-particles and their spectral weight, it is natural to introduce the quasi-particle interaction strength $g^\alpha_R=g(\Zqp)^2$. By analogy with the dilute Bose gas, we then define an effective scattering length $a^*_\alpha$ by 
\beq 
g^\alpha_R \Bigl|_{\mu=\mu_\alpha} = \frac{4\pi a^*_\alpha}{m^*_\alpha} .
\eeq 
The quasi-particle weight $\Zqp=\mlat/m^*_\alpha$ and the effective scattering length $a^*_\alpha$ as a function of $D/D_c$ are shown in Fig.~\ref{fig_mastar} for the Mott insulator $\nmi=1$. The low accuracy of $a^*_\alpha$, compared to the FRG estimate, is not surprising, as $a^*_\alpha$ is derived from the local four-point vertex and does not take into account non-local corrections resulting from the hopping term. Note that the quasi-particle weight $\Zqp$ is larger than unity and the effective mass $m^*_\alpha$ smaller than $\mlat$; this property is also valid for $\nmi\geq 2$. For the transition between the trivial Mott insulator $\nmi=0$ (vacuum) and the superfluid state, using $\mu_+=-D$, $m^*_+=\mlat$ and Eq.~(\ref{gnzero}), one finds $a^*_+=1/[8\pi(t/U+1/12)]$. Since the ground state at the QCP is the vacuum, the effective scattering length can be determined exactly by solving the two-body problem and $a^*_+$ should be compared with the scattering length
\beq 
\alat = \frac{1}{8\pi(t/U+A)} , \quad A\simeq 0.1264 ,
\label{alat}  
\eeq 
of the free bosons moving on the cubic lattice~\cite{Rancon11b}. 

We are now in a position to determine the momentum distribution $n^{\rm MI}_\k=\mean{\hat\psi^\dagger_\k\hat\psi_\k}$ in the Mott insulator. Writing the propagator as 
\beq 
G(\k,i\wn) = \frac{\calS_{\rm MI}(E^+_\k)}{i\wn-E^+_\k} + \frac{\calS_{\rm MI}(E^-_\k)}{i\wn-E^-_\k} ,
\eeq 
we obtain 
\beq 
n^{\rm MI}_\k = - \int_{-\infty}^0 d\w \, A(\k,\w) = - \calS_{\rm MI}(E^-_\k) 
\eeq 
for $T\to 0$, where 
\begin{align} 
A(\k,\w) &= -\frac{1}{\pi} \Im[G(\k,\w+i0^+)] \nonumber \\  &= \calS_{\rm MI}(E^+_\k) \delta(\w-E^+_\k) + \calS_{\rm MI}(E^-_\k) \delta(\w-E^-_\k)
\end{align}
is the spectral function and
\beq 
\begin{split}
\calS_{\rm MI}(E^+_\k) &= \frac{\delta\mu+Ux+E^+_\k}{E^+_\k-E^-_\k} , \\  
\calS_{\rm MI}(E^-_\k) &= 1 -  \calS_{\rm MI}(E^+_\k)
\end{split} 
\eeq 
are the spectral weights associated with the excitation energies $E^+_\k$ and $E^-_\k$, respectively. Note that $\calS_{\rm MI}(E^\pm_\k)$ and $n_\k^{\rm MI}$ are independent of the chemical potential.    

\subsection{Superfluid phase -- universal contact $\cuni$} 
\label{sec_bhm:subsec_sf} 

In the superfluid state, the order parameter $\phi_\r(\tau)=\phi_0$ is nonzero and the effective action is given by 
\begin{multline}
\Gamrpa[\phi^*_0,\phi_0] = \Gamloc[0,0] \\ + \beta N \Bigl\{ - [D+\Gloc^{-1}(0) ] |\phi_0|^2 + \frac{g}{2} |\phi_0|^4 \Bigr\} . 
\end{multline}
Minimizing with respect to $\phi_0$, we obtain the condensate density 
\beq 
n_0 = |\phi_0|^2 = \frac{D+\Gloc^{-1}(0)}{g} = \Zqp \frac{m^*_\alpha}{4\pi a^*_\alpha} |\mu-\mu_\alpha| . 
\label{n0} 
\eeq 
The pressure $P=-\Omega/N=-\Gamrpa[\phi^*_0,\phi_0]/\beta N$ is given by 
\begin{align}
P &= \Pmi + \frac{[D+\Gloc^{-1}(0)]^2}{2g} \nonumber\\ 
&= \Pmi + \frac{m^*_\alpha}{8\pi a^*_\alpha} (\mu -\mu_\alpha)^2 , 
\label{Psf} 
\end{align} 
with $\Pmi$ the pressure of the (unstable) Mott insulator, and the mean density reads 
\beq
n = \frac{\partial P}{\partial \mu} = \nmi + \frac{m^*_\alpha}{4\pi a^*_\alpha} (\mu -\mu_\alpha) .
\label{nSF} 
\eeq  
The last expression in~(\ref{n0}) and (\ref{Psf}) is obtained using $D+\Gloc^{-1}(0)= |\mu-\mu_\alpha|/\Zqp$ for $\mu-\mu_\alpha\to 0$ and evaluating $g$ at $\mu=\mu_\alpha$. The singular part $\Psing=P-\Pmi$ of the pressure, 
\begin{align}
\Psing(\mu-\mu_\alpha,m^*_\alpha,a^*_\alpha) &= \frac{m^*_\alpha}{8\pi a^*_\alpha} (\mu -\mu_\alpha)^2 \nonumber\\  
&= \frac{2\pi a^*_\alpha (n-\nmi)^2}{m^*_\alpha} 
\end{align} 
exhibits the standard Bogoliubov form but with the effective mass $m^*_\alpha$ and the effective scattering length $a^*_\alpha$ instead of the bare boson mass and scattering length in vacuum, and the distance $\mu-\mu_\alpha$ (or $|n-\nmi|$) to the critical point rather than the chemical potential (or the density). From~(\ref{n0}) and (\ref{nSF}), one obtains 
\beq
n_0 = \Zqp |n-\nmi| . 
\eeq  
The condensate density also takes the standard Bogoliubov form but is multiplied by the quasi-particle weight $\Zqp$. This point is further discussed in Sec.~\ref{sec_bhm:subsec_universal}. 

It is also possible to compute the superfluid density. If the order parameter $\phi_\r=\phi_0 e^{i\theta_\r}$ varies slowly in space, with a time-independent phase $\theta_\r$, the effective action increases by 
\beq
\Delta\Gamrpa[\phi^*,\phi] = \beta \frac{n_0}{2\mlat} \int d^3r \, (\nabla\theta_\r)^2 
\label{DeltaGam} 
\eeq
to leading order in derivatives (taking the continuum limit). The superfluid stiffness $\rho_s=n_0/\mlat$ defines the superfluid density $n_s$ {\it via} the relation $\rho_s=n_s/m^*_\alpha$, which gives 
\beq 
n_s = \frac{m^*_\alpha}{\mlat} n_0 = \frac{n_0}{\Zqp} = |n-\nmi| . 
\eeq 
The superfluid density is thus given by the excess density of particles (or holes) $|n-\nmi|$ with respect to the Mott insulator. Since $\Zqp=\mlat/m^*_\alpha\geq 1$ (see Fig.~\ref{fig_mastar}), the condensate density $n_0\simeq \Zqp n_s$ is larger than the superfluid density $n_s=|n-\nmi|$: The excess particles (holes) with respect to the Mott insulator drag other particles (holes) into the condensation.

These results are fully consistent with the FRG approach~\cite{Rancon12d}. In the latter, $g$ is not simply given by the four-point vertex in the local limit, but by its value at the QCP. Furthermore, the thermodynamic relations include the Lee-Huang-Yang corrections, which are absent in the strong-coupling RPA. We will return to this point in Sec.~\ref{sec_bhm:subsec_universal}. 

We can now define a universal contact by taking the derivative of the singular part of the pressure with respect to the effective scattering length $a^*_\alpha$, 
\begin{align}
\frac{\cuni}{\calV} &= 8\pi m^*_\alpha \frac{\partial}{\partial(1/a^*_\alpha)} \Psing(\mu-\mu_\alpha,m^*_\alpha,a^*_\alpha) \Bigl|_{\mu-\mu_\alpha,m^*_\alpha} \nonumber \\  &= [m^*_\alpha(\mu-\mu_\alpha)]^2  \nonumber \\  &= [4\pi a^*_\alpha(n-\nmi)]^2 ,
\label{contact} 
\end{align} 
which is analog to the result of Bogoliubov's theory for a dilute Bose gas (ignoring the Lee-Huang-Yang correction), but with the effective scattering length $a^*_\alpha$ and $|n-\nmi|$ instead of the full density. The contact~(\ref{contact}) can also be written as $\cuni/\calV=(4\pi a^*_\alpha n_s)^2$, an expression that is also valid in the dilute Bose gas where Galilean invariance implies $n_s=n$.

\subsection{Spectrum and momentum distribution} 
\label{sec_bhm:subsec_spectrum} 

The spectrum can be obtained from the poles of the propagator or, equivalently, from the zeros of the determinant of the two-point vertex
\begin{equation}
\Gamma^{(2)}(k) = \begin{pmatrix} 
\dfrac{\delta^2 \Gamrpa}{\delta\phi^*_k \delta\phi_k} & 
\dfrac{\delta^2 \Gamrpa}{\delta\phi^*_k \delta\phi^*_{-k}} \\ 
\dfrac{\delta^2 \Gamrpa}{\delta\phi_{-k} \delta\phi_k} & 
\dfrac{\delta^2 \Gamrpa}{\delta\phi_{-k} \delta\phi^*_{-k}}
\end{pmatrix} \Biggl|_{\phi^{(*)}_\r=\phi^{(*)}_0} 
\label{Gam2a} 
\end{equation} 
with
\beq 
\begin{split}
&\Gamma^{(2)}_{\phi^*\phi}(\k,i\wn) =t_\k - \Gloc^{-1}(i\wn) + 2g n_0 , \\
&\Gamma^{(2)}_{\phi^*\phi^*}(\k,i\wn) = g \phi_0^2 
\end{split} 
\label{Gam2} 
\eeq
and 
\beq 
\begin{split}
	&\Gamma^{(2)}_{\phi\phi^*}(\k,i\wn) = \Gamma^{(2)}_{\phi^*\phi}(-\k,-i\wn) , \\ 
	&\Gamma^{(2)}_{\phi\phi}(\k,i\wn) = [\Gamma^{(2)}_{\phi^*\phi^*}(\k,i\wn)]^* ,
\end{split} 
\label{Gam2b} 
\eeq
where we use the notation $k=(\k,i\wn)$, $\Gamma^{(2)}_{\phi^*\phi}=\delta^2\Gamrpa/\delta\phi^*\delta\phi$, etc. For $\nmi\neq 0$, one obtains 
\beq 
\det\, \Gamma^{(2)}(k) = \frac{(i\wn)^4 + B_\k (i\wn)^2 + C_\k}{\wn^2 + (\delta\mu+Ux)^2} ,
\label{detGam2}  
\eeq 
where 
\beq
\begin{split}
B_\k &= - \tilde A_\k^2 + 2\tilde B_\k + [D+\Gloc^{-1}(0)]^2 , \\ 
C_\k &= \tilde B_\k^2 - [D+\Gloc^{-1}(0)]^2 (\delta\mu+Ux)^2 
\end{split}
\eeq
and 
\beq
\begin{split} 
\tilde A_\k &= 2 \delta\mu -  2[D+\Gloc^{-1}(0)] - t_\k , \\
\tilde B_\k &= - (\delta\mu+Ux) \{t_\k+2[D+\Gloc^{-1}(0)] \} + \delta\mu^2 - \frac{U^2}{4} . 
\end{split}
\eeq 
After analytic continuation $i\wn\to\w+i0^+$, one finds the zeros $\pm \calE^\pm_\k$ of the determinant~(\ref{detGam2}),  
\beq 
\calE^\pm_\k = \left[ - \frac{B_\k}{2} \pm \half (B_\k^2 - 4 C_\k)^{1/2} \right]^{1/2}. 
\eeq 
By inverting the two-point vertex~(\ref{Gam2a}), one finally obtains the (normal) propagator
\begin{equation}
G(k) = \frac{(i\wn+\delta\mu+Ux)(i\wn-z_\k^+)(i\wn-z_\k^-)}{(\wn^2+\calE^+_\k{}^2)(\wn^2+\calE^-_\k{}^2)} , 
\label{Gsfdef} 
\end{equation} 
where 
\beq 
z_\k^\pm = \frac{\tilde A_\k}{2} \pm \half (\tilde A_\k^2 - 4\tilde B_\k)^{1/2} . 
\eeq 
In the superfluid state, the two bands $E_\k^\pm$ of the Mott insulator split into fours bands $\pm\calE^\pm_\k$ as shown in Fig.~\ref{fig_spectrum}. The bands $\pm\calE^-_\k$ are gapless with a linear spectrum $\pm c|\k|$ in the small-momentum limit. The velocity of the sound mode is given by 
\beq 
c = \left( \frac{2t[D+\Gloc^{-1}(0)]}{\bar a^2 + 2 \bar b[D+\Gloc^{-1}(0)] } \right)^{1/2} ,
\label{cdef} 
\eeq 
where 
\beq 
\begin{split}
\bar a &= \frac{\delta\mu^2 + 2\delta\mu Ux + U^2/4}{(\delta\mu+Ux)^2} , \\ 
\bar b &= \frac{U^2(x^2-1/4)}{(\delta\mu+Ux)^3} .
\end{split}
\eeq 
The gapped excitations $\pm\calE_\k^+$ are sometimes referred to as Higgs modes. 
In the superfluid phase near the vacuum ($\nmi=0$), one finds only two bands $\pm\calE_\k$ with $\calE_\k=[(t_\k+D)(t_\k+D+2\mu+2D)]^{1/2}$, which gives the sound-mode velocity $c=\sqrt{2t(\mu+D)}$. 

Ignoring the contribution $\delta_{\k,0}Nn_0$ of the condensate, in the generic case ($\nmi\neq 0$) the momentum distribution is given by 
\beq 
n_\k = - \int_{-\infty}^0 d\w \, A(\k,\w) = - \calS(-\calE^-_\k) - \calS(-\calE^+_\k) , 
\eeq 
where 
\begin{multline}
A(\k,\w) = \calS(\calE^+_\k) \delta(\w-\calE^+_\k) + \calS(\calE^-_\k) \delta(\w-\calE^-_\k) \\ + \calS(-\calE^+_\k) \delta(\w+\calE^+_\k) + \calS(-\calE^-_\k) \delta(\w+\calE^-_\k)
\end{multline}
is the spectral function and 
\beq 
\calS(\gamma \calE^\alpha_\k) = \alpha\gamma \frac{(\gamma \calE^\alpha_\k+\delta\mu+Ux)(\gamma \calE^\alpha_\k-z^+_\k) (\gamma \calE^\alpha_\k-z^-_\k)}{2\calE^\alpha_\k(\calE^+_\k{}^2 - \calE^-_\k{}^2)}  
\eeq  
the spectral weight associated with the pole $\gamma\calE^\alpha_\k$ ($\alpha,\gamma=\pm$) of the propagator~(\ref{Gsfdef}). The singular part of the momentum distribution,
\begin{align}
\nksing &= n_\k - \nkmi \nonumber\\ &= - \calS(-\calE^-_\k) - \calS(-\calE^+_\k) + \calS_{\rm MI}(E_\k^-) ,
\label{nksingpart}  
\end{align}
can be expressed in terms of the spectral weights in the superfluid state and Mott insulator. 

\begin{figure}
\centerline{\includegraphics[width=7.5cm]{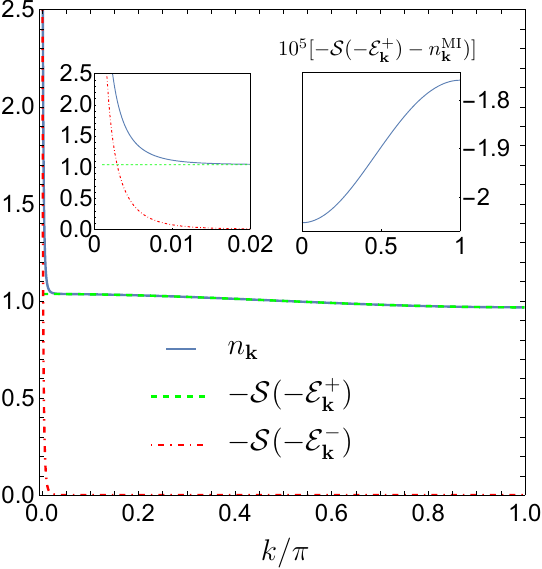}
}
\caption{Momentum distribution $n_\k=-\calS(-\calE_\k^+)-\calS(-\calE_\k^-)$, along the Brillouin zone diagonal $\k=(k,k,k)$, for the particle-doped Mott insulator $\nmi=1$: $\mu=1.000005\mu_+$ ($n=1.0003$) and $D=D_c/2$. The contribution $n_0\calV \delta_{\k,0}$ of the condensate is not taken into account. The gapless band $-\calE_\k^-$ gives a significant contribution only near $k=0$ (see left inset). For larger values of $k$, the momentum distribution is essentially due to the gapped band $-\calE_\k^+$ whose contribution is very close to $\nkmi=-\calS_{\rm MI}(E_\k^-)$ (see right inset).}
\label{fig_nk} 
\end{figure}

\begin{figure}
	\centerline{\includegraphics[width=7cm]{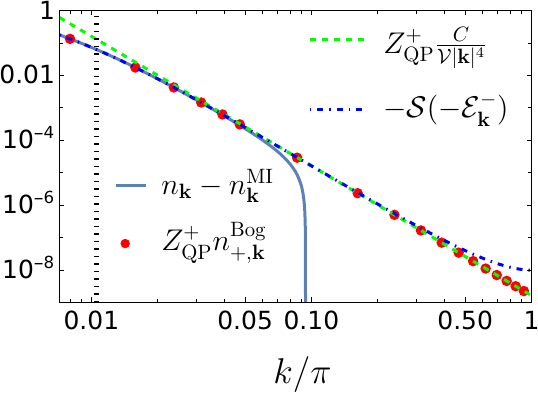}}
	\centerline{\includegraphics[width=7cm]{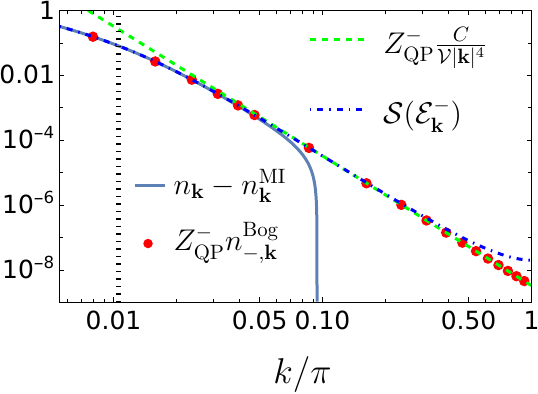}}
	\caption{Singular part $\nksing=n_\k-n_\k^{\rm MI}$ of the momentum distribution, along the Brillouin zone diagonal $\k=(k,k,k)$, for $D=D_c/2$. Top: $\mu=1.000005\mu_+$ ($n=1.0003$); Bottom: $\mu=0.99944\mu_-$ ($n=0.99972$).  The figure also shows $\Zqp n^{\rm Bog}_{\alpha,\k}$~[Eq.~(\ref{nkbog})] and $Z_{\rm QP}^+ \cuni/\calV|\k|^4$ where $\cuni=\calV[m^*_+(\mu-\mu_+)]^2$ is the contact. The dash-dotted (blue) line corresponds to the contribution of the gapless band with negative energy $-\calE^-_\k$ (top) or positive energy $\calE^-_\k$ (bottom); see Eqs.~(\ref{nksingpart}) and (\ref{nksinghole}). The vertical dotted line shows the momentum scale $k_*/\sqrt{3}$ where $k_*=2(m^*_\alpha|\mu-\mu_\alpha|)^{1/2}$.}
	\label{fig_momentum} 
\end{figure}

Let us first discuss the case of particle doping ($\mu>\mu_+$, $n>\nmi$) where the MI-SF transition occurs when the positive energy band $E_\k^+$ of the Mott insulator becomes gapless. In that case, the band $E_\k^-$ of the Mott insulator evolves into the band $-\calE_\k^+$ of the superfluid, and the band $E_\k^+$ into the band $\calE_\k^-$ (Fig.~\ref{fig_spectrum}). Two additional bands, $\calE_\k^+$ and $-\calE_\k^-$, appear in the superfluid. The band $\calE_\k^+$ carries a negligible spectral weight in the vicinity of the transition. Figure~\ref{fig_nk} shows the momentum distribution for the doped Mott insulator $\nmi=1$. The gapped band $E_\k^-$ is little affected when $\mu$ becomes larger than $\mu_+$ and $\calS(-\calE^+_\k)$ is essentially equal to $\calS_{\rm MI}(E^-_\k)$ near the transition. On the other hand, although the band $\calE_\k^-$ carries most of the spectral weight of the band $E_\k^+$ of the Mott insulator, the gapless negative energy band $-\calE^-_\k$ gives a large contribution to the momentum distribution for small momenta, as in a dilute superfluid gas. This implies that $\nksing$ [Eq.~(\ref{nksingpart})] is well approximated by $-\calS(-\calE^-_\k)$. Figure~\ref{fig_momentum} (top panel) shows that this is indeed the case for momenta that are not too large, but the agreement breaks down when $|\k|\gtrsim 0.25$ for $\k$ varying along the Brillouin zone diagonal, that is, well before reaching the Brillouin zone boundary $|\k|=\pi\sqrt{3}$; we will come back to this point later. Furthermore, we find that $-\calS(-\calE^-_\k) \simeq \Zqpp n^{\rm Bog}_{+,\k}$, where
\beq 
n^{\rm Bog}_{\alpha,\k} = - \half + \frac{\eps^\alpha_\k+|\mu-\mu_\alpha|}{2\sqrt{\eps^\alpha_\k(\eps^\alpha_\k + 2|\mu-\mu_\alpha|)}} 
\label{nkbog} 
\eeq 
is the standard Bogoliubov result for bosons (ignoring the contribution of the condensate) with dispersion $\eps^\alpha_\k=\k^2/2m^*_\alpha$ and chemical potential $|\mu-\mu_\alpha|$. When $|\k|$ is larger than the characteristic momentum scale $k_*=2(m^*_\alpha|\mu-\mu_\alpha|)^{1/2}$, $n^{\rm Bog}_{\alpha,\k}\simeq \cuni/\calV |\k|^4$ with the contact $\cuni$ defined by~(\ref{contact}) so that the momentum distribution~\cite{not2} 
\beq 
\nksing \simeq \frac{\Zqpp \cuni}{\calV|\k|^4} \qquad (k_*\ll |\k|) 
\eeq 
exhibits a high-momentum tail $\sim 1/|\k|^4$, as in a dilute Bose gas, provided the characteristic scale $k_*$ is sufficiently small. 

In the case of hole doping ($\mu<\mu_-$, $n<\nmi$), using the spectral weight normalization 
\beq 
\sum_{\alpha=\pm} [\calS(\calE^\alpha_\k) + \calS(-\calE^\alpha_\k)]= \sum_{\alpha=\pm} \calS_{\rm MI}(E^\alpha_\k)=1 , 
\eeq 
we can rewrite the momentum distribution as 
\beq 
\nksing = \calS(\calE^-_\k) + \calS(\calE^+_\k) - \calS_{\rm MI}(E_\k^+) . 
\label{nksinghole}  
\eeq 
The positive energy band $E_\k^+$ of the Mott insulator remains gapped at the transition and $\calS(\calE^+_\k)\simeq \calS_{\rm MI}(E_\k^+)$ so that $\nksing\simeq \calS(\calE_\k^-)$ is well approximated by the contribution of the gapless band $\calE^-_\k$ of the superfluid. Note that this positive energy band is not occupied but can nevertheless be considered as an occupied hole band. The agreement between $\nksing$ and $\calS(\calE^-_\k)$ over a large part of the Brillouin zone is shown in Fig.~\ref{fig_momentum} (bottom panel). We find that $\calS(\calE^-_\k)$ is well approximated by $\Zqpm n^{\rm Bog}_{-,\k}$, where $n^{\rm Bog}_{-,\k}$ is given by~(\ref{nkbog}), and we obtain $\nksing\simeq \Zqpm \cuni/\calV|\k|^4$ for $k_*\ll |\k|.$

Let us now discuss the disagreement between $\nksing$ and $n_{\alpha,\k}^{\rm Bog}$ when $|\k|\gtrsim 0.25$ in Fig.~\ref{fig_momentum}. The momentum distribution must satisfy the sum rules 
\beq
\nmi = \int \frac{d^3k}{(2\pi)^3} n^{\rm MI}_\k = \int \frac{d^3k}{(2\pi)^3} [ -\calS(E^-_\k) ]
\eeq 
and 
\begin{align} 
n &= n_0 + \int \frac{d^3k}{(2\pi)^3} n_\k \nonumber\\ 
&= n_0 + \int \frac{d^3k}{(2\pi)^3} [ -\calS(-\calE^-_\k) -\calS(-\calE^+_\k) ] , 
\end{align} 
where the density $n$ is given by~(\ref{nSF}). In the strong-coupling RPA, these sum rules are not perfectly satisfied. For example, in the case of Fig.~\ref{fig_momentum}, the integral over the momentum distribution differs from the density $n-n_0$ (or $\nmi$) by about $10^{-4}$. This implies that the spectral weights $\calS_{\rm MI}(E^\pm_\k)$ and $\calS(\pm\calE^\pm_\k)$ are probably only accurate to within $10^{-4}$. We therefore believe that the slight difference between $-\calS(-\calE_\k^+)$ and $-\calS_{\rm MI}(E_\k^-)$ at large momenta (see the right inset in Fig.~\ref{fig_nk}), which spoils the agreement between $\nksing$ and $-\calS(-\calE_\k^-)$, is an artifact of the strong-coupling RPA. We expect the agreement between these two quantities, which is observed up to $|\k|\simeq 0.25$, to extend up to the Brillouin zone boundary ---except very close to the zone boundaries where the free dispersion $t_\k$ differs from $\k^2/2\mlat-D$ due to lattice effects--- similarly to what is observed for the agreement between $-\calS(-\calE^-_\k)$ (or $\calS(\calE^-_\k)$) and $\Zqp n^{\rm Bog}_{\alpha,\k}$. This expectation is supported by a study of a hard-core boson model (see Sec.~\ref{sec_hard_core}).

\subsection{Universal thermodynamics} 
\label{sec_bhm:subsec_universal} 

The universality of the equation of state of a dilute Bose gas can be understood from the presence of a QCP at $\mu=0$ that separates the vacuum ($\mu\leq 0$) from the superfluid state ($\mu\geq 0$)~\cite{Sachdev_book,NDbook1}. In the vacuum, the one-particle excitations have energy $\w=\k^2/2m+|\mu|$ to that the correlation length $\xi=(2m|\mu|)^{-\nu}$ diverges with the exponent $\nu=1/2$ when approaching the QCP. At the QCP ($\mu=0$) the excitations are gapless, $\w=|\k|^z/2m$, with a dynamical critical exponent $z=2$. A straightforward dimensional analysis of the Gaussian action (corresponding to non-interacting bosons) shows that in $d$ dimensions the field has scaling dimension $[\psi]=(d+z-2)/2=d/2$ while $[\mu]=2$ and $[g]=2-d$, where $g$ is the strength of the two-body interaction (assumed to be local). For $d>2$, the interaction is irrelevant (in the RG sense) and $\mu$ is the only relevant variable; the Gaussian fixed point is stable and the transition is mean-field-like. Standard RG arguments~\cite{Rancon12b,NDbook1} then  imply that in three dimensions the zero-temperature pressure can be written in the scaling form 
\beq 
P = \left( \frac{m}{2\pi} \right)^{3/2} \mu^{5/2} \calG(\sqrt{ma^2\mu}) .
\eeq
The dependence of the universal scaling function $\calG(x)$ on $ma^2\mu$ is due to the interaction $g$ being dangerously irrelevant (in the RG sense) and taking the renormalized value $g_R=4\pi a/m$ (when expressed in dimensionful units)~\cite{NDbook1}. The two nonuniversal quantities that enter the equation of state, namely the mass of the particle and the scattering length, are properties of the critical excitations at the QCP (where the ground state is the vacuum with $\mu=0$). 

The density $n=\partial P/\partial\mu$ and the compressibility $\kappa=\partial n/\partial\mu$ are determined by the scaling function $\calG$ and its derivatives. The condensate density and the superfluid density also satisfy scaling forms, 
\beq 
\begin{split}
n_0 &=  \left( \frac{m\mu}{2\pi} \right)^{3/2} \calI(\sqrt{ma^2\mu}) , \\ 
n_s &=  \left( \frac{m\mu}{2\pi} \right)^{3/2} \calJ(\sqrt{ma^2\mu}) .
\end{split}
\eeq 
Since the superfluid density is given by the full density in a Galilean-invariant system, i.e. $n_s=n=\partial P/\partial\mu$, $\calJ$ is not an independent scaling function but is given by 
\begin{equation}
\calJ(x) = \frac{5}{2} \calG(x) + \frac{x}{2}\calG'(x)
 . 
\label{JvsG} 
\end{equation}
The universal scaling functions $\calG$ and $\calI$ can be computed in the limit $x\ll 1$ from Bogoliubov's theory, 
\beq 
\begin{split}
P =& \frac{m\mu^2}{8\pi a}\left( 1 - \frac{64}{15\pi} \sqrt{ma^2\mu} \right) , \\ 
n_0 =& \frac{m\mu}{4\pi a}\left( 1 - \frac{20}{3\pi} \sqrt{ma^2\mu} \right) , 
\end{split}
\eeq 
which gives 
\beq 
\begin{split}
\calG(x) &= \frac{\sqrt{\pi}}{2\sqrt{2}x} \left( 1 - \frac{64}{15\pi} x \right) , \\
\calI(x) &= \frac{\sqrt{\pi}}{\sqrt{2}x} \left( 1 - \frac{20}{3\pi} x \right),  \\
\calJ(x) &= \frac{\sqrt{\pi}}{\sqrt{2}x} \left( 1 - \frac{16}{3\pi} x \right) ,
\end{split} 
\label{scalfunc}
\eeq 
where the subleading term comes from the Lee-Huang-Yang correction~\cite{Lee57b,Dalfovo99,NDbook1}. 

A crucial property of the MI-SF transition of bosons in a periodic potential is that is belongs to the dilute-Bose-gas universality class when it is induced by a density change. This implies that in the vicinity of the transition, the singular part of physical quantities can be written in the same scaling form as in the dilute Bose gas. For instance, the pressure reads~\cite{Rancon12d}
\begin{multline}
P = P_c + n_c(\mu-\mu_\alpha)\\ + \left( \frac{m^*_\alpha}{2\pi} \right)^{3/2} |\mu-\mu_\alpha|^{5/2} \calG(\sqrt{m^*_\alpha a^*_\alpha{}^2|\mu-\mu_\alpha|}) ,
\label{Pscaling} 
\end{multline}
where $P_c$ and $n_c=\nmi$ are the pressure and density, respectively, at the QCP and $|\mu-\mu_\alpha|=\alpha(\mu-\mu_\alpha)$. The regular part of the pressure is given by $P_c+n_c(\mu-\mu_\alpha)$. The two nonuniversal parameters entering the singular part of the pressure are the effective mass $m^*_\alpha$ of the quasi-particles at the QCP and the effective scattering length $a^*_\alpha$ describing their mutual interaction. The fact that the effective chemical potential $\mu-\mu_\alpha$ enters the scaling function without an additional scale factor is a nontrivial property that follows from the invariance of the microscopic action in the semilocal (time-dependent) gauge transformation 
\beq
\begin{gathered}
\psi_\r(\tau)\to\psi_\r(\tau) e^{i\alpha(\tau)}, \quad \psi^*_\r(\tau)\to\psi^*_\r(\tau) e^{-i\alpha(\tau)}, \\ 
\mu\to \mu+i\dtau\alpha(\tau) .
\end{gathered} 
\label{semilocalT} 
\eeq
This invariance ensures in particular that the pressure is independent of the quasi-particle weight $\Zqp$~\cite{Rancon12d}. The density and superfluid density are given by 
\begin{align} 
n &= \nmi + \alpha \left( \frac{m^*_\alpha|\mu-\mu_\alpha|}{2\pi} \right)^{3/2}  \calJ(\sqrt{m^*_\alpha a^*_\alpha{}^2|\mu-\mu_\alpha|}) , \nonumber\\ 
n_s &= \left( \frac{m^*_\alpha|\mu-\mu_\alpha|}{2\pi} \right)^{3/2} \calJ(\sqrt{m^*_\alpha a^*_\alpha{}^2|\mu-\mu_\alpha|}) ,
\label{nscaling} 
\end{align}
where we have used~(\ref{JvsG}) to express $n$ in terms of $\calJ$. The superfluid density has no regular part, since it vanishes in the Mott insulator. We conclude that the superfluid density is given by the density of additional particles (or holes) introduced in the Mott insulator, 
\beq 
n_s = |n-\nmi| . 
\eeq 
This is an exact result, inherited from the Galilean invariance of the dilute Bose gas. 

The condensate density satisfies the scaling form~\cite{Rancon12d}
\beq 
n_0 = \Zqp \left( \frac{m^*_\alpha|\mu-\mu_\alpha|}{2\pi} \right)^{3/2}  \calI(\sqrt{m^*_\alpha a^*_\alpha{}^2|\mu-\mu_\alpha|}) .
\label{n0scaling} 
\eeq 
Contrary to other physical quantities, it is not invariant in the semilocal gauge transformation~(\ref{semilocalT}) and depends on the quasi-particle weight $\Zqp$. From a physical point of view, this is because Bose-Einstein condensation involves quasi-particles and not (bare) particles. Thus, if we define the quasi-particle field $\bar\psi_\r(\tau)=\psi_\r(\tau)/(\Zqp)^{1/2}$, we find that the quasi-particle condensate density $\bar n_0=n_0/\Zqp$ is independent of the quasi-particle weight. As noted in Sect.~\ref{sec_bhm:subsec_sf}, the condensate density $n_0\simeq \Zqp n_s$ is larger than the superfluid density $n_s=|n-\nmi|$; the small difference between the scaling functions $\calI$ and $\calJ$ due to the Lee-Huang-Yang correction cannot counterbalance the effect of $\Zqp$.

From~(\ref{contact}) and (\ref{Pscaling}), we obtain the scaling form of the contact, 
\beq 
\frac{\cuni}{\calV} = - \sqrt{\frac{8}{\pi}} (m^*_\alpha|\mu-\mu_\alpha|)^3 a^*_\alpha{}^2 \calG'(\sqrt{m^*_\alpha a^*_\alpha{}^2|\mu-\mu_\alpha|}) .
\label{Cscaling} 
\eeq 
Using the explicit expression of $\calG$ [Eqs.~(\ref{scalfunc})], this gives 
\beq 
\frac{\cuni}{\calV} = (m^*_\alpha|\mu-\mu_\alpha|)^2 ,
\eeq 
which is the result obtained in the strong-coupling RPA. Note that there is no correction of order $\sqrt{m^*_\alpha a^*_\alpha{}^2|\mu-\mu_\alpha|}$. From the relation~(\ref{nscaling}) between $n-\nmi$ and $\mu-\mu_\alpha$ and the explicit expression of $\calJ$, one finally obtains
\beq 
\frac{\cuni}{\calV} = [4\pi a^*_\alpha(n-\nmi)]^2 \biggl( 1 + \frac{64}{3} \sqrt{ \frac{|n-\nmi| a^*_\alpha{}^3}{\pi}} \biggr) .
\eeq 
The subleading term is due to the Lee-Huang-Yang correction in the expression of $n-\nmi$ as a function of $\mu-\mu_\alpha$.

The strong-coupling RPA agrees with the scaling forms~(\ref{Pscaling}), (\ref{nscaling}), (\ref{n0scaling}) and (\ref{Cscaling}) but the corresponding scaling functions do not include the Lee-Huang-Yang correction.

\section{Hard-core bosons}
\label{sec_hard_core} 

In this section, we consider a hard-core boson system defined on a cubic lattice within the RPA. Somewhat surprisingly, as already observed in \cite{Rancon14a}, it allows recovering the semi-classical approximation of the equivalent quantum XY model to leading order in the $1/S$ expansion, evaluated at $S=1/2$. The Hamiltonian reads
\beq
\hat H = \sum_{\r,\r'} t_{\r,\r'}  \hat\psi^\dagger_\r \hat\psi_{\r'} - \mu \sum_\r \hat\psi^\dagger_\r \hat\psi_\r .
\label{ham_hc} 
\eeq 
The hard-core constraint is enforced by restricting the Hilbert space on each site to the vacuum state $\ket{0}_\r$ and the singly-occupied state $\hat\psi^\dagger_\r\ket{0}_\r$. In this restricted Hilbert space, the boson operators satisfy the commutation relations $[\hat\psi_\r,\hat\psi_{\r'}]=[\hat\psi^\dagger_\r,\hat\psi^\dagger_{\r'}]=0$ and  
\beq 
[\hat\psi_\r,\hat\psi^\dagger_{\r'}]=\delta_{\r,\r'}(1-2\hat\psi^\dagger_\r \hat\psi_\r) .
\label{CRhc} 
\eeq 

The source-dependent partition function can be written as 
\beq
Z[J^*,J] = \Tr\Bigl\{ T_\tau e^{ -\inttau \, \tilde H(\tau) + \inttau \sum_\r [J^*_\r(\tau) \tilde\psi_\r(\tau)+ \hc ] } \Bigr\} ,
\eeq 
where $\tau$ should be understood as a formal time label allowing the imaginary-time ordering operator $T_\tau$ to appropriately interlace operators with no explicit time dependence. These operators (denoted with a tilde) should not be confused with the Heisenberg-picture operators, e.g. $\hat\psi_\r(\tau)=e^{\tau\hat H} \hat\psi_\r e^{-\tau\hat H}$. The generating functional of time-ordered correlation functions
is given by $\ln Z[J^*,J]$. In particular, the expectation value of the boson operator reads
\beq 
\begin{split}
	\phi_\r(\tau) &= \mean{\hat\psi_\r(\tau)} = \frac{\delta\ln Z[J^*,J]}{\delta J^* _\r(\tau)} , \\
	\phi^*_\r(\tau) &= \mean{\hat\psi^\dagger_\r(\tau)} = \frac{\delta\ln Z[J^*,J]}{\delta J _\r(\tau)} . 
\end{split} 
\eeq

In the RPA, we compute the partition function using a mean-field approximation for the hopping term, which amounts to replacing $t_{\r,\r '}{\tilde\psi}^\dagger_\r(\tau) \tilde\psi_{\r'}(\tau)$ by $t_{\r,\r '}[{\tilde\psi}^\dagger_\r(\tau) \phi_{\r'}(\tau) + \phi^*_\r(\tau) \tilde\psi_{\r'}(\tau)- \phi^*_\r(\tau) \phi_{\r'}(\tau)]$. This allows us to write the partition function in the form~(\ref{Zrpa}) and obtain the effective action~(\ref{gam1}) and the order parameter~(\ref{op}), so that  
we finally obtain the RPA effective action~(\ref{gam1a}). The only difference with the Bose-Hubbard model is the explicit expression of the local part. Using the results of \cite{Rancon14a}, reproduced in Appendix~\ref{sec_hardcore}, one finds the effective potential 
\begin{align}
	V(|\phi|^2) &= -D|\phi|^2 + \Vloc(|\phi|^2) \nonumber\\  
	&= -D|\phi|^2 - \frac{\mu}{2} - \frac{|\mu|}{2} \sqrt{1-4|\phi|^2} 
	\label{hc1} 
\end{align}
and the two-point vertices (for a constant field $\phi$)
\begin{align}
&\Gamma^{(2)}_{\phi^*\phi}(\k,i\wn) = - Z_C i\wn + \Vloc'(|\phi|^2) + |\phi|^2 \Vloc''(|\phi|^2) + t_\k , \nonumber\\ 
&\Gamma^{(2)}_{\phi^*\phi^*}(\k,i\wn) = \phi^2 \Vloc''(|\phi|^2) 
\label{hc2a} 
\end{align}
and 
\beq
\begin{split}
	&\Gamma^{(2)}_{\phi\phi^*}(\k,i\wn) = \Gamma^{(2)}_{\phi^*\phi}(-\k,-i\wn), \\ 
	&\Gamma^{(2)}_{\phi\phi}(\k,i\wn) =  [ \Gamma^{(2)}_{\phi^*\phi^*}(\k,i\wn) ]^*,
\end{split}
\label{hc2b} 
\eeq 
where $Z_C=-\sgn(\mu)/\sqrt{1-4|\phi|^2}$. 

By minimizing $V(|\phi|^2)$ we find that the ground state is a trivial Mott insulator (vacuum) with vanishing pressure $\Pmi=0$ and vanishing density $\nmi=0$ when $\bar\mu<-1$, and a Mott insulator with $\Pmi=\mu$ and $\nmi=1$ when $\bar\mu>1$, where $\bar\mu=\mu/D$. We denote by $\mu_+=-D$ the critical value of the chemical potential at the transition between the vacuum and the superfluid, and by $\mu_-=D$ the critical value corresponding to the transition from the superfluid to the Mott insulator with one boson per site. Contrary to Sec.~\ref{sec_bhm}, where we analyzed the upper and lower transition lines of a given Mott lobe with fixed $\nmi$, here we study the upper and lower transition line of the two different Mott phases. The system is superfluid when $-1<\bar\mu<1$ with a condensate density $n_0=|\phi|^2=\quarter(1-\bar\mu^2)$. The pressure is given by 
\beq 
P = - V(n_0) = \frac{(\mu+D)^2}{4D} , 
\label{hc3} 
\eeq 
and the density by
\beq 
n=\frac{\partial P}{\partial \mu} = \frac{\mu+D}{2D} . 
\label{hc4} 
\eeq 
Near the transition to the Mott insulator ($\mu\to\mu_\alpha$), the condensate density
\beq 
n_0 = |n-\nmi| 
\eeq 
is equal to the excess density of particles (or holes) with respect to the Mott insulator.

\subsection{Mott insulator}

When $\phi=0$, the boson propagator is given by 
\beq 
G(\k,i\wn) = \frac{-\sgn(\mu)}{i\wn-E_\k} , 
\label{hc5}
\eeq 
where 
\beq 
E_\k = -\mu - \sgn(\mu) t_\k . 
\eeq 
For $\mu$ near $\mu_\alpha$, it takes the quasi-particle form~(\ref{Gqp}) with $\alpha=-\sgn(\mu)$ and  
\beq 
\Zqp = 1 , \quad 
m^*_\alpha = \mlat , \quad  
\Delta_\alpha = |\mu|-D . 
\eeq 
At the QCP, the quasi-particles have a quadratic dispersion with effective mass $\mlat=1/2t$ and unit spectral weight. This effective mass is also obtained in the Bose-Hubbard model in the low-density limit and for the hole-doped Mott insulator $\nmi=1$ in the limit $t/U\to 0$ (Fig.~\ref{fig_mastar}). Note that the total spectral weight in~(\ref{hc5}) is equal to $-1$ in the Mott insulator $\nmi=1$ as particle excitations are suppressed by the hard-core constraint. The momentum distribution is simply $\nkmi=\nmi$. 

The quasi-particle interaction strength $g$ is defined by the static limit of the four-point vertex $\Gamma^{(4)}$. The latter can be obtained from the effective potential, i.e.  
\beq 
g = \half \frac{\partial^4 V(|\phi|^2)}{\partial\phi^*{}^2 \partial\phi^2} \biggl|_{\phi^*=\phi=0} = 2 |\mu| . 
\eeq 
This result can also be obtained from the expression obtained in the Bose-Hubbard model in the limit $U\to\infty$, Eq.~(\ref{ghardcore}), for $\mu<0$ ($\mu>0$) in the case $\nmi=0$ ($\nmi=1$). Since $\Zqp=1$, the effective scattering length is defined by $g|_{\mu=\mu_\alpha}=4\pi a^*/\mlat$, 
\beq 
a^* = \frac{D\mlat}{2\pi} = \frac{3}{2\pi} , 
\label{astarhc} 
\eeq
which, as expected, agrees with the infinite-$U$ limit of $a^*_+$ obtained in the Bose-Hubbard model when $\nmi=0$. The infinite-$U$ limit of $a^*_-$ obtained in the strong-coupling RPA is different from~(\ref{astarhc}), but the result obtained from the nonperturbative FRG~\cite{Rancon12d} agrees with $a^*$ in the limit $t/U\ll 1$ (Fig.~\ref{fig_mastar}); including Gaussian fluctuations about the RPA calculation is sufficient to recover the exact $s$-wave scattering length, given by Eq.~\eqref{alat} in the limit $t/U\to0$~\cite{Rancon22}.

\subsection{Superfluid state} 

The pressure~(\ref{hc3}) and the density~(\ref{hc4}) can be written as
\beq
\begin{split} 
P &= \Pmi+  \frac{\mlat}{8\pi a^*} (\mu-\mu_\alpha)^2 , \\ 
n &= \nmi + \frac{\mlat}{4\pi a^*} (\mu-\mu_\alpha) .
\end{split}
\label{hc8}
\eeq 
When the order parameter $\phi_\r=\phi_0 e^{i\theta_\r}$ varies slowly in space, with a time-independent phase $\theta_\r$, the variation of the effective action~(\ref{gam1a}) is entirely due to the hopping part since $\Gamloc[\phi^*,\phi]$ does not change. Thus Eq.~(\ref{DeltaGam}) still holds and we deduce the superfluid density 
\beq 
n_s = n_0 = |n-\nmi| . 
\eeq 
From~(\ref{hc8}), we obtain the universal contact 
\begin{align}
\frac{\cuni}{\calV} 
&= [\mlat(\mu-\mu_\alpha)]^2 \nonumber\\
&= [4\pi a^*(n-\nmi)]^2 ,
\label{contacthc}
\end{align}   
in agreement with what was found in the Bose-Hubbard model. 

The zeros of the determinant of the two-point vertex $\Gamma^{(2)}(\k,\w+i0^+)$ give the spectrum $\w=\pm\calE_\k$, where 
\beq 
\calE_\k = \{(t_\k+D) [\bar\mu^2 (t_\k+D) + D(1-\bar\mu^2)] \}^{1/2} .
\eeq
When $\mu\to -D$, one recovers the dispersion and the velocity $c=\sqrt{2t(\mu+D)}$ near the vacuum state obtained in the Bose-Hubbard model. The (normal) propagator is obtained by inverting $\Gamma^{(2)}(\k,i\wn)$, 
\beq 
G(\k,i\wn) = \frac{i\wn \bar\mu -D(1-\bar\mu^2)/2-\bar\mu^2 (t_\k+D)}{\wn^2 + \calE^2_\k} ,
\eeq 
which gives the spectral function 
\beq 
A(\k,\w) = \calS(\calE_\k) \delta(\w-\calE_\k) + \calS(-\calE_\k) \delta(\w+\calE_\k) , 
\eeq 
with 
\beq 
\calS(\pm \calE_\k) = - \frac{\bar\mu}{2} \pm \frac{D(1-\bar\mu^2)+2\bar\mu^2(t_\k+D)}{4\calE_\k} . 
\label{weigthEcalk}
\eeq 
The spectral function satisfies the sum rule 
\beq 
\intinf d\w\, A(\k,\w) = -\bar\mu ,
\label{Anorm} 
\eeq 
which is consistent with the general result~\cite{NDbook1}
\beq 
\intinf d\w\, A(\k,\w) = \mean{[\hat\psi_\k , \hat\psi^\dagger_\k ]} ,
\eeq 
given the commutation relations~(\ref{CRhc}). We recover the usual normalization to unity only for $\bar\mu\to-1$ (and in the trivial Mott insulator $\nmi=0$). When $\bar\mu>-1$, part of the spectral weight is suppressed by the hard-core constraint that prohibits excitations with two or more bosons on the same site. 

The momentum distribution is given by (ignoring the contribution of the condensate) 
\beq 
n_\k = -\int_{-\infty}^0 d\w\, A(\k,\w) = - \calS(-\calE_\k) . 
\eeq
This expression, with Eq.~(\ref{weigthEcalk}), was previously obtained from a spin-wave analysis of the equivalent XY model~\cite{Coletta12}. For $\mu$ near $\mu_+=-D$ (low-density limit), there is no regular part since $n_\k$ vanishes in the trivial Mott insulator so that $\nksing=n_\k$. For $\mu$ near $\mu_-=D$, we use the normalization condition~(\ref{Anorm}) to write the momentum distribution as $n_\k=\calS(\calE_\k)+\bar\mu$ and consider the band $\calE_\k$ as an occupied hole band. This leads us to define the singular part of the momentum distribution as
\begin{align}
	\nksing &= -\calS(-\calE_\k) - \bar\mu \theta(\bar\mu) \nonumber\\  
	&= \calS(\calE_\k) + \bar\mu [1 - \theta(\bar\mu) ] .
	\label{momentdist_hc} 
\end{align}
It is identical for the particle-doped Mott insulator $\nmi=0$ and the hole-doped Mott insulator $\nmi=1$, as required by particle-hole symmetry, 
\begin{align}
	\mean{\hat\psi^\dagger_\r \hat\psi_\r}\bigl|_{\mu=\mu_++\Delta\mu} &= 
	\mean{\hat\psi_\r \hat\psi^\dagger_\r}\bigl|_{\mu=\mu_--\Delta\mu}  \nonumber\\ 
	&= 
	\mean{\hat\psi^\dagger_\r \hat\psi_\r}\bigl|_{\mu=\mu_--\Delta\mu} - \bar\mu ,
\end{align}
using~(\ref{CRhc}), i.e. 
\beq 
n_\k\bigl|_{\mu=\mu_++\Delta\mu} = n_\k\bigl|_{\mu=\mu_--\Delta\mu} - \bar\mu , 
\eeq
where $0\leq\Delta\mu<D$. 

\begin{figure}
	\centerline{\includegraphics[width=7cm]{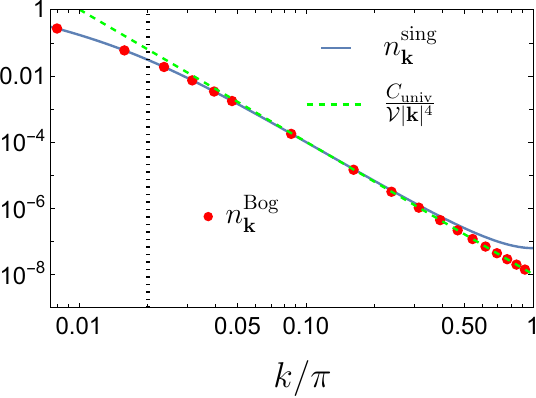}}
	\caption{Same as Fig.~\ref{fig_momentum} but for hard-core bosons near the transition between the superfluid state and the Mott insulator (with $\nmi=0$ or $\nmi=1$): $|\mu-\mu_\alpha|=10^{-3}\,D$. The Bogoliubov distribution $n_{\k}^{\rm Bog}$ is given by~(\ref{hc6}).}
	\label{fig_momentum_hardcore} 
\end{figure} 

When $|\mu-\mu_\alpha|\ll D$, we recover the Bogoliubov expression, 
\begin{equation} 
\nksing \simeq  - \half + \frac{t_\k+D + |\mu-\mu_\alpha|}{2[(t_\k+D)(t_\k+D+2|\mu-\mu_\alpha|]^{1/2}} ,  
\end{equation} 
of the momentum distribution of bosons with free dispersion $t_\k+D$ and chemical potential $|\mu-\mu_\alpha|$. For $|\k|$ not too close to the Brillouin zone boundary, we can approximate $t_\k+D$ by the quadratic dispersion $\epsk=\k^2/2\mlat$, so that 
\begin{align}
\nksing &\simeq - \half + \frac{\epsk + |\mu-\mu_\alpha|}{2[\epsk(\epsk+2|\mu-\mu_\alpha|)]^{1/2}} \nonumber \\ 
&\simeq \frac{\cuni}{\calV|\k|^4} \quad (|\k|\gg k_*) \label{hc6} ,
\end{align}
where $k_*=2(\mlat|\mu-\mu_\alpha|)^{1/2}$ and $\cuni$ is the universal contact defined in~(\ref{contacthc}). In Fig.~\ref{fig_momentum_hardcore}, we show that the momentum distribution~(\ref{momentdist_hc}) is well approximated by the Bogoliubov form~(\ref{hc6}) and exhibits the $\cuni/\calV|\k|^4$ tail over a large momentum range. The momentum distribution does not perfectly satisfy the sum rule 
\beq 
n = n_0 + \int \frac{d^3k}{(2\pi)^3} n_\k 
\eeq 
in the superfluid phase~\cite{Coletta12} but, contrary to the case of the Bose-Hubbard model, this has no dramatic consequence for hard-core bosons since $n_\k$ is fully determined by $\calS(-\calE_\k)\equiv\calS(-\calE_\k^-)$ due to the absence of the bands $\pm\calE_\k^+$. 

Thus, the hard-core boson model reproduces the thermodynamics of the Bose-Hubbard model in the low-density limit and for the hole-doped Mott insulator $\nmi=1$ in the limit $t/U\to 0$. The dispersion, spectral weight, effective mass and effective scattering length of the critical quasi-particles are also identical in the two models. Of course, this is expected on physical grounds. The fact that the hard-core boson model clearly shows that the singular momentum distribution $\nksing$, near the superfluid--Mott-insulator transition, exhibits a high-momentum tail $\cuni/\calV|\k|^4$ with a strength determined by the universal contact strongly supports the conclusion reached in the Bose-Hubbard model.

\section{Bosons in an optical lattice}
\label{sec_optical_lattice} 

A dilute Bose gas in an optical lattice is described by the effective low-energy  Hamiltonian 
\begin{align}
\hat H ={}& \int d^3r \biggl\{ \hat\psi^\dagger(\r) \biggl[ -\frac{\nablabf^2}{2m} + V_{\rm lat}(\r) \biggr] \hat\psi(\r) \nonumber\\ 
&+ \half \frac{4\pi a}{m} \hat\psi^\dagger(\r) \hat\psi^\dagger(\r) \hat\psi(\r) \hat\psi(\r) \biggr\} ,
\label{hamopt} 
\end{align}
where $V_{\rm lat}(\r)$ is the optical lattice potential with period $\ell$ (we will not set $\ell$ to unity in this section). The interaction potential between atoms is approximated by a short-range pseudo-potential with $a$ the $s$-wave scattering length in vacuum and $m$ the boson mass. The single-particle eigenstates are Bloch wave functions, and an appropriate superposition of Bloch states yields a set of Wannier functions that are well localized on the individual lattice sites. The single-band Bose-Hubbard Hamiltonian~(\ref{hamBHM}) provides us with an effective model that is valid when excitations to the second band can be neglected~\cite{Jaksch98}. The hopping amplitude $t$ and on-site interaction $U$, 
\begin{align}
&t = - \int d^3r \, w^*(\r-\r_i) \left[ - \frac{\nablabf^2}{2m} + V_{\rm lat}(\r) \right] w(\r-\r_j) , \nonumber\\ 
&U = \frac{4\pi a}{m}  \frac{W}{\ell^3} , \qquad W = \ell^3 \int d^3r \, |w(\r)|^4 ,
\label{tUdef} 
\end{align}
where $\{\r_i\}$ denote the lattice sites (i.e. the minima of the optical potential) and $\r_i,\r_j$ are nearest neighbors, can be expressed in terms of the Wannier function $w(\r)$ associated with the lowest-energy band. 

Contrary to the Bose-Hubbard model, which describes the system only at length scales larger than the optical lattice spacing $\ell$, the Hamiltonian~(\ref{hamopt}) can be used to understand the physics at length scales smaller than $\ell$. At these length scales, the boson system should be seen as a dilute gas subjected to a periodic potential whose period $\ell$ is much larger than typical microscopic length scales such as the $s$-wave scattering length $a$ (typically of the order of the van der Waals length~\cite{Braaten06}). At short distances ($\ll\ell$), we therefore expect the contact to be defined by the scattering length in vacuum $a$ rather than the effective $s$-wave scattering length $a^*_\alpha$ introduced in Sec.~\ref{sec_bhm}. This leads us to consider the short-distance universal contact
\beq 
\frac{\csd}{\calV} = 8\pi m \frac{\partial \Psing(\mu,m,a)}{\partial(1/a)} \biggl|_{\mu,m} ,
\label{Cshort} 
\eeq   
defined with the boson mass $m$ and a derivative with respect to $1/a$. We expect the gas to exhibit a high-momentum tail $\nksing\simeq \csd/\calV|\k|^4$ in the singular momentum distribution when $|\k|\gg 1/\ell$, with an amplitude set by $\csd$. In the Mott insulator, there are no singular parts of the pressure and momentum distribution, and $\csd$ vanishes. A calculation of the momentum distribution in this momentum range appears very difficult as it would require to include many energy bands of the optical lattice. However, the contact can be directly computed from~(\ref{Cshort}) since the pressure of the gas can be obtained from the Bose-Hubbard model. 

The reason for defining the contact in the Bose-Hubbard model from the singular part of the pressure is that only this part is associated with weakly interacting quasi-particles and thus takes the same form as the pressure of a dilute Bose gas provided we replace $m$ and $a$ by $m^*_\alpha$ and $a^*_\alpha$. The full pressure is dominated by the regular part $\Pmi$ and the momentum distribution $n_\k$ does not exhibit a $1/|\k|^4$ tail (Fig.~\ref{fig_nk}). However, in the continuum model, since the gas behaves as a dilute Bose gas at short distances, we expect a $1/|\k|^4$ tail in the momentum distribution $n_\k$ for $|\k|\gg 1/\ell$ even in the Mott insulator. This leads us to consider the full contact
\beq 
\frac{\cfull}{\calV} = 8\pi m \frac{\partial P(\mu,m,a)}{\partial(1/a)} \biggl|_{\mu,m} ,
\label{Cfull} 
\eeq    
defined in the usual way, i.e. from a derivative of the total pressure with respect to $1/a$~\cite{[{A general discussion of the critical behavior of the contact (defined in the usual way, i.e. from~(\ref{Cfull})) near phase transitions can be found in }]Chen14}. 
  
In the following, we denote by $\bar n$ the mean density of bosons and by $n=\bar n\ell^3$ (or $\nmi$) the mean number of bosons per site (i.e. per minimum of the optical lattice).

\subsection{Short-distance universal contact $\csd$}
\label{subsec_sdc}

In this section, we compute $\csd$ considering first the low-density limit and then the hole-doped Mott insulator $\nmi=1$. 

\subsubsection{Low-density limit} 

At low density, near the vacuum-superfluid transition, the chemical potential is close to $\mu_+=-D$ and the pressure is given by 
\beq 
P = \frac{\mlat}{8\pi \alat}(\mu-\mu_+)^2 , 
\eeq 
since $m^*_+=\mlat=1/2t\ell^2$ and $a^*_+=\alat=\ell/[8\pi(t/U+A)]$ [Eq.~(\ref{alat})]. The density is given by 
\beq 
\bar n = \frac{\partial P}{\partial \mu} = \frac{\mlat}{4\pi \alat}(\mu-\mu_+) .
\eeq 
Using 
\beq 
\frac{\partial}{\partial a} = \frac{\partial U}{\partial a} \frac{\partial}{\partial U} = \frac{U}{a}  \frac{\partial}{\partial U} ,
\eeq 
we obtain the two-body contact 
\begin{align}
\frac{\csd}{\calV} &= [\mlat(\mu-\mu_+)]^2 64\pi^2 W \frac{a^2t^2}{\ell^2 U^2} \nonumber\\
&= (4\pi a\bar n)^2 \frac{W}{(1+AU/t)^2} . 
\end{align}
Thus $\csd/\calV$ is equal to the contact $(4\pi a\bar n)^2$ in the absence of the optical lattice, corrected by a function of $t/U$ and a geometrical dimensionless factor $W$, which depends on the Wannier function of the first energy band of the optical lattice potential [Eq.~(\ref{tUdef})]. The limiting values of the contact, for weak and strong on-site interactions, are
\beq 
\frac{\csd}{\calV} = \llbrace 
\begin{array}{lll} (4\pi a\bar n)^2 W & \mbox{if} & U\ll t , \\ 
	(4\pi a\bar n)^2 W \frac{t^2}{A^2U^2} & \mbox{if} & U\gg t .
\end{array}
\right. 
\eeq  
As expected, the short-distance universal contact is determined by the scattering length in vacuum $a$. 

Since the pressure has no regular part in the low-density limit ($P$ vanishes in the trivial Mott insulator $\nmi=0$), the short-distance universal contact $\csd$ is equal to the full contact $C$.

\subsubsection{Hole-doped Mott insulator $\nmi=1$} 

For a doped Mott insulator with an arbitrary value of $\nmi$, the two-body contact is given by 
\begin{align} 
\frac{\csd}{\calV} ={}& 8\pi m \frac{\partial}{\partial(1/a)} \frac{m^*_\alpha}{8\pi a^*_\alpha}(\mu-\mu_\alpha)^2 \biggl|_{\mu,m} \nonumber\\
={}& (\mu-\mu_\alpha)^2 a^2 4\pi \frac{W}{\ell^3} \left( \frac{m^*_\alpha}{a^*_\alpha{}^2} \frac{\partial a^*_\alpha}{\partial U} - \frac{1}{a^*_\alpha}  \frac{\partial m^*_\alpha}{\partial U} \right) \nonumber\\ 
& + (\mu-\mu_\alpha)8\pi m^*_\alpha \frac{W a^2}{\ell^3 a^*_\alpha}  
\frac{\partial \mu_\alpha}{\partial U} .
\end{align}
In the following, we consider the hole-doped Mott insulator $\nmi=1$ in the limit $D\ll D_c$. The FRG shows that $a^*_-\simeq \alat$ (Fig.~\ref{fig_mastar}), while  $m^*_-\simeq \mlat(1-4D/U)$ [Eq.~(\ref{m1})]. This leads to 
\begin{align}
\frac{\csd}{\calV} ={}& \left[4\pi a \frac{(n-\nmi)}{\ell^3}\right]^2 W \left( \frac{t}{U} \right)^2 \left( \frac{1}{A^2}-\frac{24}{A} \right) \nonumber\\
& + \left( 16\pi^2 a^2 \frac{|n-\nmi|}{\ell^6} \right) 144 W \frac{t^2}{U^2} .
\label{Cuniv1}
\end{align}
Note that the first contribution is negative since $1/A^2-24/A<0$. The second, dominant (for small $|n-\nmi|$), contribution comes from the dependence of $\mu_-$ on $a$.

\subsection{Full contact $\cfull$} 
\label{subsec_Cfull} 

When computing the full contact away from the low-density limit, we can ignore the singular part of the pressure, which gives a subleading contribution, and thus assume that the system is in the Mott insulating phase. As we will see, the strong-coupling RPA gives a contact which vanishes when $\nmi=1$. To obtain a non-vanishing result in that case, we must include Gaussian fluctuations of the hopping term. The calculation, detailed in Appendix~\ref{sec_pressure}, gives the pressure 
\beq 
P = \frac{1}{\ell^3} \left[ \mu \nmi - \frac{U}{2} \nmi(\nmi-1) + 6 \nmi (\nmi+1) \frac{t^2}{U} \right] 
\label{Pmott} 
\eeq 
to order $t^2/U$, where the first two terms come from the mean-field (RPA) pressure and the last one from Gaussian fluctuations. We deduce the contact 
\begin{align}
\frac{\cfull}{\calV} ={}& 16\pi^2 a^2 \frac{\nmi(\nmi-1)}{\ell^6} W \nonumber\\ 
&+ 48\pi \nmi(\nmi+1)  \frac{mat^2}{\ell^3 U} .
\end{align} 
The last term can be rewritten by introducing the recoil energy $E_R=k^2/2m$ with $k=2\pi/\lambda$ and $\lambda=2\ell$ the wavelength of the laser light creating the optical lattice,
\begin{align}
	\frac{\cfull}{\calV} ={}& 16\pi^2 a^2 \frac{\nmi(\nmi-1)}{\ell^6} W \nonumber\\ 
	&+ 24\pi^3 \nmi(\nmi+1) \frac{a}{\ell^5} \frac{t^2}{UE_R} .
\label{Cfull1} 
\end{align}
When $\nmi=1$, the contact is fully determined by quantum fluctuations associated with the hopping term. In other cases, the ratio between the second and first terms in~(\ref{Cfull1}) is of order 
\beq 
\frac{\ell U}{aE_R} \frac{t^2}{U^2} . 
\label{ratio1}
\eeq 
Although $U\ell/aE_R$ is typically in the range $10-100$ for small values of $\nmi$~\cite{Jaksch98}, the smallness of $t/U\leq (1/6)(2\nmi+1-2\sqrt{\nmi^2+\nmi})$ ensures that the ratio~(\ref{ratio1}) is much smaller than unity; the contact is dominated by the RPA (mean-field) contribution. 

Comparing the short-distance universal contact to the full contact, we obtain 
\beq 
\frac{\csd}{\cfull} \sim \llbrace 
\begin{array}{lll}
	|n-\nmi| \frac{t^2}{U^2} & \mbox{if} & \nmi\geq 2 , \\ 
	|n-\nmi| W \frac{aE_R}{\ell U} & \mbox{if} & \nmi=1 ,
\end{array} 
\right. 
\eeq 
i.e. $\csd/\cfull\ll |n-\nmi|$, assuming that Eq.~(\ref{Cuniv1}), which was derived for $\nmi=1$, gives the correct order of magnitude for any $\nmi\geq 2$. As expected, the short-distance contact associated with the singular part of the pressure is small compared to the full contact.

\section{Conclusion} 

The strong-coupling RPA theory of the Bose-Hubbard model is based on a mean-field treatment of the hopping term, while onsite fluctuations are taken into account exactly. Although we have focused on the generic Mott transition, it also applies to the transition induced by a change of the ratio $t/U$ at fixed density. In Refs.~\cite{Rancon11a,Rancon11b,Rancon12a,Rancon12b,Rancon12d,Rancon12d,Rancon13b}, the strong-coupling RPA theory was used as the initial condition of the flow in the nonperturbative FRG approach. In this paper, we have shown that many qualitative results can be obtained without integrating the nonperturbative flow equations. In particular, the strong-coupling RPA captures the universal behavior of the superfluid state near the phase transition to the Mott insulator. Moreover, it allows one to define a universal contact that controls the high-momentum tail $\cuni/|\k|^4$ of the singular part of the momentum distribution function, a physical quantity which is not readily available~\cite{not4} from the nonperturbative FRG approach of Refs.~\cite{Rancon11a,Rancon11b,Rancon12a,Rancon12b,Rancon12d,Rancon12d}. The strong-coupling RPA also applies to hard-core bosons, giving results similar to those of the Bose-Hubbard model. 

The existence of a (universal) two-body contact in a strongly correlated Bose gas near the Mott transition is due to the excess of particles, with respect to the Mott insulator, behaving as a dilute Bose gas, which is a consequence of the superfluid--Mott-insulator transition belonging to the dilute-Bose-gas universality class. In a continuum model of bosons in an optical lattice, the definition of the universal contact can be extended to length scales shorter than the lattice spacing $\ell$; in addition, one can define a full contact that controls the $1/|\k|^4$ tail of the full momentum distribution function in the range $|\k|\gg 1/\ell$. In a companion paper~\cite{Bhateja25}, we have argued that universal and full contacts can be measured in Bose gases in optical lattices, and in magnetic insulators.

\section*{Acknowledgment} 

We thank D. Cl\'ement and T. Chalopin for useful discussions and comments, and B. Capogrosso-Sansone for providing us with the QMC data~\cite{Capogrosso07} shown in Fig.~\ref{fig_mastar}. 
MB was supported by a Charpak fellowship and the Labex CEMPI (ANR-11-LABX-0007-01). AR is supported in part by an IEA CNRS project, and by the “PHC COGITO” program (Project No.: 49149VE) funded by the French Ministry for Europe and Foreign Affairs, the French Ministry for Higher Education and Research, and the Croatian Ministry of Science and Education.

\appendix

\section{Hubbard-Stratonovich transformation}
\label{app_HST} 

In this Appendix, we show how Eqs.~(\ref{Srpa}) and (\ref{Zrpa}) can be obtained from a Hubbard-Stratonovich transformation. We start from the action~(\ref{Sbhm}) and decouple the hopping term by means of an auxiliary complex field $\varphi_\r$. We thus rewrite the partition function as 
\begin{widetext}
\beq 
Z[J^*,J] = \calN \int \calD[\psi^*,\psi,\varphi^*,\varphi] \, e^{- \Sloc[\psi^*,\psi] -  \inttau \{ \sum_{\r,\r'} \varphi^*_\r t^{-1}_{\r,\r'} \varphi_{\r'} - i \sum_\r (\varphi^*_\r \psi_\r + \cc) - \sum_\r (J^*_\r \psi_\r + \cc) \} } ,
\label{Zdef} 
\eeq
where $\calN=\det(t^{-1})$. By performing the Gaussian functional integration over the field $\varphi$, we recover the original action $S[\psi^*,\psi]$ of the Bose-Hubbard model. This integration can be carried out only if $t^{-1}_{\r,\r'}$ is a positive matrix, which is not the case since the matrix $t_{\r,\r'}$ has eigenvalues $t_\k=-2t(\cos k_x+\cos k_y+\cos k_z)$. These eigenvalues can be made positive by adding a (local) term $C\psi^*_\r\psi_\r$ with $C>6t$ in the hopping part of the action and subtracting this term from the local action $\Sloc$ in order to leave the complete action unchanged. We ignore this issue which is not relevant to the discussion that follows. 

In the RPA, the functional integral on $\varphi$ is realized {\it via} a saddle-point approximation (the constant $\calN$ can then be omitted), 
\beq 
\Zrpa[J^*,J] = \int \calD[\psi^*,\psi] \, e^{- \Sloc[\psi^*,\psi] -  \inttau \{ \sum_{\r,\r'} \varphi^*_\r t^{-1}_{\r,\r'} \varphi_{\r'} - i \sum_\r (\varphi^*_\r \psi_\r + \cc) - \sum_\r (J^*_\r \psi_\r + \cc) \} } ,
\label{Zrpa2} 
\eeq  
\end{widetext}
where the value of the auxiliary field is obtained from the saddle-point equations
\beq 
\frac{\delta \ln \Zrpa[J^*,J]}{\delta\varphi^*_\r(\tau)} =
\frac{\delta \ln \Zrpa[J^*,J]}{\delta\varphi_\r(\tau)} = 0 , 
\eeq 
i.e. 
\beq
\varphi_\r = i \sum_{\r'} t_{\r,\r'} \phi_{\r'} , \quad
\varphi^*_\r = i \sum_{\r'} t_{\r,\r'} \phi^*_{\r'} ,
\label{varphiSP} 
\eeq 
where $\phi^{(*)}_\r=\mean{\psi^{(*)}_\r}$. Note that the field $i\varphi_\r^{(*)}$ is real, which ensures that the hopping part of the action in~(\ref{Zrpa2}) is real. Inserting~(\ref{varphiSP}) into Eq.~(\ref{Zrpa2}), we recover the RPA action~(\ref{Srpa}).

\section{Local vertices $\Gamma_{\rm loc}^{(2)}$ and $\Gamma_{\rm loc}^{(4)}$} 
\label{sec_local} 

In the local limit $t=0$, the one- and two-particle Green functions can be calculated considering a single site and the states $|p\rangle=(p!)^{-1/2}(\hat\psi^\dagger)^p|0\rangle$ ($p\geq 0$ integer) which are eigenstates of the local Hamiltonian $\Hloc$ with eigenvalues $\epsilon_p=-\mu p+(U/2)p(p-1)$~\cite{Sengupta05}. The two-point local vertex $\Gamma^{(2)}_{\rm loc}(i\wn)$ is equal to $-1/\Gloc(i\wn)$ with $\Gloc$ given by~(\ref{Gloc}). 

The static limit of the (connected) two-particle Green function is given by~\cite{Sengupta05}
\begin{align}
	\bar G_{\rm loc}^{(4)}  
	={}& - \frac{4(\nmi+1)(\nmi+2)}{[2\mu-(2\nmi+1)U](U\nmi-\mu)^2} \nonumber \\ &
	-  \frac{4\nmi(\nmi-1)}{[\mu-U(\nmi
		-1)]^2[U(2\nmi-3)-2\mu]} \nonumber \\ &
	+ \frac{4\nmi(\nmi+1)}{(\mu-U\nmi)[-\mu+U(\nmi-1)]^2} \nonumber \\ & 
	+ \frac{4\nmi(\nmi+1)}{(\mu-U\nmi)^2[-\mu+U(\nmi-1)]} \nonumber \\ &
	+ \frac{4\nmi^2}{[-\mu+U(\nmi-1)]^3}
	+ \frac{4(\nmi+1)^2}{(\mu-U\nmi)^3} .
\end{align}
The static limit $\bar\Gamma^{(4)}_{\rm loc}$ of the four-point vertex is equal
to $-\bar G^{(4)}_{\rm loc}/\Gloc(i\wn=0)^4$. In the trivial Mott insulator $\nmi=0$ (i.e. the vacuum), one has 
\beq 
g = \half \bar\Gamma^{(4)}_{\rm loc} = \frac{2\mu U}{2\mu - U} \qquad (\nmi=0) . 
\label{gnzero} 
\eeq 
In the low-density limit, where $\mu\simeq -6t$, one deduces $g\simeq U$ if $U\ll t$. 

In the hard-core limit $U\to\infty$, the two possible Mott phases are $\nmi=0$ and $\nmi=1$ for $\mu<0$ and $\mu>0$, respectively. The static local Green function becomes $\Gloc(i\wn=0)=-1/|\mu|$, while $\bar G_{\rm loc}^{(4)}=-4/|\mu|^3$. One has
\beq 
g = 2|\mu|  ,
\label{ghardcore} 
\eeq 
in agreement with (\ref{gnzero}) in the same limit.

\section{Excitation gap and effective mass in the Mott insulator} 
\label{app_Delta_mstar}  

The excitation gap obtained from the pole of the propagator is given by~(\ref{Delta2}). Since $\Gloc(i\wn)$ is actually a function of $i\wn+\mu$, $\dmu\Gloc(i\w)=\partial_{i\w}G(i\w)$. It follows that 
\begin{align} 
	1+D\Gloc(0) \simeq{}& 1+D\Gloc(0)|_{\delta\mu=\delta\mu_\alpha} \nonumber\\ & + D (\delta\mu-\delta\mu_\alpha)\dmu\Gloc(0) \nonumber \\ 
	\simeq{}& D (\delta\mu-\delta\mu_\alpha) \Gloc'(0) 
\end{align}
for $\delta\mu$ close to $\delta\mu_\alpha$ (for a given $\alpha$). We conclude that the expression of $\Delta_\alpha$ in~(\ref{Delta2}) agrees with~(\ref{Delta1}) in the limit $\delta\mu-\delta\mu_\alpha\to 0$. 

At the transition, $1+D\Gloc(0)\to 0$ so that the effective mass~(\ref{m2}) obtained from the pole of the propagator satisfies  
\beq
\frac{\mlat}{m^*_\alpha} \simeq - \alpha \frac{\Gloc(0)^2}{\Gloc'(0)} . 
\eeq 
Using 
\beq 
\Gloc(0) = \frac{\delta\mu+Ux}{\delta\mu^2- U^2/4} 
\eeq
and 
\beq 
\Gloc'(0) = - \frac{\delta\mu^2 + 2 \delta\mu U x + U^2/4}{(\delta\mu^2 - U^2/4)^2} ,
\eeq 
we obtain 
\beq 
\frac{\mlat}{m^*_\alpha} = \alpha \frac{(\delta\mu_\alpha+Ux)^2}{\delta\mu_\alpha^2 + 2 \delta\mu_\alpha U x + U^2/4 } 
\eeq 
for $\delta\mu\to \delta\mu_\alpha$. On the other hand, Eq.~(\ref{deltamupm}) implies
\beq 
\begin{gathered}
\delta\mu_\alpha + Ux = \half ( \alpha A + B ) , \\ 
\delta\mu_\alpha^2 + 2 \delta\mu_\alpha U x + \frac{U^2}{4} = \frac{A}{2} (A+ \alpha B) , 
\end{gathered}
\eeq 
where $A=(D^2-4UDx+U^2)^{1/2}$ and $B=2Ux-D$, and we finally obtain 
\beq 
\frac{\mlat}{m^*_\alpha} = \frac{\alpha}{2} \left( 1 + \alpha \frac{B}{A} \right) 
\eeq 
in agreement with the expression~(\ref{m1}) deduced from the energy $E^\alpha_\alpha$.

\section{Hard-core bosons: local limit} 
\label{sec_hardcore} 
 
In this appendix, we discuss the local limit ($t=0$) of the hard-core boson model defined by~(\ref{ham_hc}). In the presence of a time-independent source, the single-site Hamiltonian reads 
\beq 
\Hloc = -\mu \hat\psi^\dagger \hat\psi - J^* \hat\psi - J\hat\psi^\dagger 
\eeq 
and the Hilbert space is restricted to the vacuum state $\ket{0}$ and the singly-occupied state $\ket{1}=\hat\psi^\dagger\ket{0}$. The two eigenstates are given by 
\beq 
\ket{\pm} = \frac{|J|^2}{|J|^2+ E_\pm^2} \left( \ket{0} - \frac{E_\pm}{J^*} \ket{1} \right)
\eeq 
with eigenenergies 
\beq 
E_\pm = - \frac{\mu}{2} \pm \half \sqrt{\mu^2 + 4|J|^2} . 
\label{Epmdef} 
\eeq 
In the zero-temperature limit, the expectation value $\phi=\mean{\hat\psi}$ of the boson operator is given by 
\beq 
\phi = - \frac{\partial E_-}{\partial J^*} = \frac{J}{\sqrt{\mu^2 + 4|J|^2}} . 
\label{phivsJ} 
\eeq 
The effective potential is given by the Legendre transform of the ground-state energy, 
\begin{align}
\Vloc(|\phi|^2) &= E_- + J^* \phi + \phi^* J \nonumber\\ 
&= - \frac{\mu}{2} - \frac{|\mu|}{2} \sqrt{1-4|\phi|^2} ,
\label{Vloc} 
\end{align}
where we have used~(\ref{phivsJ}) to express $J^{(*)}$ as a function of $\phi^{(*)}$. 

Normal and anomalous propagators are defined by
\beq 
\begin{split}
G_{\rm n}(\tau) &= - \mean{T_\tau \hat\psi(\tau) \hat\psi^\dagger(0)} + |\mean{\hat\psi}|^2 , \\ 
G_{\rm an}(\tau) &= - \mean{T_\tau \hat\psi(\tau) \hat\psi(0)} + \mean{\hat\psi}^2  .
\end{split}
\eeq 
A straightforward calculation gives 
\begin{align}
G_{\rm n}(i\wn) &= - \frac{|A_{+-}|^2}{i\wn+E_+-E_-} +  \frac{|A_{-+}|^2}{i\wn-E_+ +E_-} , \nonumber\\ 
G_{\rm an}(i\wn) &= - A_{+-} A_{-+} \frac{2(E_+ - E_-)}{\wn^2+(E_+-E_-)^2} ,
\end{align}
where 
\beq 
\begin{split}
A_{+-} = \bra{+} \hat\psi \ket{-} = - E_- \frac{\phi}{|\mu|} \sqrt{ |\phi|^{-2}-4 } , \\
A_{-+} = \bra{-} \hat\psi \ket{+} = - E_+ \frac{\phi}{|\mu|} \sqrt{ |\phi|^{-2}-4 } . 
\end{split}
\label{Ahc} 
\eeq 
By inverting the matrix 
\beq
- 
\begin{pmatrix}
	G_{\rm n}(i\wn) & G_{\rm an}(i\wn) \\ 
	G_{\rm an}(i\wn)^* & G_{\rm n}(-i\wn) 
\end{pmatrix} ,
\eeq 
one obtains the two-point vertices 
\begin{align}
&\Gamma^{(2)}_{\phi^*\phi}(i\wn) = \frac{|A_{+-}|^2(i\wn+\Delta E) -|A_{-+}|^2(i\wn-\Delta E) }{ 1-4|\phi|^2} , \nonumber\\   
&\Gamma^{(2)}_{\phi^*\phi^*}(i\wn) = - A_{+-} A_{-+} \frac{2\Delta E}{1-4|\phi|^2} 
\end{align}
and
\beq 
\begin{split}
	&\Gamma^{(2)}_{\phi\phi^*}(i\wn) = \Gamma^{(2)}_{\phi^*\phi}(-i\wn) , \\  
	&\Gamma^{(2)}_{\phi\phi}(i\wn) =  [\Gamma^{(2)}_{\phi^*\phi^*}(i\wn)]^*, 
\end{split}
\eeq 
with $\Delta E=E_+-E_-$. Using~(\ref{Epmdef}), (\ref{phivsJ}) and (\ref{Ahc}), we finally obtain
\beq 
\begin{split}
	&\Gamma^{(2)}_{\phi^*\phi}(i\wn) =  i\wn \frac{\sgn(\mu)}{\sqrt{1-4|\phi|^2}} + \frac{\partial^2 \Vloc(|\phi|^2)}{\partial \phi^* \partial\phi}, \\ 
	&\Gamma^{(2)}_{\phi^*\phi^*}(i\wn) =  \frac{\partial^2 \Vloc(|\phi|^2)}{\partial \phi^*{}^2},
\end{split}
\label{G2loc1}
\eeq 
where 
\beq 
\begin{split}
\frac{\partial^2 \Vloc(|\phi|^2)}{\partial \phi^* \partial\phi} &= \Vloc'(|\phi|^2) +
|\phi|^2  \Vloc''(|\phi|^2) , \\ 
\frac{\partial^2 \Vloc(|\phi|^2)}{\partial \phi^*{}^2} &= \phi^2  \Vloc''(|\phi|^2) . 
\end{split}
\label{G2loc2}
\eeq 
Equations~(\ref{Vloc}) and (\ref{G2loc1},\ref{G2loc2}) agree with the results of Ref.~\cite{Rancon14a}.

\section{Pressure of the Mott insulator}
\label{sec_pressure} 

In this appendix, we compute the pressure in the Mott insulator by including fluctuations of the auxiliary field about its mean-field value $\varphi_\r=0$. This can be done by starting from the partition function~(\ref{Zdef}) with $J^*=J=0$ and integrating out the $\psi$ field in a cumulant expansion, 
\begin{widetext}
\begin{align}
Z &= \calN \int \calD[\psi^*,\psi,\varphi^*,\varphi] \, e^{- \Sloc[\psi^*,\psi] - \inttau  \sum_{\r,\r'} \varphi^*_\r t^{-1}_{\r,\r'} \varphi_{\r'} - S'[\psi^*,\psi,\varphi^*,\varphi]}  \nonumber\\ 
&= \calN \Zloc \int \calD[\varphi^*,\varphi] \, e^{ -  \inttau  \sum_{\r,\r'}\varphi^*_\r t^{-1}_{\r,\r'} \varphi_{\r'} + \half \mean{S'[\psi^*,\psi,\varphi^*,\varphi]^2}_{\rm loc}}  \nonumber\\
&=  \calN \Zloc \int \calD[\varphi^*,\varphi] \, e^{ - \sum_{\k,\wn} |\varphi_\k(i\wn)|^2 [t_\k^{-1} - \Gloc(i\wn)] } \nonumber\\ 
&= \Zloc \prod_{\k,\wn} [1-t_\k \Gloc(i\wn)]^{-1} , 
\end{align}
\end{widetext}
where 
\beq 
S'[\psi^*,\psi,\varphi^*,\varphi] = - i  \inttau \sum_\r (\varphi^*_\r \psi_\r + \cc) 
\eeq 
and $\Zloc$ is the partition function in the local limit. We have used the fact that the first cumulant $\mean{S'}_{\rm loc}$ vanishes. We thus obtain the grand potential
\beq
\Omega = \Omega_{\rm loc} + \frac{1}{\beta} \sum_{\k,\wn} \ln[1-t_\k \Gloc(i\wn)] e^{i\wn 0^+}
\eeq
where the Matsubara sum can be written as 
\beq
\frac{1}{\beta} \sum_{\k,\wn} \ln \left[ \frac{(i\wn-E^+_\k)(i\wn-E^-_\k)}{\bigl(i\wn+\delta\mu+\frac{U}{2}\bigr) \bigl(i\wn+\delta\mu-\frac{U}{2}\bigr)} \right] e^{i\wn 0^+}
\eeq
and we have added the usual convergence factor $e^{i\wn 0^+}$. Using 
\beq 
\begin{array}{lll} 
\dfrac{1}{\beta} \dsum_{\wn} \ln(-i\wn+a) e^{i\wn 0^+} = \frac{1}{\beta} \ln(1-e^{-\beta a}) & \mbox{if} & a>0 ,  \\ 
\dfrac{1}{\beta} \dsum_{\wn} \ln(i\wn-a) e^{i\wn 0^+} = \frac{1}{\beta} \ln(e^{-\beta a}-1) & \mbox{if} & a<0 ,
\end{array}
\eeq 
we obtain the pressure $P=-\Omega/N$ in the zero-temperature limit, 
\begin{align}
P ={}& \mu \nmi - \frac{U}{2} \nmi(\nmi -1) \nonumber\\ & 
+ \frac{1}{\calV} \sum_\k \left(E^-_\k + \delta\mu + \frac{U}{2} \right) ,
\end{align}
which yields~(\ref{Pmott}) to order $t^2/U$. 

%


\begin{thebibliography}{81}%
\makeatletter
\providecommand \@ifxundefined [1]{%
 \@ifx{#1\undefined}
}%
\providecommand \@ifnum [1]{%
 \ifnum #1\expandafter \@firstoftwo
 \else \expandafter \@secondoftwo
 \fi
}%
\providecommand \@ifx [1]{%
 \ifx #1\expandafter \@firstoftwo
 \else \expandafter \@secondoftwo
 \fi
}%
\providecommand \natexlab [1]{#1}%
\providecommand \enquote  [1]{``#1''}%
\providecommand \bibnamefont  [1]{#1}%
\providecommand \bibfnamefont [1]{#1}%
\providecommand \citenamefont [1]{#1}%
\providecommand \href@noop [0]{\@secondoftwo}%
\providecommand \href [0]{\begingroup \@sanitize@url \@href}%
\providecommand \@href[1]{\@@startlink{#1}\@@href}%
\providecommand \@@href[1]{\endgroup#1\@@endlink}%
\providecommand \@sanitize@url [0]{\catcode `\\12\catcode `\$12\catcode
  `\&12\catcode `\#12\catcode `\^12\catcode `\_12\catcode `\%12\relax}%
\providecommand \@@startlink[1]{}%
\providecommand \@@endlink[0]{}%
\providecommand \url  [0]{\begingroup\@sanitize@url \@url }%
\providecommand \@url [1]{\endgroup\@href {#1}{\urlprefix }}%
\providecommand \urlprefix  [0]{URL }%
\providecommand \Eprint [0]{\href }%
\providecommand \doibase [0]{https://doi.org/}%
\providecommand \selectlanguage [0]{\@gobble}%
\providecommand \bibinfo  [0]{\@secondoftwo}%
\providecommand \bibfield  [0]{\@secondoftwo}%
\providecommand \translation [1]{[#1]}%
\providecommand \BibitemOpen [0]{}%
\providecommand \bibitemStop [0]{}%
\providecommand \bibitemNoStop [0]{.\EOS\space}%
\providecommand \EOS [0]{\spacefactor3000\relax}%
\providecommand \BibitemShut  [1]{\csname bibitem#1\endcsname}%
\let\auto@bib@innerbib\@empty
\bibitem [{\citenamefont {Greiner}\ \emph {et~al.}(2002)\citenamefont
  {Greiner}, \citenamefont {Mandel}, \citenamefont {Esslinger}, \citenamefont
  {H\"ansch},\ and\ \citenamefont {Bloch}}]{Greiner02}%
  \BibitemOpen
  \bibfield  {author} {\bibinfo {author} {\bibfnamefont {M.}~\bibnamefont
  {Greiner}}, \bibinfo {author} {\bibfnamefont {O.}~\bibnamefont {Mandel}},
  \bibinfo {author} {\bibfnamefont {T.}~\bibnamefont {Esslinger}}, \bibinfo
  {author} {\bibfnamefont {T.~W.}\ \bibnamefont {H\"ansch}},\ and\ \bibinfo
  {author} {\bibfnamefont {I.}~\bibnamefont {Bloch}},\ }\bibfield  {title}
  {\bibinfo {title} {{Quantum phase transition from a superfluid to a Mott
  insulator in a gas of ultracold atoms}},\ }\href
  {https://doi.org/doi:10.1038/415039a} {\bibfield  {journal} {\bibinfo
  {journal} {Nature}\ }\textbf {\bibinfo {volume} {415}},\ \bibinfo {pages}
  {39} (\bibinfo {year} {2002})}\BibitemShut {NoStop}%
\bibitem [{\citenamefont {Jim\'enez-Garc\'\i{}a}\ \emph
  {et~al.}(2010)\citenamefont {Jim\'enez-Garc\'\i{}a}, \citenamefont {Compton},
  \citenamefont {Lin}, \citenamefont {Phillips}, \citenamefont {Porto},\ and\
  \citenamefont {Spielman}}]{Jimenez10}%
  \BibitemOpen
  \bibfield  {author} {\bibinfo {author} {\bibfnamefont {K.}~\bibnamefont
  {Jim\'enez-Garc\'\i{}a}}, \bibinfo {author} {\bibfnamefont {R.~L.}\
  \bibnamefont {Compton}}, \bibinfo {author} {\bibfnamefont {Y.-J.}\
  \bibnamefont {Lin}}, \bibinfo {author} {\bibfnamefont {W.~D.}\ \bibnamefont
  {Phillips}}, \bibinfo {author} {\bibfnamefont {J.~V.}\ \bibnamefont
  {Porto}},\ and\ \bibinfo {author} {\bibfnamefont {I.~B.}\ \bibnamefont
  {Spielman}},\ }\bibfield  {title} {\bibinfo {title} {{Phases of a
  Two-Dimensional Bose Gas in an Optical Lattice}},\ }\href
  {https://doi.org/10.1103/PhysRevLett.105.110401} {\bibfield  {journal}
  {\bibinfo  {journal} {Phys. Rev. Lett.}\ }\textbf {\bibinfo {volume} {105}},\
  \bibinfo {pages} {110401} (\bibinfo {year} {2010})}\BibitemShut {NoStop}%
\bibitem [{\citenamefont {Becker}\ \emph {et~al.}(2010)\citenamefont {Becker},
  \citenamefont {Soltan-Panahi}, \citenamefont {Kronj\"ager}, \citenamefont
  {D\"orscher}, \citenamefont {Bongs},\ and\ \citenamefont
  {Sengstock}}]{Becker10}%
  \BibitemOpen
  \bibfield  {author} {\bibinfo {author} {\bibfnamefont {C.}~\bibnamefont
  {Becker}}, \bibinfo {author} {\bibfnamefont {P.}~\bibnamefont
  {Soltan-Panahi}}, \bibinfo {author} {\bibfnamefont {J.}~\bibnamefont
  {Kronj\"ager}}, \bibinfo {author} {\bibfnamefont {S.}~\bibnamefont
  {D\"orscher}}, \bibinfo {author} {\bibfnamefont {K.}~\bibnamefont {Bongs}},\
  and\ \bibinfo {author} {\bibfnamefont {K.}~\bibnamefont {Sengstock}},\
  }\bibfield  {title} {\bibinfo {title} {Ultracold quantum gases in triangular
  optical lattices},\ }\href {https://doi.org/10.1088/1367-2630/12/6/065025}
  {\bibfield  {journal} {\bibinfo  {journal} {New J. Phys.}\ }\textbf {\bibinfo
  {volume} {12}},\ \bibinfo {pages} {065025} (\bibinfo {year}
  {2010})}\BibitemShut {NoStop}%
\bibitem [{\citenamefont {Trotzky}\ \emph {et~al.}(2010)\citenamefont
  {Trotzky}, \citenamefont {Pollet}, \citenamefont {Gerbier}, \citenamefont
  {Schnorrberger}, \citenamefont {Bloch}, \citenamefont {Prokof'ev},
  \citenamefont {Svistunov},\ and\ \citenamefont {Troyer}}]{Trotzky10}%
  \BibitemOpen
  \bibfield  {author} {\bibinfo {author} {\bibfnamefont {S.}~\bibnamefont
  {Trotzky}}, \bibinfo {author} {\bibfnamefont {L.}~\bibnamefont {Pollet}},
  \bibinfo {author} {\bibfnamefont {F.}~\bibnamefont {Gerbier}}, \bibinfo
  {author} {\bibfnamefont {U.}~\bibnamefont {Schnorrberger}}, \bibinfo {author}
  {\bibfnamefont {I.}~\bibnamefont {Bloch}}, \bibinfo {author} {\bibfnamefont
  {N.~V.}\ \bibnamefont {Prokof'ev}}, \bibinfo {author} {\bibfnamefont
  {B.}~\bibnamefont {Svistunov}},\ and\ \bibinfo {author} {\bibfnamefont
  {M.}~\bibnamefont {Troyer}},\ }\bibfield  {title} {\bibinfo {title}
  {{Suppression of the critical temperature for superfluidity near the Mott
  transition}},\ }\href {https://doi.org/doi:10.1038/nphys1799} {\bibfield
  {journal} {\bibinfo  {journal} {Nat. Phys.}\ }\textbf {\bibinfo {volume}
  {6}},\ \bibinfo {pages} {998} (\bibinfo {year} {2010})}\BibitemShut {NoStop}%
\bibitem [{\citenamefont {Mark}\ \emph {et~al.}(2011)\citenamefont {Mark},
  \citenamefont {Haller}, \citenamefont {Lauber}, \citenamefont {Danzl},
  \citenamefont {Daley},\ and\ \citenamefont {N\"agerl}}]{Mark11}%
  \BibitemOpen
  \bibfield  {author} {\bibinfo {author} {\bibfnamefont {M.~J.}\ \bibnamefont
  {Mark}}, \bibinfo {author} {\bibfnamefont {E.}~\bibnamefont {Haller}},
  \bibinfo {author} {\bibfnamefont {K.}~\bibnamefont {Lauber}}, \bibinfo
  {author} {\bibfnamefont {J.~G.}\ \bibnamefont {Danzl}}, \bibinfo {author}
  {\bibfnamefont {A.~J.}\ \bibnamefont {Daley}},\ and\ \bibinfo {author}
  {\bibfnamefont {H.-C.}\ \bibnamefont {N\"agerl}},\ }\bibfield  {title}
  {\bibinfo {title} {{Precision Measurements on a Tunable Mott Insulator of
  Ultracold Atoms}},\ }\href {https://doi.org/10.1103/PhysRevLett.107.175301}
  {\bibfield  {journal} {\bibinfo  {journal} {Phys. Rev. Lett.}\ }\textbf
  {\bibinfo {volume} {107}},\ \bibinfo {pages} {175301} (\bibinfo {year}
  {2011})}\BibitemShut {NoStop}%
\bibitem [{\citenamefont {Thomas}\ \emph {et~al.}(2017)\citenamefont {Thomas},
  \citenamefont {Barter}, \citenamefont {Leung}, \citenamefont {Okano},
  \citenamefont {Jo}, \citenamefont {Guzman}, \citenamefont {Kimchi},
  \citenamefont {Vishwanath},\ and\ \citenamefont {Stamper-Kurn}}]{Thomas17a}%
  \BibitemOpen
  \bibfield  {author} {\bibinfo {author} {\bibfnamefont {C.~K.}\ \bibnamefont
  {Thomas}}, \bibinfo {author} {\bibfnamefont {T.~H.}\ \bibnamefont {Barter}},
  \bibinfo {author} {\bibfnamefont {T.-H.}\ \bibnamefont {Leung}}, \bibinfo
  {author} {\bibfnamefont {M.}~\bibnamefont {Okano}}, \bibinfo {author}
  {\bibfnamefont {G.-B.}\ \bibnamefont {Jo}}, \bibinfo {author} {\bibfnamefont
  {J.}~\bibnamefont {Guzman}}, \bibinfo {author} {\bibfnamefont
  {I.}~\bibnamefont {Kimchi}}, \bibinfo {author} {\bibfnamefont
  {A.}~\bibnamefont {Vishwanath}},\ and\ \bibinfo {author} {\bibfnamefont
  {D.~M.}\ \bibnamefont {Stamper-Kurn}},\ }\bibfield  {title} {\bibinfo {title}
  {{Mean-Field Scaling of the Superfluid to Mott Insulator Transition in a 2D
  Optical Superlattice}},\ }\href
  {https://doi.org/10.1103/PhysRevLett.119.100402} {\bibfield  {journal}
  {\bibinfo  {journal} {Phys. Rev. Lett.}\ }\textbf {\bibinfo {volume} {119}},\
  \bibinfo {pages} {100402} (\bibinfo {year} {2017})}\BibitemShut {NoStop}%
\bibitem [{\citenamefont {Hercé}\ \emph {et~al.}(2021)\citenamefont {Hercé},
  \citenamefont {Carcy}, \citenamefont {Tenart}, \citenamefont {Bureik},
  \citenamefont {Dareau}, \citenamefont {Clément},\ and\ \citenamefont
  {Roscilde}}]{Herce21}%
  \BibitemOpen
  \bibfield  {author} {\bibinfo {author} {\bibfnamefont {G.}~\bibnamefont
  {Hercé}}, \bibinfo {author} {\bibfnamefont {C.}~\bibnamefont {Carcy}},
  \bibinfo {author} {\bibfnamefont {A.}~\bibnamefont {Tenart}}, \bibinfo
  {author} {\bibfnamefont {J.-P.}\ \bibnamefont {Bureik}}, \bibinfo {author}
  {\bibfnamefont {A.}~\bibnamefont {Dareau}}, \bibinfo {author} {\bibfnamefont
  {D.}~\bibnamefont {Clément}},\ and\ \bibinfo {author} {\bibfnamefont
  {T.}~\bibnamefont {Roscilde}},\ }\bibfield  {title} {\bibinfo {title}
  {{Studying the low-entropy Mott transition of bosons in a three-dimensional
  optical lattice by measuring the full momentum-space density}},\ }\href
  {https://doi.org/10.1103/physreva.104.l011301} {\ \textbf {\bibinfo {volume}
  {104}},\ \bibinfo {pages} {l011301} (\bibinfo {year} {2021})}\BibitemShut
  {NoStop}%
\bibitem [{\citenamefont {Fisher}\ \emph {et~al.}(1989)\citenamefont {Fisher},
  \citenamefont {Weichman}, \citenamefont {Grinstein},\ and\ \citenamefont
  {Fisher}}]{Fisher89}%
  \BibitemOpen
  \bibfield  {author} {\bibinfo {author} {\bibfnamefont {M.~P.~A.}\
  \bibnamefont {Fisher}}, \bibinfo {author} {\bibfnamefont {P.~B.}\
  \bibnamefont {Weichman}}, \bibinfo {author} {\bibfnamefont {G.}~\bibnamefont
  {Grinstein}},\ and\ \bibinfo {author} {\bibfnamefont {D.~S.}\ \bibnamefont
  {Fisher}},\ }\bibfield  {title} {\bibinfo {title} {{Boson localization and
  the superfluid-insulator transition}},\ }\href
  {https://doi.org/10.1103/PhysRevB.40.546} {\bibfield  {journal} {\bibinfo
  {journal} {Phys. Rev. B}\ }\textbf {\bibinfo {volume} {40}},\ \bibinfo
  {pages} {546} (\bibinfo {year} {1989})}\BibitemShut {NoStop}%
\bibitem [{\citenamefont {Jaksch}\ \emph {et~al.}(1998)\citenamefont {Jaksch},
  \citenamefont {Bruder}, \citenamefont {Cirac}, \citenamefont {Gardiner},\
  and\ \citenamefont {Zoller}}]{Jaksch98}%
  \BibitemOpen
  \bibfield  {author} {\bibinfo {author} {\bibfnamefont {D.}~\bibnamefont
  {Jaksch}}, \bibinfo {author} {\bibfnamefont {C.}~\bibnamefont {Bruder}},
  \bibinfo {author} {\bibfnamefont {J.~I.}\ \bibnamefont {Cirac}}, \bibinfo
  {author} {\bibfnamefont {C.~W.}\ \bibnamefont {Gardiner}},\ and\ \bibinfo
  {author} {\bibfnamefont {P.}~\bibnamefont {Zoller}},\ }\bibfield  {title}
  {\bibinfo {title} {{Cold Bosonic Atoms in Optical Lattices}},\ }\href
  {https://doi.org/10.1103/PhysRevLett.81.3108} {\bibfield  {journal} {\bibinfo
   {journal} {Phys. Rev. Lett.}\ }\textbf {\bibinfo {volume} {81}},\ \bibinfo
  {pages} {3108} (\bibinfo {year} {1998})}\BibitemShut {NoStop}%
\bibitem [{\citenamefont {Sheshadri}\ \emph {et~al.}(1993)\citenamefont
  {Sheshadri}, \citenamefont {Krishnamurthy}, \citenamefont {Pandit},\ and\
  \citenamefont {Ramakrishnan}}]{Sheshadri93}%
  \BibitemOpen
  \bibfield  {author} {\bibinfo {author} {\bibfnamefont {K.}~\bibnamefont
  {Sheshadri}}, \bibinfo {author} {\bibfnamefont {H.~R.}\ \bibnamefont
  {Krishnamurthy}}, \bibinfo {author} {\bibfnamefont {R.}~\bibnamefont
  {Pandit}},\ and\ \bibinfo {author} {\bibfnamefont {T.~V.}\ \bibnamefont
  {Ramakrishnan}},\ }\bibfield  {title} {\bibinfo {title} {{Superfluid and
  Insulating Phases in an Interacting-Boson Model: Mean-Field Theory and the
  RPA}},\ }\href {https://doi.org/10.1209/0295-5075/22/4/004} {\bibfield
  {journal} {\bibinfo  {journal} {Europhys. Lett.}\ }\textbf {\bibinfo {volume}
  {22}},\ \bibinfo {pages} {257} (\bibinfo {year} {1993})}\BibitemShut
  {NoStop}%
\bibitem [{\citenamefont {van Oosten}\ \emph {et~al.}(2001)\citenamefont {van
  Oosten}, \citenamefont {van~der Straten},\ and\ \citenamefont
  {Stoof}}]{Oosten01}%
  \BibitemOpen
  \bibfield  {author} {\bibinfo {author} {\bibfnamefont {D.}~\bibnamefont {van
  Oosten}}, \bibinfo {author} {\bibfnamefont {P.}~\bibnamefont {van~der
  Straten}},\ and\ \bibinfo {author} {\bibfnamefont {H.~T.~C.}\ \bibnamefont
  {Stoof}},\ }\bibfield  {title} {\bibinfo {title} {{Quantum phases in an
  optical lattice}},\ }\href {https://doi.org/10.1103/PhysRevA.63.053601}
  {\bibfield  {journal} {\bibinfo  {journal} {Phys. Rev. A}\ }\textbf {\bibinfo
  {volume} {63}},\ \bibinfo {pages} {053601} (\bibinfo {year}
  {2001})}\BibitemShut {NoStop}%
\bibitem [{\citenamefont {Sengupta}\ and\ \citenamefont
  {Dupuis}(2005)}]{Sengupta05}%
  \BibitemOpen
  \bibfield  {author} {\bibinfo {author} {\bibfnamefont {K.}~\bibnamefont
  {Sengupta}}\ and\ \bibinfo {author} {\bibfnamefont {N.}~\bibnamefont
  {Dupuis}},\ }\bibfield  {title} {\bibinfo {title}
  {{Mott-insulator--to--superfluid transition in the Bose-Hubbard model: A
  strong-coupling approach}},\ }\href
  {https://doi.org/10.1103/PhysRevA.71.033629} {\bibfield  {journal} {\bibinfo
  {journal} {Phys. Rev. A}\ }\textbf {\bibinfo {volume} {71}},\ \bibinfo
  {pages} {033629} (\bibinfo {year} {2005})}\BibitemShut {NoStop}%
\bibitem [{\citenamefont {Ohashi}\ \emph {et~al.}(2006)\citenamefont {Ohashi},
  \citenamefont {Kitaura},\ and\ \citenamefont {Matsumoto}}]{Ohashi06}%
  \BibitemOpen
  \bibfield  {author} {\bibinfo {author} {\bibfnamefont {Y.}~\bibnamefont
  {Ohashi}}, \bibinfo {author} {\bibfnamefont {M.}~\bibnamefont {Kitaura}},\
  and\ \bibinfo {author} {\bibfnamefont {H.}~\bibnamefont {Matsumoto}},\
  }\bibfield  {title} {\bibinfo {title} {{Itinerant-localized dual character of
  a strongly correlated superfluid Bose gas in an optical lattice}},\ }\href
  {https://doi.org/10.1103/PhysRevA.73.033617} {\bibfield  {journal} {\bibinfo
  {journal} {Phys. Rev. A}\ }\textbf {\bibinfo {volume} {73}},\ \bibinfo
  {pages} {033617} (\bibinfo {year} {2006})}\BibitemShut {NoStop}%
\bibitem [{\citenamefont {Menotti}\ and\ \citenamefont
  {Trivedi}(2008)}]{Menotti08}%
  \BibitemOpen
  \bibfield  {author} {\bibinfo {author} {\bibfnamefont {C.}~\bibnamefont
  {Menotti}}\ and\ \bibinfo {author} {\bibfnamefont {N.}~\bibnamefont
  {Trivedi}},\ }\bibfield  {title} {\bibinfo {title} {{Spectral weight
  redistribution in strongly correlated bosons in optical lattices}},\ }\href
  {https://doi.org/10.1103/PhysRevB.77.235120} {\bibfield  {journal} {\bibinfo
  {journal} {Phys. Rev. B}\ }\textbf {\bibinfo {volume} {77}},\ \bibinfo
  {pages} {235120} (\bibinfo {year} {2008})}\BibitemShut {NoStop}%
\bibitem [{\citenamefont {Freericks}\ \emph {et~al.}(2009)\citenamefont
  {Freericks}, \citenamefont {Krishnamurthy}, \citenamefont {Kato},
  \citenamefont {Kawashima},\ and\ \citenamefont {Trivedi}}]{Freericks09}%
  \BibitemOpen
  \bibfield  {author} {\bibinfo {author} {\bibfnamefont {J.~K.}\ \bibnamefont
  {Freericks}}, \bibinfo {author} {\bibfnamefont {H.~R.}\ \bibnamefont
  {Krishnamurthy}}, \bibinfo {author} {\bibfnamefont {Y.}~\bibnamefont {Kato}},
  \bibinfo {author} {\bibfnamefont {N.}~\bibnamefont {Kawashima}},\ and\
  \bibinfo {author} {\bibfnamefont {N.}~\bibnamefont {Trivedi}},\ }\bibfield
  {title} {\bibinfo {title} {{Strong-coupling expansion for the momentum
  distribution of the Bose-Hubbard model with benchmarking against exact
  numerical results}},\ }\href {https://doi.org/10.1103/PhysRevA.79.053631}
  {\bibfield  {journal} {\bibinfo  {journal} {Phys. Rev. A}\ }\textbf {\bibinfo
  {volume} {79}},\ \bibinfo {pages} {053631} (\bibinfo {year}
  {2009})}\BibitemShut {NoStop}%
\bibitem [{\citenamefont {Teichmann}\ \emph
  {et~al.}(2009{\natexlab{a}})\citenamefont {Teichmann}, \citenamefont
  {Hinrichs}, \citenamefont {Holthaus},\ and\ \citenamefont
  {Eckardt}}]{Teichmann09a}%
  \BibitemOpen
  \bibfield  {author} {\bibinfo {author} {\bibfnamefont {N.}~\bibnamefont
  {Teichmann}}, \bibinfo {author} {\bibfnamefont {D.}~\bibnamefont {Hinrichs}},
  \bibinfo {author} {\bibfnamefont {M.}~\bibnamefont {Holthaus}},\ and\
  \bibinfo {author} {\bibfnamefont {A.}~\bibnamefont {Eckardt}},\ }\bibfield
  {title} {\bibinfo {title} {{Bose-Hubbard phase diagram with arbitrary integer
  filling}},\ }\href {https://doi.org/10.1103/PhysRevB.79.100503} {\bibfield
  {journal} {\bibinfo  {journal} {Phys. Rev. B}\ }\textbf {\bibinfo {volume}
  {79}},\ \bibinfo {pages} {100503} (\bibinfo {year}
  {2009}{\natexlab{a}})}\BibitemShut {NoStop}%
\bibitem [{\citenamefont {Teichmann}\ \emph
  {et~al.}(2009{\natexlab{b}})\citenamefont {Teichmann}, \citenamefont
  {Hinrichs}, \citenamefont {Holthaus},\ and\ \citenamefont
  {Eckardt}}]{Teichmann09b}%
  \BibitemOpen
  \bibfield  {author} {\bibinfo {author} {\bibfnamefont {N.}~\bibnamefont
  {Teichmann}}, \bibinfo {author} {\bibfnamefont {D.}~\bibnamefont {Hinrichs}},
  \bibinfo {author} {\bibfnamefont {M.}~\bibnamefont {Holthaus}},\ and\
  \bibinfo {author} {\bibfnamefont {A.}~\bibnamefont {Eckardt}},\ }\bibfield
  {title} {\bibinfo {title} {{Process-chain approach to the Bose-Hubbard model:
  Ground-state properties and phase diagram}},\ }\href
  {https://doi.org/10.1103/PhysRevB.79.224515} {\bibfield  {journal} {\bibinfo
  {journal} {Phys. Rev. B}\ }\textbf {\bibinfo {volume} {79}},\ \bibinfo
  {pages} {224515} (\bibinfo {year} {2009}{\natexlab{b}})}\BibitemShut
  {NoStop}%
\bibitem [{\citenamefont {Wang}\ \emph {et~al.}(2018)\citenamefont {Wang},
  \citenamefont {Zhang}, \citenamefont {Hou}, \citenamefont {Eggert},\ and\
  \citenamefont {Pelster}}]{Wang18}%
  \BibitemOpen
  \bibfield  {author} {\bibinfo {author} {\bibfnamefont {T.}~\bibnamefont
  {Wang}}, \bibinfo {author} {\bibfnamefont {X.-F.}\ \bibnamefont {Zhang}},
  \bibinfo {author} {\bibfnamefont {C.-F.}\ \bibnamefont {Hou}}, \bibinfo
  {author} {\bibfnamefont {S.}~\bibnamefont {Eggert}},\ and\ \bibinfo {author}
  {\bibfnamefont {A.}~\bibnamefont {Pelster}},\ }\bibfield  {title} {\bibinfo
  {title} {{High-order symbolic strong-coupling expansion for the Bose-Hubbard
  model}},\ }\href {https://doi.org/10.1103/physrevb.98.245107} {\bibfield
  {journal} {\bibinfo  {journal} {Phys. Rev. B}\ }\textbf {\bibinfo {volume}
  {98}},\ \bibinfo {pages} {245107} (\bibinfo {year} {2018})}\BibitemShut
  {NoStop}%
\bibitem [{\citenamefont {Kübler}\ \emph {et~al.}(2019)\citenamefont
  {Kübler}, \citenamefont {Sant’Ana}, \citenamefont {{dos Santos}},\ and\
  \citenamefont {Pelster}}]{Kuebler19}%
  \BibitemOpen
  \bibfield  {author} {\bibinfo {author} {\bibfnamefont {M.}~\bibnamefont
  {Kübler}}, \bibinfo {author} {\bibfnamefont {F.~T.}\ \bibnamefont
  {Sant’Ana}}, \bibinfo {author} {\bibfnamefont {F.~E.~A.}\ \bibnamefont
  {{dos Santos}}},\ and\ \bibinfo {author} {\bibfnamefont {A.}~\bibnamefont
  {Pelster}},\ }\bibfield  {title} {\bibinfo {title} {{Improving mean-field
  theory for bosons in optical lattices via degenerate perturbation theory}},\
  }\href {https://doi.org/10.1103/physreva.99.063603} {\bibfield  {journal}
  {\bibinfo  {journal} {Phys. Rev. A}\ }\textbf {\bibinfo {volume} {99}},\
  \bibinfo {pages} {063603} (\bibinfo {year} {2019})}\BibitemShut {NoStop}%
\bibitem [{\citenamefont {Sant’Ana}\ \emph {et~al.}(2019)\citenamefont
  {Sant’Ana}, \citenamefont {Pelster},\ and\ \citenamefont {dos
  Santos}}]{SantAna19}%
  \BibitemOpen
  \bibfield  {author} {\bibinfo {author} {\bibfnamefont {F.~T.}\ \bibnamefont
  {Sant’Ana}}, \bibinfo {author} {\bibfnamefont {A.}~\bibnamefont
  {Pelster}},\ and\ \bibinfo {author} {\bibfnamefont {F.~E.~A.}\ \bibnamefont
  {dos Santos}},\ }\bibfield  {title} {\bibinfo {title} {Finite-temperature
  degenerate perturbation theory for bosons in optical lattices},\ }\href
  {https://doi.org/10.1103/physreva.100.043609} {\bibfield  {journal} {\bibinfo
   {journal} {Phys. Rev. A}\ }\textbf {\bibinfo {volume} {100}},\ \bibinfo
  {pages} {043609} (\bibinfo {year} {2019})}\BibitemShut {NoStop}%
\bibitem [{\citenamefont {Sajna}\ \emph {et~al.}(2015)\citenamefont {Sajna},
  \citenamefont {Polak}, \citenamefont {Micnas},\ and\ \citenamefont
  {Rożek}}]{Sajna15}%
  \BibitemOpen
  \bibfield  {author} {\bibinfo {author} {\bibfnamefont {A.~S.}\ \bibnamefont
  {Sajna}}, \bibinfo {author} {\bibfnamefont {T.~P.}\ \bibnamefont {Polak}},
  \bibinfo {author} {\bibfnamefont {R.}~\bibnamefont {Micnas}},\ and\ \bibinfo
  {author} {\bibfnamefont {P.}~\bibnamefont {Rożek}},\ }\bibfield  {title}
  {\bibinfo {title} {{Ground-state and finite-temperature properties of
  correlated ultracold bosons on optical lattices}},\ }\href
  {https://doi.org/10.1103/physreva.92.013602} {\bibfield  {journal} {\bibinfo
  {journal} {Phys. Rev. A}\ }\textbf {\bibinfo {volume} {92}},\ \bibinfo
  {pages} {013602} (\bibinfo {year} {2015})}\BibitemShut {NoStop}%
\bibitem [{\citenamefont {Polak}\ and\ \citenamefont {Kopeć}(2007)}]{Polak07}%
  \BibitemOpen
  \bibfield  {author} {\bibinfo {author} {\bibfnamefont {T.~P.}\ \bibnamefont
  {Polak}}\ and\ \bibinfo {author} {\bibfnamefont {T.~K.}\ \bibnamefont
  {Kopeć}},\ }\bibfield  {title} {\bibinfo {title} {{Quantum rotor description
  of the Mott-insulator transition in the Bose-Hubbard model}},\ }\href
  {https://doi.org/10.1103/physrevb.76.094503} {\bibfield  {journal} {\bibinfo
  {journal} {Phys. Rev. B}\ }\textbf {\bibinfo {volume} {76}},\ \bibinfo
  {pages} {094503} (\bibinfo {year} {2007})}\BibitemShut {NoStop}%
\bibitem [{\citenamefont {Polak}\ and\ \citenamefont {Kopeć}(2009)}]{Polak09}%
  \BibitemOpen
  \bibfield  {author} {\bibinfo {author} {\bibfnamefont {T.~P.}\ \bibnamefont
  {Polak}}\ and\ \bibinfo {author} {\bibfnamefont {T.~K.}\ \bibnamefont
  {Kopeć}},\ }\bibfield  {title} {\bibinfo {title} {{Finite-temperature
  effects on the superfluid Bose–Einstein condensation of confined ultracold
  atoms in three-dimensional optical lattices}},\ }\href
  {https://doi.org/10.1088/0953-4075/42/9/095302} {\bibfield  {journal}
  {\bibinfo  {journal} {J. Phys. B: At. Mol. Opt. Phys.}\ }\textbf {\bibinfo
  {volume} {42}},\ \bibinfo {pages} {095302} (\bibinfo {year}
  {2009})}\BibitemShut {NoStop}%
\bibitem [{\citenamefont {Zaleski}\ and\ \citenamefont
  {Kopeć}(2011)}]{Zaleski11}%
  \BibitemOpen
  \bibfield  {author} {\bibinfo {author} {\bibfnamefont {T.~A.}\ \bibnamefont
  {Zaleski}}\ and\ \bibinfo {author} {\bibfnamefont {T.~K.}\ \bibnamefont
  {Kopeć}},\ }\bibfield  {title} {\bibinfo {title} {Atom-atom correlations in
  time-of-flight imaging of ultracold bosons in optical lattices},\ }\href
  {https://doi.org/10.1103/physreva.84.053613} {\bibfield  {journal} {\bibinfo
  {journal} {Phys. Rev. A}\ }\textbf {\bibinfo {volume} {84}},\ \bibinfo
  {pages} {053613} (\bibinfo {year} {2011})}\BibitemShut {NoStop}%
\bibitem [{\citenamefont {Krzywicka}\ and\ \citenamefont
  {Polak}(2022)}]{Krzywicka22}%
  \BibitemOpen
  \bibfield  {author} {\bibinfo {author} {\bibfnamefont {A.}~\bibnamefont
  {Krzywicka}}\ and\ \bibinfo {author} {\bibfnamefont {T.}~\bibnamefont
  {Polak}},\ }\bibfield  {title} {\bibinfo {title} {Coexistence of two kinds of
  superfluidity at finite temperatures in optical lattices},\ }\href
  {https://doi.org/10.1016/j.aop.2022.168973} {\bibfield  {journal} {\bibinfo
  {journal} {Ann. Phys.}\ }\textbf {\bibinfo {volume} {443}},\ \bibinfo {pages}
  {168973} (\bibinfo {year} {2022})}\BibitemShut {NoStop}%
\bibitem [{\citenamefont {Krzywicka}\ and\ \citenamefont
  {Polak}(2024)}]{Krzywicka24}%
  \BibitemOpen
  \bibfield  {author} {\bibinfo {author} {\bibfnamefont {A.}~\bibnamefont
  {Krzywicka}}\ and\ \bibinfo {author} {\bibfnamefont {T.~P.}\ \bibnamefont
  {Polak}},\ }\bibfield  {title} {\bibinfo {title} {Reentrant phase behavior in
  systems with density-induced tunneling},\ }\href
  {https://doi.org/10.1038/s41598-024-60955-1} {\bibfield  {journal} {\bibinfo
  {journal} {Sci. Rep.}\ }\textbf {\bibinfo {volume} {14}},\ \bibinfo {pages}
  {10364} (\bibinfo {year} {2024})}\BibitemShut {NoStop}%
\bibitem [{\citenamefont {Dickerscheid}\ \emph {et~al.}(2003)\citenamefont
  {Dickerscheid}, \citenamefont {van Oosten}, \citenamefont {Denteneer},\ and\
  \citenamefont {Stoof}}]{Dickerscheid03}%
  \BibitemOpen
  \bibfield  {author} {\bibinfo {author} {\bibfnamefont {D.~B.~M.}\
  \bibnamefont {Dickerscheid}}, \bibinfo {author} {\bibfnamefont
  {D.}~\bibnamefont {van Oosten}}, \bibinfo {author} {\bibfnamefont {P.~J.~H.}\
  \bibnamefont {Denteneer}},\ and\ \bibinfo {author} {\bibfnamefont {H.~T.~C.}\
  \bibnamefont {Stoof}},\ }\bibfield  {title} {\bibinfo {title} {{Ultracold
  atoms in optical lattices}},\ }\href
  {https://doi.org/10.1103/PhysRevA.68.043623} {\bibfield  {journal} {\bibinfo
  {journal} {Phys. Rev. A}\ }\textbf {\bibinfo {volume} {68}},\ \bibinfo
  {pages} {043623} (\bibinfo {year} {2003})}\BibitemShut {NoStop}%
\bibitem [{\citenamefont {Yu}\ and\ \citenamefont {Chui}(2005)}]{Yu05}%
  \BibitemOpen
  \bibfield  {author} {\bibinfo {author} {\bibfnamefont {Y.}~\bibnamefont
  {Yu}}\ and\ \bibinfo {author} {\bibfnamefont {S.~T.}\ \bibnamefont {Chui}},\
  }\bibfield  {title} {\bibinfo {title} {{Phase diagram of ultracold atoms in
  optical lattices: Comparative study of slave fermion and slave boson
  approaches to Bose-Hubbard model}},\ }\href
  {https://doi.org/10.1103/physreva.71.033608} {\bibfield  {journal} {\bibinfo
  {journal} {Phys. Rev. A}\ }\textbf {\bibinfo {volume} {71}},\ \bibinfo
  {pages} {033608} (\bibinfo {year} {2005})}\BibitemShut {NoStop}%
\bibitem [{\citenamefont {Huber}\ \emph {et~al.}(2007)\citenamefont {Huber},
  \citenamefont {Altman}, \citenamefont {B\"uchler},\ and\ \citenamefont
  {Blatter}}]{Huber07}%
  \BibitemOpen
  \bibfield  {author} {\bibinfo {author} {\bibfnamefont {S.~D.}\ \bibnamefont
  {Huber}}, \bibinfo {author} {\bibfnamefont {E.}~\bibnamefont {Altman}},
  \bibinfo {author} {\bibfnamefont {H.~P.}\ \bibnamefont {B\"uchler}},\ and\
  \bibinfo {author} {\bibfnamefont {G.}~\bibnamefont {Blatter}},\ }\bibfield
  {title} {\bibinfo {title} {{Dynamical properties of ultracold bosons in an
  optical lattice}},\ }\href {https://doi.org/10.1103/PhysRevB.75.085106}
  {\bibfield  {journal} {\bibinfo  {journal} {Phys. Rev. B}\ }\textbf {\bibinfo
  {volume} {75}},\ \bibinfo {pages} {085106} (\bibinfo {year}
  {2007})}\BibitemShut {NoStop}%
\bibitem [{\citenamefont {Fr\'erot}\ and\ \citenamefont
  {Roscilde}(2016)}]{Frerot16a}%
  \BibitemOpen
  \bibfield  {author} {\bibinfo {author} {\bibfnamefont {I.}~\bibnamefont
  {Fr\'erot}}\ and\ \bibinfo {author} {\bibfnamefont {T.}~\bibnamefont
  {Roscilde}},\ }\bibfield  {title} {\bibinfo {title} {{Entanglement Entropy
  across the Superfluid-Insulator Transition: A Signature of Bosonic
  Criticality}},\ }\href {https://doi.org/10.1103/PhysRevLett.116.190401}
  {\bibfield  {journal} {\bibinfo  {journal} {Phys. Rev. Lett.}\ }\textbf
  {\bibinfo {volume} {116}},\ \bibinfo {pages} {190401} (\bibinfo {year}
  {2016})}\BibitemShut {NoStop}%
\bibitem [{\citenamefont {Krutitsky}\ and\ \citenamefont
  {Navez}(2011)}]{Krutitsky11}%
  \BibitemOpen
  \bibfield  {author} {\bibinfo {author} {\bibfnamefont {K.~V.}\ \bibnamefont
  {Krutitsky}}\ and\ \bibinfo {author} {\bibfnamefont {P.}~\bibnamefont
  {Navez}},\ }\bibfield  {title} {\bibinfo {title} {{Excitation dynamics in a
  lattice Bose gas within the time-dependent Gutzwiller mean-field approach}},\
  }\href {https://doi.org/10.1103/physreva.84.033602} {\bibfield  {journal}
  {\bibinfo  {journal} {Phys. Rev. A}\ }\textbf {\bibinfo {volume} {84}},\
  \bibinfo {pages} {033602} (\bibinfo {year} {2011})}\BibitemShut {NoStop}%
\bibitem [{\citenamefont {Di~Liberto}\ \emph {et~al.}(2018)\citenamefont
  {Di~Liberto}, \citenamefont {Recati}, \citenamefont {Trivedi}, \citenamefont
  {Carusotto},\ and\ \citenamefont {Menotti}}]{DiLiberto18}%
  \BibitemOpen
  \bibfield  {author} {\bibinfo {author} {\bibfnamefont {M.}~\bibnamefont
  {Di~Liberto}}, \bibinfo {author} {\bibfnamefont {A.}~\bibnamefont {Recati}},
  \bibinfo {author} {\bibfnamefont {N.}~\bibnamefont {Trivedi}}, \bibinfo
  {author} {\bibfnamefont {I.}~\bibnamefont {Carusotto}},\ and\ \bibinfo
  {author} {\bibfnamefont {C.}~\bibnamefont {Menotti}},\ }\bibfield  {title}
  {\bibinfo {title} {{Particle-Hole Character of the Higgs and Goldstone Modes
  in Strongly Interacting Lattice Bosons}},\ }\href
  {https://doi.org/10.1103/PhysRevLett.120.073201} {\bibfield  {journal}
  {\bibinfo  {journal} {Phys. Rev. Lett.}\ }\textbf {\bibinfo {volume} {120}},\
  \bibinfo {pages} {073201} (\bibinfo {year} {2018})}\BibitemShut {NoStop}%
\bibitem [{\citenamefont {Caleffi}\ \emph {et~al.}(2020)\citenamefont
  {Caleffi}, \citenamefont {Capone}, \citenamefont {Menotti}, \citenamefont
  {Carusotto},\ and\ \citenamefont {Recati}}]{Caleffi20}%
  \BibitemOpen
  \bibfield  {author} {\bibinfo {author} {\bibfnamefont {F.}~\bibnamefont
  {Caleffi}}, \bibinfo {author} {\bibfnamefont {M.}~\bibnamefont {Capone}},
  \bibinfo {author} {\bibfnamefont {C.}~\bibnamefont {Menotti}}, \bibinfo
  {author} {\bibfnamefont {I.}~\bibnamefont {Carusotto}},\ and\ \bibinfo
  {author} {\bibfnamefont {A.}~\bibnamefont {Recati}},\ }\bibfield  {title}
  {\bibinfo {title} {{Quantum fluctuations beyond the Gutzwiller approximation
  in the Bose-Hubbard model}},\ }\href
  {https://doi.org/10.1103/physrevresearch.2.033276} {\bibfield  {journal}
  {\bibinfo  {journal} {Phys. Rev. Research}\ }\textbf {\bibinfo {volume}
  {2}},\ \bibinfo {pages} {033276} (\bibinfo {year} {2020})}\BibitemShut
  {NoStop}%
\bibitem [{\citenamefont {Byczuk}\ and\ \citenamefont
  {Vollhardt}(2008)}]{Byczuk08}%
  \BibitemOpen
  \bibfield  {author} {\bibinfo {author} {\bibfnamefont {K.}~\bibnamefont
  {Byczuk}}\ and\ \bibinfo {author} {\bibfnamefont {D.}~\bibnamefont
  {Vollhardt}},\ }\bibfield  {title} {\bibinfo {title} {{Correlated bosons on a
  lattice: Dynamical mean-field theory for Bose-Einstein condensed and normal
  phases}},\ }\href {https://doi.org/10.1103/PhysRevB.77.235106} {\bibfield
  {journal} {\bibinfo  {journal} {Phys. Rev. B}\ }\textbf {\bibinfo {volume}
  {77}},\ \bibinfo {pages} {235106} (\bibinfo {year} {2008})}\BibitemShut
  {NoStop}%
\bibitem [{\citenamefont {Hu}\ and\ \citenamefont {Tong}(2009)}]{Hu09}%
  \BibitemOpen
  \bibfield  {author} {\bibinfo {author} {\bibfnamefont {W.-J.}\ \bibnamefont
  {Hu}}\ and\ \bibinfo {author} {\bibfnamefont {N.-H.}\ \bibnamefont {Tong}},\
  }\bibfield  {title} {\bibinfo {title} {{Dynamical mean-field theory for the
  Bose-Hubbard model}},\ }\href {https://doi.org/10.1103/physrevb.80.245110}
  {\bibfield  {journal} {\bibinfo  {journal} {Phys. Rev. B}\ }\textbf {\bibinfo
  {volume} {80}},\ \bibinfo {pages} {245110} (\bibinfo {year}
  {2009})}\BibitemShut {NoStop}%
\bibitem [{\citenamefont {Anders}\ \emph {et~al.}(2010)\citenamefont {Anders},
  \citenamefont {Gull}, \citenamefont {Pollet}, \citenamefont {Troyer},\ and\
  \citenamefont {Werner}}]{Anders10}%
  \BibitemOpen
  \bibfield  {author} {\bibinfo {author} {\bibfnamefont {P.}~\bibnamefont
  {Anders}}, \bibinfo {author} {\bibfnamefont {E.}~\bibnamefont {Gull}},
  \bibinfo {author} {\bibfnamefont {L.}~\bibnamefont {Pollet}}, \bibinfo
  {author} {\bibfnamefont {M.}~\bibnamefont {Troyer}},\ and\ \bibinfo {author}
  {\bibfnamefont {P.}~\bibnamefont {Werner}},\ }\bibfield  {title} {\bibinfo
  {title} {{Dynamical Mean Field Solution of the Bose-Hubbard Model}},\ }\href
  {https://doi.org/10.1103/PhysRevLett.105.096402} {\bibfield  {journal}
  {\bibinfo  {journal} {Phys. Rev. Lett.}\ }\textbf {\bibinfo {volume} {105}},\
  \bibinfo {pages} {096402} (\bibinfo {year} {2010})}\BibitemShut {NoStop}%
\bibitem [{\citenamefont {Anders}\ \emph {et~al.}(2011)\citenamefont {Anders},
  \citenamefont {Gull}, \citenamefont {Pollet}, \citenamefont {Troyer},\ and\
  \citenamefont {Werner}}]{Anders11}%
  \BibitemOpen
  \bibfield  {author} {\bibinfo {author} {\bibfnamefont {P.}~\bibnamefont
  {Anders}}, \bibinfo {author} {\bibfnamefont {E.}~\bibnamefont {Gull}},
  \bibinfo {author} {\bibfnamefont {L.}~\bibnamefont {Pollet}}, \bibinfo
  {author} {\bibfnamefont {M.}~\bibnamefont {Troyer}},\ and\ \bibinfo {author}
  {\bibfnamefont {P.}~\bibnamefont {Werner}},\ }\bibfield  {title} {\bibinfo
  {title} {{Dynamical mean-field theory for bosons}},\ }\href
  {https://doi.org/10.1088/1367-2630/13/7/075013} {\bibfield  {journal}
  {\bibinfo  {journal} {New J. Phys.}\ }\textbf {\bibinfo {volume} {13}},\
  \bibinfo {pages} {075013} (\bibinfo {year} {2011})}\BibitemShut {NoStop}%
\bibitem [{\citenamefont {Panas}\ \emph {et~al.}(2015)\citenamefont {Panas},
  \citenamefont {Kauch}, \citenamefont {Kune\ifmmode~\check{s}\else
  \v{s}\fi{}}, \citenamefont {Vollhardt},\ and\ \citenamefont
  {Byczuk}}]{Panas15}%
  \BibitemOpen
  \bibfield  {author} {\bibinfo {author} {\bibfnamefont {J.}~\bibnamefont
  {Panas}}, \bibinfo {author} {\bibfnamefont {A.}~\bibnamefont {Kauch}},
  \bibinfo {author} {\bibfnamefont {J.}~\bibnamefont
  {Kune\ifmmode~\check{s}\else \v{s}\fi{}}}, \bibinfo {author} {\bibfnamefont
  {D.}~\bibnamefont {Vollhardt}},\ and\ \bibinfo {author} {\bibfnamefont
  {K.}~\bibnamefont {Byczuk}},\ }\bibfield  {title} {\bibinfo {title}
  {{Numerical calculation of spectral functions of the Bose-Hubbard model using
  bosonic dynamical mean-field theory}},\ }\href
  {https://doi.org/10.1103/PhysRevB.92.045102} {\bibfield  {journal} {\bibinfo
  {journal} {Phys. Rev. B}\ }\textbf {\bibinfo {volume} {92}},\ \bibinfo
  {pages} {045102} (\bibinfo {year} {2015})}\BibitemShut {NoStop}%
\bibitem [{\citenamefont {Capello}\ \emph {et~al.}(2007)\citenamefont
  {Capello}, \citenamefont {Becca}, \citenamefont {Fabrizio},\ and\
  \citenamefont {Sorella}}]{Capello07}%
  \BibitemOpen
  \bibfield  {author} {\bibinfo {author} {\bibfnamefont {M.}~\bibnamefont
  {Capello}}, \bibinfo {author} {\bibfnamefont {F.}~\bibnamefont {Becca}},
  \bibinfo {author} {\bibfnamefont {M.}~\bibnamefont {Fabrizio}},\ and\
  \bibinfo {author} {\bibfnamefont {S.}~\bibnamefont {Sorella}},\ }\bibfield
  {title} {\bibinfo {title} {{Superfluid to Mott-insulator transition in
  Bose-Hubbard models}},\ }\href
  {https://doi.org/10.1103/physrevlett.99.056402} {\bibfield  {journal}
  {\bibinfo  {journal} {Phys. Rev. Lett.}\ }\textbf {\bibinfo {volume} {99}},\
  \bibinfo {pages} {056402} (\bibinfo {year} {2007})}\BibitemShut {NoStop}%
\bibitem [{\citenamefont {Koller}\ and\ \citenamefont
  {Dupuis}(2006)}]{Koller06}%
  \BibitemOpen
  \bibfield  {author} {\bibinfo {author} {\bibfnamefont {W.}~\bibnamefont
  {Koller}}\ and\ \bibinfo {author} {\bibfnamefont {N.}~\bibnamefont
  {Dupuis}},\ }\bibfield  {title} {\bibinfo {title} {{Variational cluster
  perturbation theory for Bose–Hubbard models}},\ }\href
  {https://doi.org/10.1088/0953-8984/18/41/019} {\bibfield  {journal} {\bibinfo
   {journal} {J. Phys. Condens. Matter}\ }\textbf {\bibinfo {volume} {18}},\
  \bibinfo {pages} {9525} (\bibinfo {year} {2006})}\BibitemShut {NoStop}%
\bibitem [{\citenamefont {Knap}\ \emph {et~al.}(2010)\citenamefont {Knap},
  \citenamefont {Arrigoni},\ and\ \citenamefont {von~der Linden}}]{Knap10}%
  \BibitemOpen
  \bibfield  {author} {\bibinfo {author} {\bibfnamefont {M.}~\bibnamefont
  {Knap}}, \bibinfo {author} {\bibfnamefont {E.}~\bibnamefont {Arrigoni}},\
  and\ \bibinfo {author} {\bibfnamefont {W.}~\bibnamefont {von~der Linden}},\
  }\bibfield  {title} {\bibinfo {title} {Spectral properties of strongly
  correlated bosons in two-dimensional optical lattices},\ }\href
  {https://doi.org/10.1103/PhysRevB.81.024301} {\bibfield  {journal} {\bibinfo
  {journal} {Phys. Rev. B}\ }\textbf {\bibinfo {volume} {81}},\ \bibinfo
  {pages} {024301} (\bibinfo {year} {2010})}\BibitemShut {NoStop}%
\bibitem [{\citenamefont {Knap}\ \emph {et~al.}(2011)\citenamefont {Knap},
  \citenamefont {Arrigoni},\ and\ \citenamefont {von~der Linden}}]{Knap11}%
  \BibitemOpen
  \bibfield  {author} {\bibinfo {author} {\bibfnamefont {M.}~\bibnamefont
  {Knap}}, \bibinfo {author} {\bibfnamefont {E.}~\bibnamefont {Arrigoni}},\
  and\ \bibinfo {author} {\bibfnamefont {W.}~\bibnamefont {von~der Linden}},\
  }\bibfield  {title} {\bibinfo {title} {Variational cluster approach for
  strongly correlated lattice bosons in the superfluid phase},\ }\href
  {https://doi.org/10.1103/PhysRevB.83.134507} {\bibfield  {journal} {\bibinfo
  {journal} {Phys. Rev. B}\ }\textbf {\bibinfo {volume} {83}},\ \bibinfo
  {pages} {134507} (\bibinfo {year} {2011})}\BibitemShut {NoStop}%
\bibitem [{\citenamefont {Arrigoni}\ \emph {et~al.}(2011)\citenamefont
  {Arrigoni}, \citenamefont {Knap},\ and\ \citenamefont {von~der
  Linden}}]{Arrigoni11}%
  \BibitemOpen
  \bibfield  {author} {\bibinfo {author} {\bibfnamefont {E.}~\bibnamefont
  {Arrigoni}}, \bibinfo {author} {\bibfnamefont {M.}~\bibnamefont {Knap}},\
  and\ \bibinfo {author} {\bibfnamefont {W.}~\bibnamefont {von~der Linden}},\
  }\bibfield  {title} {\bibinfo {title} {Extended self-energy functional
  approach for strongly correlated lattice bosons in the superfluid phase},\
  }\href {https://doi.org/10.1103/physrevb.84.014535} {\bibfield  {journal}
  {\bibinfo  {journal} {Phys. Rev. B}\ }\textbf {\bibinfo {volume} {84}},\
  \bibinfo {pages} {014535} (\bibinfo {year} {2011})}\BibitemShut {NoStop}%
\bibitem [{\citenamefont {Ran\c{c}on}\ and\ \citenamefont
  {Dupuis}(2011{\natexlab{a}})}]{Rancon11a}%
  \BibitemOpen
  \bibfield  {author} {\bibinfo {author} {\bibfnamefont {A.}~\bibnamefont
  {Ran\c{c}on}}\ and\ \bibinfo {author} {\bibfnamefont {N.}~\bibnamefont
  {Dupuis}},\ }\bibfield  {title} {\bibinfo {title} {{Nonperturbative
  renormalization group approach to the Bose-Hubbard model}},\ }\href
  {https://doi.org/10.1103/PhysRevB.83.172501} {\bibfield  {journal} {\bibinfo
  {journal} {Phys. Rev. B}\ }\textbf {\bibinfo {volume} {83}},\ \bibinfo
  {pages} {172501} (\bibinfo {year} {2011}{\natexlab{a}})}\BibitemShut
  {NoStop}%
\bibitem [{\citenamefont {Ran\c{c}on}\ and\ \citenamefont
  {Dupuis}(2011{\natexlab{b}})}]{Rancon11b}%
  \BibitemOpen
  \bibfield  {author} {\bibinfo {author} {\bibfnamefont {A.}~\bibnamefont
  {Ran\c{c}on}}\ and\ \bibinfo {author} {\bibfnamefont {N.}~\bibnamefont
  {Dupuis}},\ }\bibfield  {title} {\bibinfo {title} {{Nonperturbative
  renormalization group approach to strongly correlated lattice bosons}},\
  }\href {https://doi.org/10.1103/PhysRevB.84.174513} {\bibfield  {journal}
  {\bibinfo  {journal} {Phys. Rev. B}\ }\textbf {\bibinfo {volume} {84}},\
  \bibinfo {pages} {174513} (\bibinfo {year} {2011}{\natexlab{b}})}\BibitemShut
  {NoStop}%
\bibitem [{\citenamefont {Ran\c{c}on}\ and\ \citenamefont
  {Dupuis}(2012{\natexlab{a}})}]{Rancon12a}%
  \BibitemOpen
  \bibfield  {author} {\bibinfo {author} {\bibfnamefont {A.}~\bibnamefont
  {Ran\c{c}on}}\ and\ \bibinfo {author} {\bibfnamefont {N.}~\bibnamefont
  {Dupuis}},\ }\bibfield  {title} {\bibinfo {title} {{Quantum criticality of a
  Bose gas in an optical lattice near the Mott transition}},\ }\href
  {https://doi.org/10.1103/PhysRevA.85.011602} {\bibfield  {journal} {\bibinfo
  {journal} {Phys. Rev. A}\ }\textbf {\bibinfo {volume} {85}},\ \bibinfo
  {pages} {011602(R)} (\bibinfo {year} {2012}{\natexlab{a}})}\BibitemShut
  {NoStop}%
\bibitem [{\citenamefont {Ran\c{c}on}\ and\ \citenamefont
  {Dupuis}(2012{\natexlab{b}})}]{Rancon12d}%
  \BibitemOpen
  \bibfield  {author} {\bibinfo {author} {\bibfnamefont {A.}~\bibnamefont
  {Ran\c{c}on}}\ and\ \bibinfo {author} {\bibfnamefont {N.}~\bibnamefont
  {Dupuis}},\ }\bibfield  {title} {\bibinfo {title} {{Thermodynamics of a Bose
  gas near the superfluid--Mott-insulator transition}},\ }\href
  {https://doi.org/10.1103/PhysRevA.86.043624} {\bibfield  {journal} {\bibinfo
  {journal} {Phys. Rev. A}\ }\textbf {\bibinfo {volume} {86}},\ \bibinfo
  {pages} {043624} (\bibinfo {year} {2012}{\natexlab{b}})}\BibitemShut
  {NoStop}%
\bibitem [{\citenamefont {Ran\c{c}on}\ and\ \citenamefont
  {Dupuis}(2013)}]{Rancon13b}%
  \BibitemOpen
  \bibfield  {author} {\bibinfo {author} {\bibfnamefont {A.}~\bibnamefont
  {Ran\c{c}on}}\ and\ \bibinfo {author} {\bibfnamefont {N.}~\bibnamefont
  {Dupuis}},\ }\bibfield  {title} {\bibinfo {title} {{Quantum XY criticality in
  a two-dimensional Bose gas near the Mott transition}},\ }\href
  {https://doi.org/10.1209/0295-5075/104/16002} {\bibfield  {journal} {\bibinfo
   {journal} {Europhys. Lett.}\ }\textbf {\bibinfo {volume} {104}},\ \bibinfo
  {pages} {16002} (\bibinfo {year} {2013})}\BibitemShut {NoStop}%
\bibitem [{\citenamefont {Kopeć}(2024)}]{Kopec24}%
  \BibitemOpen
  \bibfield  {author} {\bibinfo {author} {\bibfnamefont {T.~K.}\ \bibnamefont
  {Kopeć}},\ }\bibfield  {title} {\bibinfo {title} {{Interaction-driven
  crossover from Mott insulator–superfluid criticality to Bose-Einstein
  condensation in a simple cubic lattice}},\ }\href
  {https://doi.org/10.1103/physrevb.110.144517} {\bibfield  {journal} {\bibinfo
   {journal} {Phys. Rev. B}\ }\textbf {\bibinfo {volume} {110}},\ \bibinfo
  {pages} {144517} (\bibinfo {year} {2024})}\BibitemShut {NoStop}%
\bibitem [{\citenamefont {Krauth}\ and\ \citenamefont
  {Trivedi}(1991)}]{Krauth91}%
  \BibitemOpen
  \bibfield  {author} {\bibinfo {author} {\bibfnamefont {W.}~\bibnamefont
  {Krauth}}\ and\ \bibinfo {author} {\bibfnamefont {N.}~\bibnamefont
  {Trivedi}},\ }\bibfield  {title} {\bibinfo {title} {{Mott and superfluid
  transitions in a strongly interacting lattice boson system}},\ }\href
  {https://doi.org/10.1209/0295-5075/14/7/003} {\bibfield  {journal} {\bibinfo
  {journal} {Europhys. Lett.}\ }\textbf {\bibinfo {volume} {14}},\ \bibinfo
  {pages} {627} (\bibinfo {year} {1991})}\BibitemShut {NoStop}%
\bibitem [{\citenamefont {Capogrosso-Sansone}\ \emph
  {et~al.}(2007)\citenamefont {Capogrosso-Sansone}, \citenamefont {Prokof'ev},\
  and\ \citenamefont {Svistunov}}]{Capogrosso07}%
  \BibitemOpen
  \bibfield  {author} {\bibinfo {author} {\bibfnamefont {B.}~\bibnamefont
  {Capogrosso-Sansone}}, \bibinfo {author} {\bibfnamefont {N.~V.}\ \bibnamefont
  {Prokof'ev}},\ and\ \bibinfo {author} {\bibfnamefont {B.~V.}\ \bibnamefont
  {Svistunov}},\ }\bibfield  {title} {\bibinfo {title} {{Phase diagram and
  thermodynamics of the three-dimensional Bose-Hubbard model}},\ }\href
  {https://doi.org/10.1103/PhysRevB.75.134302} {\bibfield  {journal} {\bibinfo
  {journal} {Phys. Rev. B}\ }\textbf {\bibinfo {volume} {75}},\ \bibinfo
  {pages} {134302} (\bibinfo {year} {2007})}\BibitemShut {NoStop}%
\bibitem [{\citenamefont {Capogrosso-Sansone}\ \emph
  {et~al.}(2008)\citenamefont {Capogrosso-Sansone}, \citenamefont {S\"oyler},
  \citenamefont {Prokof'ev},\ and\ \citenamefont {Svistunov}}]{Capogrosso08}%
  \BibitemOpen
  \bibfield  {author} {\bibinfo {author} {\bibfnamefont {B.}~\bibnamefont
  {Capogrosso-Sansone}}, \bibinfo {author} {\bibfnamefont {S.~G.}\ \bibnamefont
  {S\"oyler}}, \bibinfo {author} {\bibfnamefont {N.}~\bibnamefont
  {Prokof'ev}},\ and\ \bibinfo {author} {\bibfnamefont {B.}~\bibnamefont
  {Svistunov}},\ }\bibfield  {title} {\bibinfo {title} {{Monte Carlo study of
  the two-dimensional Bose-Hubbard model}},\ }\href
  {https://doi.org/10.1103/PhysRevA.77.015602} {\bibfield  {journal} {\bibinfo
  {journal} {Phys. Rev. A}\ }\textbf {\bibinfo {volume} {77}},\ \bibinfo
  {pages} {015602} (\bibinfo {year} {2008})}\BibitemShut {NoStop}%
\bibitem [{\citenamefont {Kato}\ and\ \citenamefont
  {Kawashima}(2009)}]{Kato09}%
  \BibitemOpen
  \bibfield  {author} {\bibinfo {author} {\bibfnamefont {Y.}~\bibnamefont
  {Kato}}\ and\ \bibinfo {author} {\bibfnamefont {N.}~\bibnamefont
  {Kawashima}},\ }\bibfield  {title} {\bibinfo {title} {{Quantum Monte Carlo
  method for the Bose-Hubbard model with harmonic confining potential}},\
  }\href {https://doi.org/10.1103/physreve.79.021104} {\bibfield  {journal}
  {\bibinfo  {journal} {Phys. Rev. E}\ }\textbf {\bibinfo {volume} {79}},\
  \bibinfo {pages} {021104} (\bibinfo {year} {2009})}\BibitemShut {NoStop}%
\bibitem [{\citenamefont {Pollet}\ and\ \citenamefont
  {Prokof'ev}(2012)}]{Pollet12}%
  \BibitemOpen
  \bibfield  {author} {\bibinfo {author} {\bibfnamefont {L.}~\bibnamefont
  {Pollet}}\ and\ \bibinfo {author} {\bibfnamefont {N.}~\bibnamefont
  {Prokof'ev}},\ }\bibfield  {title} {\bibinfo {title} {{Higgs Mode in a
  Two-Dimensional Superfluid}},\ }\href
  {https://doi.org/10.1103/PhysRevLett.109.010401} {\bibfield  {journal}
  {\bibinfo  {journal} {Phys. Rev. Lett.}\ }\textbf {\bibinfo {volume} {109}},\
  \bibinfo {pages} {010401} (\bibinfo {year} {2012})}\BibitemShut {NoStop}%
\bibitem [{\citenamefont {Ran\c{c}on}\ and\ \citenamefont
  {Dupuis}(2012{\natexlab{c}})}]{Rancon12b}%
  \BibitemOpen
  \bibfield  {author} {\bibinfo {author} {\bibfnamefont {A.}~\bibnamefont
  {Ran\c{c}on}}\ and\ \bibinfo {author} {\bibfnamefont {N.}~\bibnamefont
  {Dupuis}},\ }\bibfield  {title} {\bibinfo {title} {{Universal thermodynamics
  of a two-dimensional Bose gas}},\ }\href
  {https://doi.org/10.1103/PhysRevA.85.063607} {\bibfield  {journal} {\bibinfo
  {journal} {Phys. Rev. A}\ }\textbf {\bibinfo {volume} {85}},\ \bibinfo
  {pages} {063607} (\bibinfo {year} {2012}{\natexlab{c}})}\BibitemShut
  {NoStop}%
\bibitem [{not({\natexlab{a}})}]{not7}%
  \BibitemOpen
   \bibinfo {note} {For a discussion of the two-body contact
  in dilute weakly interacting gases, see
  Refs.~\cite{Tan08a,Tan08b,Tan08c,Olshanii03,Braaten08,Zhang09,Combescot09,Valiente12,Werner12a,Werner12}.}\BibitemShut
  {Stop}%
\bibitem [{\citenamefont {Bhateja}\ \emph {et~al.}()\citenamefont {Bhateja},
  \citenamefont {Dupuis},\ and\ \citenamefont {Rançon}}]{Bhateja25}%
  \BibitemOpen
  \bibfield  {author} {\bibinfo {author} {\bibfnamefont {M.}~\bibnamefont
  {Bhateja}}, \bibinfo {author} {\bibfnamefont {N.}~\bibnamefont {Dupuis}},\
  and\ \bibinfo {author} {\bibfnamefont {A.}~\bibnamefont {Rançon}},\
  }\bibfield  {title} {\bibinfo {title} {{Two-body contact of a Bose gas near
  the superfluid--Mott-insulator transition}},\ }\href@noop {} {\ }\Eprint
  {https://arxiv.org/abs/2501.14884} {arXiv:2501.14884} \BibitemShut {NoStop}%
\bibitem [{not({\natexlab{b}})}]{not4}%
  \BibitemOpen
    \bibinfo {note} {The two-body contact can be computed
  from the nonperturbative FRG but this requires to extend the approach of
  Refs.~\cite{Rancon11a,Rancon11b} by including an external source $h_\k(\tau)$
  that couples to the $\psi^*_\k(\tau)\psi_\k(\tau)$; see, e.g.
  Ref.~\cite{Rose15} for an example.}\BibitemShut {Stop}%
\bibitem [{\citenamefont {Coletta}\ \emph {et~al.}(2012)\citenamefont
  {Coletta}, \citenamefont {Laflorencie},\ and\ \citenamefont
  {Mila}}]{Coletta12}%
  \BibitemOpen
  \bibfield  {author} {\bibinfo {author} {\bibfnamefont {T.}~\bibnamefont
  {Coletta}}, \bibinfo {author} {\bibfnamefont {N.}~\bibnamefont
  {Laflorencie}},\ and\ \bibinfo {author} {\bibfnamefont {F.}~\bibnamefont
  {Mila}},\ }\bibfield  {title} {\bibinfo {title} {Semiclassical approach to
  ground-state properties of hard-core bosons in two dimensions},\ }\href
  {https://doi.org/10.1103/physrevb.85.104421} {\bibfield  {journal} {\bibinfo
  {journal} {Phys. Rev. B}\ }\textbf {\bibinfo {volume} {85}},\ \bibinfo
  {pages} {104421} (\bibinfo {year} {2012})}\BibitemShut {NoStop}%
\bibitem [{not({\natexlab{c}})}]{not3}%
  \BibitemOpen
   \bibinfo {note} {Note that the functional derivative with
  respect to $J^{(*)}_\r$ in Eqs.~(\ref{op}) should also act on the
  source-dependent order parameter $\phi^{(*)}_\r$. This, however, gives a
  vanishing contribution due to the relation
  $\phi_\r^{(*)}=\mean{\psi_\r^{(*)}}$.}\BibitemShut {Stop}%
\bibitem [{not({\natexlab{d}})}]{not1}%
  \BibitemOpen
   \bibinfo {note} {Note that the hopping matrix
  $t_{\r,\r'}$ is defined with the opposite sign in
  Ref.~\cite{Sengupta05}.}\BibitemShut {Stop}%
\bibitem [{\citenamefont {Sachdev}(2011)}]{Sachdev_book}%
  \BibitemOpen
  \bibfield  {author} {\bibinfo {author} {\bibfnamefont {S.}~\bibnamefont
  {Sachdev}},\ }\href@noop {} {\emph {\bibinfo {title} {{Quantum Phase
  Transitions}}}},\ \bibinfo {edition} {2nd}\ ed.\ (\bibinfo  {publisher}
  {Cambridge University Press},\ \bibinfo {address} {Cambridge, England},\
  \bibinfo {year} {2011})\BibitemShut {NoStop}%
\bibitem [{\citenamefont {Dupuis}(2023)}]{NDbook1}%
  \BibitemOpen
  \bibfield  {author} {\bibinfo {author} {\bibfnamefont {N.}~\bibnamefont
  {Dupuis}},\ }\href {https://doi.org/10.1142/q0409} {\emph {\bibinfo {title}
  {Field Theory of Condensed Matter and Ultracold Gases}}},\ Vol.~\bibinfo
  {volume} {1}\ (\bibinfo  {publisher} {World Scientific Europe, London},\
  \bibinfo {year} {2023})\BibitemShut {NoStop}%
\bibitem [{not({\natexlab{e}})}]{not2}%
  \BibitemOpen
  \bibinfo {note} {The volume in the numerator comes from
  the normalization of the momentum distribution: $\sum_\k
  n_\k=n\calV$.}\BibitemShut {Stop}%
\bibitem [{\citenamefont {Lee}\ \emph {et~al.}(1957)\citenamefont {Lee},
  \citenamefont {Huang},\ and\ \citenamefont {Yang}}]{Lee57b}%
  \BibitemOpen
  \bibfield  {author} {\bibinfo {author} {\bibfnamefont {T.~D.}\ \bibnamefont
  {Lee}}, \bibinfo {author} {\bibfnamefont {K.}~\bibnamefont {Huang}},\ and\
  \bibinfo {author} {\bibfnamefont {C.~N.}\ \bibnamefont {Yang}},\ }\bibfield
  {title} {\bibinfo {title} {{Eigenvalues and Eigenfunctions of a Bose System
  of Hard Spheres and its Low-Temperature Properties}},\ }\href
  {https://doi.org/10.1103/PhysRev.106.1135} {\bibfield  {journal} {\bibinfo
  {journal} {Phys. Rev.}\ }\textbf {\bibinfo {volume} {106}},\ \bibinfo {pages}
  {1135} (\bibinfo {year} {1957})}\BibitemShut {NoStop}%
\bibitem [{\citenamefont {Dalfovo}\ \emph {et~al.}(1999)\citenamefont
  {Dalfovo}, \citenamefont {Giorgini}, \citenamefont {Pitaevskii},\ and\
  \citenamefont {Stringari}}]{Dalfovo99}%
  \BibitemOpen
  \bibfield  {author} {\bibinfo {author} {\bibfnamefont {F.}~\bibnamefont
  {Dalfovo}}, \bibinfo {author} {\bibfnamefont {S.}~\bibnamefont {Giorgini}},
  \bibinfo {author} {\bibfnamefont {L.~P.}\ \bibnamefont {Pitaevskii}},\ and\
  \bibinfo {author} {\bibfnamefont {S.}~\bibnamefont {Stringari}},\ }\bibfield
  {title} {\bibinfo {title} {{Theory of Bose-Einstein condensation in trapped
  gases}},\ }\href {https://doi.org/10.1103/RevModPhys.71.463} {\bibfield
  {journal} {\bibinfo  {journal} {Rev. Mod. Phys.}\ }\textbf {\bibinfo {volume}
  {71}},\ \bibinfo {pages} {463} (\bibinfo {year} {1999})}\BibitemShut
  {NoStop}%
\bibitem [{\citenamefont {Ran\c{c}on}(2014)}]{Rancon14a}%
  \BibitemOpen
  \bibfield  {author} {\bibinfo {author} {\bibfnamefont {A.}~\bibnamefont
  {Ran\c{c}on}},\ }\bibfield  {title} {\bibinfo {title} {{Nonperturbative
  renormalization group approach to quantum $XY$ spin models}},\ }\href
  {https://doi.org/10.1103/PhysRevB.89.214418} {\bibfield  {journal} {\bibinfo
  {journal} {Phys. Rev. B}\ }\textbf {\bibinfo {volume} {89}},\ \bibinfo
  {pages} {214418} (\bibinfo {year} {2014})}\BibitemShut {NoStop}%
\bibitem [{\citenamefont {Ran\c{c}on}(2022)}]{Rancon22}%
  \BibitemOpen
  \bibfield  {author} {\bibinfo {author} {\bibfnamefont {A.}~\bibnamefont
  {Ran\c{c}on}},\ }\bibfield  {title} {\bibinfo {title} {{Comment on
  “Universal and Non-Universal Correction Terms of Bose Gases in Dilute
  Region: A Quantum Monte Carlo Study”}},\ }\href
  {https://doi.org/10.7566/jpsj.91.066001} {\bibfield  {journal} {\bibinfo
  {journal} {J. Phys. Soc. Jpn.}\ }\textbf {\bibinfo {volume} {91}},\ \bibinfo
  {pages} {066001} (\bibinfo {year} {2022})}\BibitemShut {NoStop}%
\bibitem [{\citenamefont {Braaten}\ and\ \citenamefont
  {Hammer}(2006)}]{Braaten06}%
  \BibitemOpen
  \bibfield  {author} {\bibinfo {author} {\bibfnamefont {E.}~\bibnamefont
  {Braaten}}\ and\ \bibinfo {author} {\bibfnamefont {H.-W.}\ \bibnamefont
  {Hammer}},\ }\bibfield  {title} {\bibinfo {title} {{Universality in few-body
  systems with large scattering length}},\ }\href {https://doi.org/DOI:
  10.1016/j.physrep.2006.03.001} {\bibfield  {journal} {\bibinfo  {journal}
  {Phys. Rep.}\ }\textbf {\bibinfo {volume} {428}},\ \bibinfo {pages} {259 }
  (\bibinfo {year} {2006})}\BibitemShut {NoStop}%
\bibitem [{\citenamefont {Chen}\ \emph {et~al.}(2014)\citenamefont {Chen},
  \citenamefont {Jiang}, \citenamefont {Guan},\ and\ \citenamefont
  {Zhou}}]{Chen14}%
  \BibitemOpen
  \bibfield  {author} {\bibinfo {author} {\bibfnamefont {Y.-Y.}\ \bibnamefont
  {Chen}}, \bibinfo {author} {\bibfnamefont {Y.-Z.}\ \bibnamefont {Jiang}},
  \bibinfo {author} {\bibfnamefont {X.-W.}\ \bibnamefont {Guan}},\ and\
  \bibinfo {author} {\bibfnamefont {Q.}~\bibnamefont {Zhou}},\ }\bibfield
  {title} {\bibinfo {title} {Critical behaviours of contact near phase
  transitions},\ }\href {https://doi.org/10.1038/ncomms6140} {\bibfield
  {journal} {\bibinfo  {journal} {Nat. Commun.}\ }\textbf {\bibinfo {volume}
  {5}},\ \bibinfo {pages} {5140} (\bibinfo {year} {2014})}\BibitemShut
  {NoStop}%
\bibitem [{\citenamefont {Tan}(2008{\natexlab{a}})}]{Tan08a}%
  \BibitemOpen
  \bibfield  {author} {\bibinfo {author} {\bibfnamefont {S.}~\bibnamefont
  {Tan}},\ }\bibfield  {title} {\bibinfo {title} {{Energetics of a strongly
  correlated Fermi gas}},\ }\href
  {https://doi.org/https://doi.org/10.1016/j.aop.2008.03.004} {\bibfield
  {journal} {\bibinfo  {journal} {Ann. Phys.}\ }\textbf {\bibinfo {volume}
  {323}},\ \bibinfo {pages} {2952} (\bibinfo {year}
  {2008}{\natexlab{a}})}\BibitemShut {NoStop}%
\bibitem [{\citenamefont {Tan}(2008{\natexlab{b}})}]{Tan08b}%
  \BibitemOpen
  \bibfield  {author} {\bibinfo {author} {\bibfnamefont {S.}~\bibnamefont
  {Tan}},\ }\bibfield  {title} {\bibinfo {title} {{Generalized virial theorem
  and pressure relation for a strongly correlated Fermi gas}},\ }\href
  {https://doi.org/https://doi.org/10.1016/j.aop.2008.03.003} {\bibfield
  {journal} {\bibinfo  {journal} {Ann. Phys.}\ }\textbf {\bibinfo {volume}
  {323}},\ \bibinfo {pages} {2987} (\bibinfo {year}
  {2008}{\natexlab{b}})}\BibitemShut {NoStop}%
\bibitem [{\citenamefont {Tan}(2008{\natexlab{c}})}]{Tan08c}%
  \BibitemOpen
  \bibfield  {author} {\bibinfo {author} {\bibfnamefont {S.}~\bibnamefont
  {Tan}},\ }\bibfield  {title} {\bibinfo {title} {{Large momentum part of a
  strongly correlated Fermi gas}},\ }\href
  {https://doi.org/https://doi.org/10.1016/j.aop.2008.03.005} {\bibfield
  {journal} {\bibinfo  {journal} {Ann. Phys.}\ }\textbf {\bibinfo {volume}
  {323}},\ \bibinfo {pages} {2971} (\bibinfo {year}
  {2008}{\natexlab{c}})}\BibitemShut {NoStop}%
\bibitem [{\citenamefont {Olshanii}\ and\ \citenamefont
  {Dunjko}(2003)}]{Olshanii03}%
  \BibitemOpen
  \bibfield  {author} {\bibinfo {author} {\bibfnamefont {M.}~\bibnamefont
  {Olshanii}}\ and\ \bibinfo {author} {\bibfnamefont {V.}~\bibnamefont
  {Dunjko}},\ }\bibfield  {title} {\bibinfo {title} {{Short-distance
  correlation properties of the Lieb-Liniger system and momentum distributions
  of trapped one-dimensional atomic gases}},\ }\href
  {https://doi.org/10.1103/physrevlett.91.090401} {\bibfield  {journal}
  {\bibinfo  {journal} {Phys. Rev. Lett.}\ }\textbf {\bibinfo {volume} {91}},\
  \bibinfo {pages} {090401} (\bibinfo {year} {2003})}\BibitemShut {NoStop}%
\bibitem [{\citenamefont {Braaten}\ and\ \citenamefont
  {Platter}(2008)}]{Braaten08}%
  \BibitemOpen
  \bibfield  {author} {\bibinfo {author} {\bibfnamefont {E.}~\bibnamefont
  {Braaten}}\ and\ \bibinfo {author} {\bibfnamefont {L.}~\bibnamefont
  {Platter}},\ }\bibfield  {title} {\bibinfo {title} {{Exact Relations for a
  Strongly Interacting Fermi Gas from the Operator Product Expansion}},\ }\href
  {https://doi.org/10.1103/PhysRevLett.100.205301} {\bibfield  {journal}
  {\bibinfo  {journal} {Phys. Rev. Lett.}\ }\textbf {\bibinfo {volume} {100}},\
  \bibinfo {pages} {205301} (\bibinfo {year} {2008})}\BibitemShut {NoStop}%
\bibitem [{\citenamefont {Zhang}\ and\ \citenamefont
  {Leggett}(2009)}]{Zhang09}%
  \BibitemOpen
  \bibfield  {author} {\bibinfo {author} {\bibfnamefont {S.}~\bibnamefont
  {Zhang}}\ and\ \bibinfo {author} {\bibfnamefont {A.~J.}\ \bibnamefont
  {Leggett}},\ }\bibfield  {title} {\bibinfo {title} {{Universal properties of
  the ultracold Fermi gas}},\ }\href
  {https://doi.org/10.1103/PhysRevA.79.023601} {\bibfield  {journal} {\bibinfo
  {journal} {Phys. Rev. A}\ }\textbf {\bibinfo {volume} {79}},\ \bibinfo
  {pages} {023601} (\bibinfo {year} {2009})}\BibitemShut {NoStop}%
\bibitem [{\citenamefont {Combescot}\ \emph {et~al.}(2009)\citenamefont
  {Combescot}, \citenamefont {Alzetto},\ and\ \citenamefont
  {Leyronas}}]{Combescot09}%
  \BibitemOpen
  \bibfield  {author} {\bibinfo {author} {\bibfnamefont {R.}~\bibnamefont
  {Combescot}}, \bibinfo {author} {\bibfnamefont {F.}~\bibnamefont {Alzetto}},\
  and\ \bibinfo {author} {\bibfnamefont {X.}~\bibnamefont {Leyronas}},\
  }\bibfield  {title} {\bibinfo {title} {Particle distribution tail and related
  energy formula},\ }\href {https://doi.org/10.1103/PhysRevA.79.053640}
  {\bibfield  {journal} {\bibinfo  {journal} {Phys. Rev. A}\ }\textbf {\bibinfo
  {volume} {79}},\ \bibinfo {pages} {053640} (\bibinfo {year}
  {2009})}\BibitemShut {NoStop}%
\bibitem [{\citenamefont {Valiente}\ \emph {et~al.}(2012)\citenamefont
  {Valiente}, \citenamefont {Zinner},\ and\ \citenamefont
  {M{\o}lmer}}]{Valiente12}%
  \BibitemOpen
  \bibfield  {author} {\bibinfo {author} {\bibfnamefont {M.}~\bibnamefont
  {Valiente}}, \bibinfo {author} {\bibfnamefont {N.~T.}\ \bibnamefont
  {Zinner}},\ and\ \bibinfo {author} {\bibfnamefont {K.}~\bibnamefont
  {M{\o}lmer}},\ }\bibfield  {title} {\bibinfo {title} {{Universal properties
  of Fermi gases in arbitrary dimensions}},\ }\href
  {https://doi.org/10.1103/physreva.86.043616} {\bibfield  {journal} {\bibinfo
  {journal} {Phys. Rev. A}\ }\textbf {\bibinfo {volume} {86}},\ \bibinfo
  {pages} {043616} (\bibinfo {year} {2012})}\BibitemShut {NoStop}%
\bibitem [{\citenamefont {Werner}\ and\ \citenamefont
  {Castin}(2012{\natexlab{a}})}]{Werner12a}%
  \BibitemOpen
  \bibfield  {author} {\bibinfo {author} {\bibfnamefont {F.}~\bibnamefont
  {Werner}}\ and\ \bibinfo {author} {\bibfnamefont {Y.}~\bibnamefont
  {Castin}},\ }\bibfield  {title} {\bibinfo {title} {General relations for
  quantum gases in two and three dimensions: Two-component fermions},\ }\href
  {https://doi.org/10.1103/PhysRevA.86.013626} {\bibfield  {journal} {\bibinfo
  {journal} {Phys. Rev. A}\ }\textbf {\bibinfo {volume} {86}},\ \bibinfo
  {pages} {013626} (\bibinfo {year} {2012}{\natexlab{a}})}\BibitemShut
  {NoStop}%
\bibitem [{\citenamefont {Werner}\ and\ \citenamefont
  {Castin}(2012{\natexlab{b}})}]{Werner12}%
  \BibitemOpen
  \bibfield  {author} {\bibinfo {author} {\bibfnamefont {F.}~\bibnamefont
  {Werner}}\ and\ \bibinfo {author} {\bibfnamefont {Y.}~\bibnamefont
  {Castin}},\ }\bibfield  {title} {\bibinfo {title} {General relations for
  quantum gases in two and three dimensions. ii. bosons and mixtures},\ }\href
  {https://doi.org/10.1103/PhysRevA.86.053633} {\bibfield  {journal} {\bibinfo
  {journal} {Phys. Rev. A}\ }\textbf {\bibinfo {volume} {86}},\ \bibinfo
  {pages} {053633} (\bibinfo {year} {2012}{\natexlab{b}})}\BibitemShut
  {NoStop}%
\bibitem [{\citenamefont {Rose}\ \emph {et~al.}(2015)\citenamefont {Rose},
  \citenamefont {L\'eonard},\ and\ \citenamefont {Dupuis}}]{Rose15}%
  \BibitemOpen
  \bibfield  {author} {\bibinfo {author} {\bibfnamefont {F.}~\bibnamefont
  {Rose}}, \bibinfo {author} {\bibfnamefont {F.}~\bibnamefont {L\'eonard}},\
  and\ \bibinfo {author} {\bibfnamefont {N.}~\bibnamefont {Dupuis}},\
  }\bibfield  {title} {\bibinfo {title} {{Higgs amplitude mode in the vicinity
  of a $(2+1)$-dimensional quantum critical point: A nonperturbative
  renormalization-group approach}},\ }\href
  {https://doi.org/10.1103/PhysRevB.91.224501} {\bibfield  {journal} {\bibinfo
  {journal} {Phys. Rev. B}\ }\textbf {\bibinfo {volume} {91}},\ \bibinfo
  {pages} {224501} (\bibinfo {year} {2015})}\BibitemShut {NoStop}%
\end{thebibliography}

\end{document}